\documentclass[12pt]{iopart}
\expandafter\let\csname equation*\endcsname\relax
\expandafter\let\csname endequation*\endcsname\relax
\usepackage{graphicx, caption}
\usepackage{amsmath}
\usepackage{booktabs}
\usepackage{float}
\usepackage{nicematrix}
\usepackage{blindtext, rotating}
\usepackage{titlesec}

\begin{document}

\title{Microscopic origin of local electric polarization in NiPS\textsubscript{3}}

\author{Hyeon Jung Kim$^1$ and Ki-Seok Kim$^1$$^,$$^2$}
\address{$^1$Department of Physics, POSTECH, Pohang, Gyeongbuk 37673, Korea}
\address{$^2$Asia Pacific Center for Theoretical Physics (APCTP), Pohang, Gyeongbuk 37673, Korea}

\ead{hkim7218@postech.ac.kr and tkfkd@postech.ac.kr}
\vspace{10pt}

\begin{abstract}
Recently, Zhang-Rice triplet to singlet excitations have been measured experimentally and verified numerically in a van der Waals antiferromagnet NiPS\textsubscript{3}, which reveals a collective local change of an electronic structure. In particular, such numerical simulations predicted that these electronic excitations occur simultaneously with local electric polarizations. In this study, we uncover the microscopic origin of this local electric polarization in the Zhang-Rice triplet to singlet excitation. Our lattice-model calculation predicts that the electric polarization can be controlled by applied magnetic fields, where the atomic spin-orbit coupling plays an important role. We speculate emergence of real space Berry curvature to describe the electric polarization in this strongly correlated system.
\end{abstract}

\section{Introduction}

It is an interesting feature of strong-correlation physics the emergence of low-energy multi-particle composite basis in an effective Hamiltonian. Zhang-Rice singlet formation in 3d transition metal oxides \cite{Zhang_Rice_Singlet_CuO}, effective Kitaev-type models in 5d spin-orbit coupled metal oxides \cite{Kitaev-type_Models_IrO}, and emergent spin currents and magnetoelectric effects in noncollinear magnets \cite{Spin_Current_Polarization} are such prominent examples in actual materials. In particular, the Zhang-Rice singlet state in CuO has been measured spectroscopically \cite{Zhang_Rice_Singlet_CuO_Observation}.

Recently, Zhang-Rice triplet to singlet excitations have been measured in a van der Waals antiferromagnet NiPS\textsubscript{3} \cite{brec1986review,grasso1988electronic,foot1980optical,kurita1989band1,kurita1989band2} based on several spectroscopic measurements combined with numerical simulations, which reveals a collective local change of an electronic structure \cite{NiPS3_Nature}. In particular, such numerical simulations predicted that these spin-coherent exciton excitations occur simultaneously with local electric polarizations. Unfortunately, it remains unrevealed the microscopic mechanism of the local electric polarization coupled with such Zhang-Rice triplet to singlet excitations in NiPS\textsubscript{3}.

In this study, we describe the Zhang-Rice triplet to singlet exciton state based on a microscopic lattice Hamiltonian for the NiS\textsubscript{6} unit structure. Furthermore, we reveal that this spin-exciton state gives rise to the local electric polarization, taking into account hopping from the centered NiS\textsubscript{6} unit to three nearest neighbor NiS\textsubscript{6} units in an antiferromagnetic phase of NiPS\textsubscript{3}. In particular, our lattice-model calculation predicts that the electric polarization can be controlled by applied magnetic fields. Here, we reveal the role of the atomic spin-orbit coupling. In conclusion, we speculate emergence of local Berry curvature to describe the electric polarization in this strongly correlated system. 

Although NiPS\textsubscript{3} belongs to the transition metal trichalcogenides family, which includes MnPS\textsubscript{3}, FePS\textsubscript{3}, CoPS\textsubscript{3}, and NiPSe\textsubscript{3}, its two distinctive features, the Zhang-Rice triplet nature and an anti-ferromagnetic zig-zag pattern give rise to the local electric polarization beyond other transition metal trichalcogenides. For example, we can verify that the electric polarization of MnPS\textsubscript{3} and FePS\textsubscript{3} cannot be found by the same procedure of the present study, where the antiferromagnetic ordering pattern does not match \cite{gu2019ni}. CoPS\textsubscript{3} has three holes in the CoS\textsubscript{6} cluster \cite{CoPs3}. This ground state does not allow the Zhang-rice triplet state, where only two holes are necessary. NiPSe\textsubscript{3} has two holes in the NiSe\textsubscript{6} cluster, but Se has the 4p orbital, whose hole energy is lower than that of the 3d orbital of Nickel. The single cluster of this material cannot make a self-doped system, where one hole of Ni$^{2+}$ has to be transferred to the nearby ligands in the ground state.

\section{Zhang-Rice triplet to singlet excitations in NiPS\textsubscript{3}}
\begin{figure}[ht]
\centering
\includegraphics[scale=0.26]{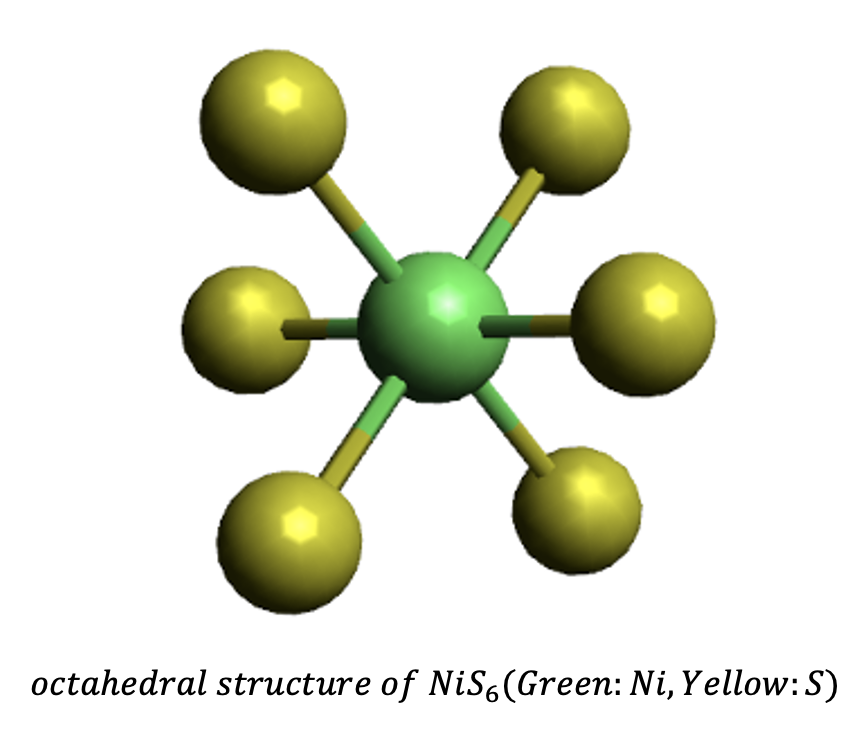}
\includegraphics[scale=0.24]{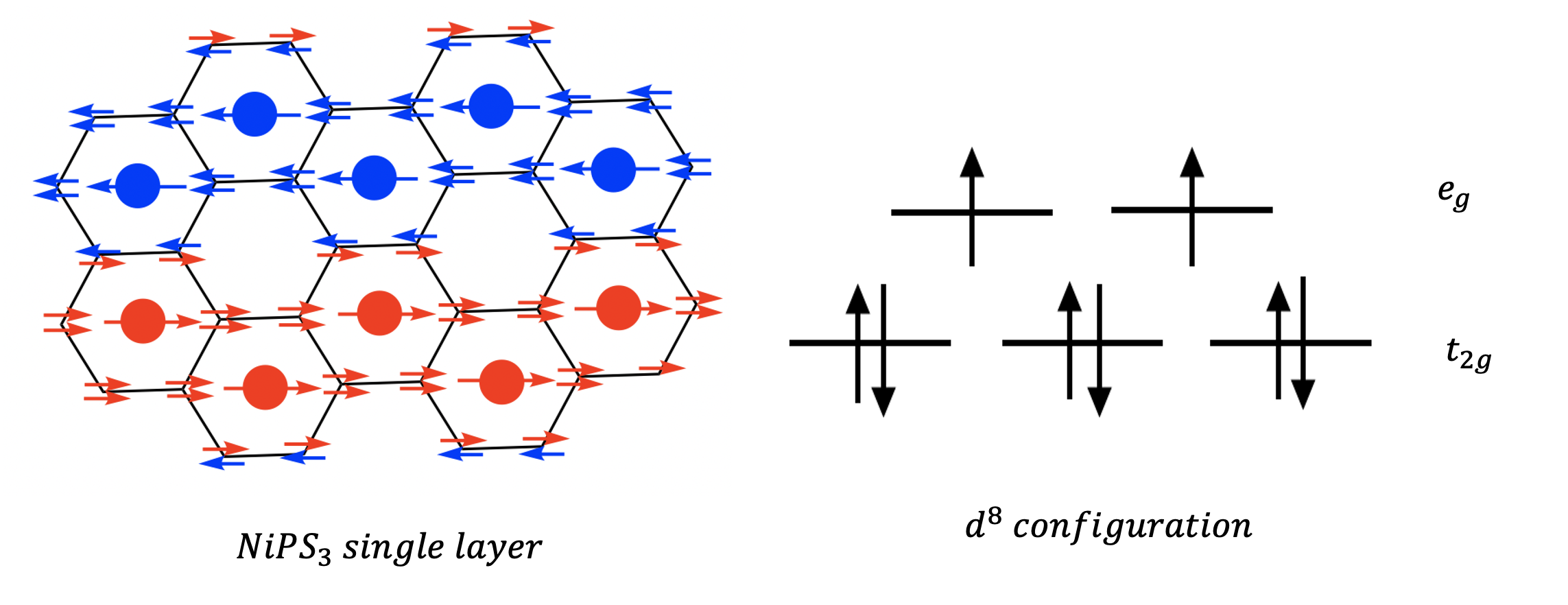}
\caption{Structure of NiPS\textsubscript{3}. Red and blue circles represent Ni ions with a zigzag antiferromagnetic ordering pattern. Vertices denote S ions, where arrows show spin up and down.}
\label{fig:structures}
\end{figure}

\begin{figure}[ht]
\centering
\includegraphics[scale=0.2]{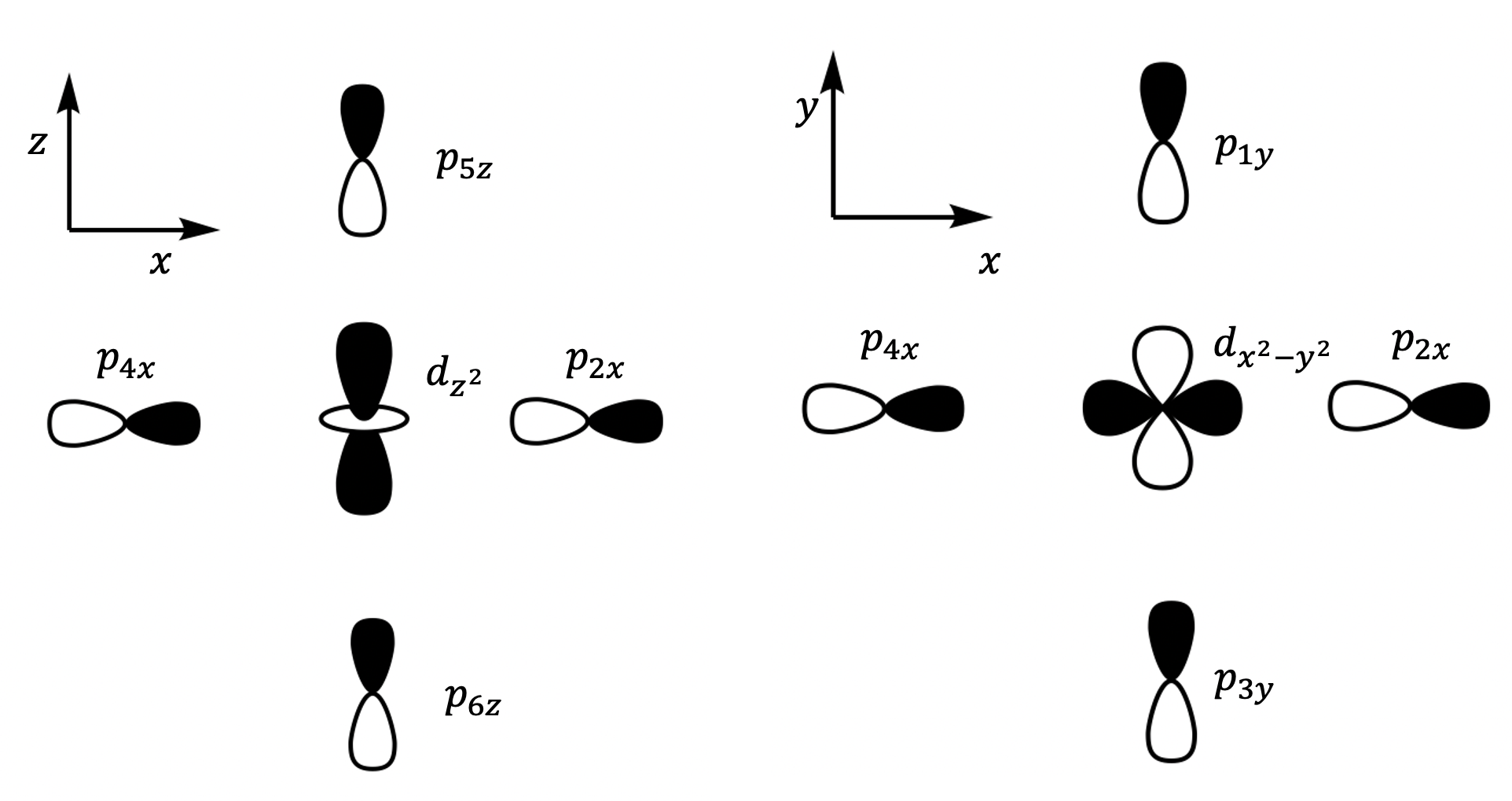}
\caption{Labeling of orbitals.
\label{fig:singlelattice}}
\end{figure}

NiPS\textsubscript{3} is a quasi-two dimensional van der Waals material \cite{coleman2011two,kuo2016exfoliation} that shows a zigzag anti-ferromagnetic ordering pattern \cite{sivadas2015magnetic,wildes2015magnetic,chittari2016electronic}. See Figure \ref{fig:structures}. The Ni ion has a d\textsuperscript{8} configuration in an octahedral structure of S. There are two holes in the e\textsubscript{g} orbital. A recent study showed that the ground state of the NiS\textsubscript{6} single unit is given by the fact that one hole in the Ni ion is self-doped to the ligand p-orbital \cite{kim2018charge}. See Figure \ref{fig:singlelattice} for labeling of ligands.

To construct an effective lattice Hamiltonian for the NiS\textsubscript{6} single unit, let $d^\dagger_{m\sigma}$ be a creation operator of a hole in the Ni ion's m-orbital ($x^2-y^2, 3z^2-r^2$) with spin $\sigma$ and $p^\dagger_{nm\sigma}$ be a creation operator of a hole in the n-th ligand's m-orbital with spin $\sigma$. We define a vacuum state as the absence of holes in the single NiS\textsubscript{6} unit. Then, we propose an effective lattice Hamiltonian for the NiS\textsubscript{6} single unit as follows 

\begin{equation}
\resizebox{\textwidth}{!}{$%
\begin{aligned}
H_K=&U\sum_{m}d^\dagger_{m\uparrow}d_{m\uparrow}d^\dagger_{m\downarrow}d_{m\downarrow}+(U-2J)\sum_{m\neq m'}d^\dagger_{m\uparrow} d_{m\uparrow} d^\dagger_{m'\downarrow} d_{m'\downarrow}+(U-3J)\sum_{m<m',\sigma}d^\dagger_{m\sigma} d_{m\sigma} d^\dagger_{m'\sigma}d_{m'\sigma}\\
 &-J\sum_{m\neq m'}d^\dagger_{m\uparrow}d_{m\downarrow}d^\dagger_{m'\downarrow}d_{m'\uparrow}+J\sum_{m\neq m'}d^\dagger_{m\uparrow}d^\dagger_{m\downarrow}d_{m'\downarrow}d_{m'\uparrow}+\epsilon_d\sum_{m\sigma}d^{\dagger}_{m\sigma}d_{m\sigma}+\epsilon_p\sum_{m\sigma}p^{\dagger}_{m\sigma}p_{m\sigma}
 \end{aligned}$%
 }
 \label{eq:KM}
\end{equation}

 \begin{equation}
 \resizebox{\textwidth}{!}{$%
\begin{aligned}
H_t= t \sum_{\sigma} \Biggl\{\frac{\sqrt{3}}{2}d^\dagger_{x^2-y^2\sigma}(p_{1y\sigma}-p_{2x\sigma}-p_{3y\sigma}+p_{4x\sigma})+d^\dagger_{z^2\sigma} \left[\frac{1}{2}(p_{1y\sigma}+p_{2x\sigma}-p_{3y\sigma}-p_{4x\sigma})-p_{5z\sigma}+p_{6z\sigma}\right]\Biggr\} + h.c.
 \end{aligned}$%
 }
\label{eq:hopping}
\end{equation}

$H_K$ (\ref{eq:KM}) is an effective local-interaction Hamiltonian \cite{georges2013strong} and $H_t$ (\ref{eq:hopping}) is an effective hopping Hamiltonian \cite{slater1954simplified}. Here, $U$ is an on-site Coulomb interaction, $J$ is an effective spin-exchange interaction, and $t$ is a hopping parameter. Recalling that the ground state of NiPS\textsubscript{3} is given by self-doping, we obtain two constraints for parameters as $U-3J+2\epsilon_d>\epsilon_d+\epsilon_p$ and $\epsilon_p>\epsilon_d$, where $\epsilon_d$ ($\epsilon_p$) is an energy level of the Ni (ligand) ion hole.

Based on this effective Hamiltonian $H=H_K+H_t$, we perform the perturbation analysis around the ground state of the NiS\textsubscript{6} single unit. Let $P_0$ be a projection operator to $d^\dagger p^\dagger$, i.e., $P_0=|dp\rangle\langle dp|$, and $P_1$ be a projection operator to $d^\dagger d^\dagger$ and $p^\dagger p^\dagger$, i.e., $P_1=|pp\rangle\langle pp| + |dd\rangle\langle dd|$. Then, $P_0+P_1$ makes a complete Hilbert space for the effective Hamiltonian $H$. Using these projection operators, we represent the effective Hamiltonian in the following way

\begin{equation}
H \doteq
\begin{pmatrix}  P_0HP_0  &  P_0HP_1 \\ P_1HP_0  &  P_1HP_1 \end{pmatrix} .
 \label{eq:mt}
\end{equation}
As a result, the second-order perturbation theory with respect to $H_t$ gives rise to
\begin{equation}
\left(P_0HP_0+P_0HP_1\frac{ 1 }{E-P_1HP_1}P_1HP_0\right)P_0\Psi = EP_0\Psi .
\end{equation}
Here, $P_0\Psi$ is a state, projected to the $d^\dagger p^\dagger$ Hilbert space with an eigenvalue $E$.

It is straightforward to solve this Schrodinger equation although the procedure is rather tedious. When we write $|0\rangle$ as a vacuum state, the lowest energy eigenstate is given by a Zhang-Rice triplet state

\begin{equation}
\resizebox{\textwidth}{!}{$%
\begin{aligned}
|ZRT, \sigma \rangle =&\frac{1}{\sqrt{6}}\Biggl\{-\frac{\sqrt{3}}{2}d^\dagger_{z^2\sigma} (p^\dagger_{1y\sigma}-p^\dagger_{2x\sigma}-p^\dagger_{3y\sigma}+p^\dagger_{4x\sigma})+d^\dagger_{x^2-y^2\sigma} \left[\frac{1}{2}(p^\dagger_{1y\sigma}+p^\dagger_{2x\sigma}-p^\dagger_{3y\sigma}-p^\dagger_{4x\sigma})-p^\dagger_{5z\sigma}+p^\dagger_{6z\sigma}\right] \Biggr\}|0\rangle,\\
|ZRT, 0 \rangle =\frac{1}{2\sqrt{3}}&\Biggl\{ \left(-\frac{\sqrt{3}}{2}d^\dagger_{z^2\uparrow}(p^\dagger_{1y\downarrow}-p^\dagger_{2x\downarrow}-p^\dagger_{3y\downarrow}+p^\dagger_{4x\downarrow}) +d^\dagger_{x^2-y^2\downarrow}\left[\frac{1}{2}(p^\dagger_{1y\downarrow}+p^\dagger_{2x\downarrow}-p^\dagger_{3y\downarrow}-p^\dagger_{4x\downarrow}) -p^\dagger_{5z\downarrow}+p^\dagger_{6z\downarrow}\right]\right)\\
&+\left(-\frac{\sqrt{3}}{2}d^\dagger_{z^2\downarrow} (p^\dagger_{1y\uparrow}-p^\dagger_{2x\uparrow}-p^\dagger_{3y\uparrow}+p^\dagger_{4x\uparrow}) +d^\dagger_{x^2-y^2\downarrow}\left[\frac{1}{2}(p^\dagger_{1y\uparrow}+p^\dagger_{2x\uparrow}-p^\dagger_{3y\uparrow}-p^\dagger_{4x\uparrow}) -p^\dagger_{5z\uparrow}+p^\dagger_{6z\uparrow}\right]\right)\Biggr\}|0\rangle
 \end{aligned}$%
 }
\label{eq:zrt}
\end{equation}
\\
with an eigenenergy $E_t=\epsilon_p+\epsilon_d-\frac{6t^2(U-3J)}{(\epsilon_p-\epsilon_d)(U-3J-\epsilon_p+\epsilon_d)}$. This lowest energy can be decomposed into $\frac{6t^2}{(\epsilon_p-\epsilon_d)}$ and $\frac{6t^2}{(U-3J-\epsilon_p+\epsilon_d)}$. It means that hybridization via hopping between the transition metal d-orbital and the ligand p-orbitals lowers the energy eigenvalue, forming a collective local spin-triplet state in the NiPS\textsubscript{6} unit.

The next excited eigenstate is given by a Zhang-Rice singlet state

\begin{equation}
\resizebox{\textwidth}{!}{$%
\begin{aligned}
|ZRS\rangle=\frac{1}{2\sqrt{3}}&\Biggl\{\left(\frac{\sqrt{3}}{2}d^\dagger_{z^2\uparrow}(p^\dagger_{1y\downarrow}-p^\dagger_{2x\downarrow}-p^\dagger_{3y\downarrow}+p^\dagger_{4x\downarrow})+d^\dagger_{x^2-y^2\uparrow}\left[\frac{1}{2}(p^\dagger_{1y\downarrow}+p^\dagger_{2x\downarrow}-p^\dagger_{3y\downarrow}-p^\dagger_{4x\downarrow})-p^\dagger_{5z\downarrow}+p^\dagger_{6z\downarrow}\right]\right)\\
&-\left(\frac{\sqrt{3}}{2}d^\dagger_{z^2\downarrow}(p^\dagger_{1y\uparrow}-p^\dagger_{2x\uparrow}-p^\dagger_{3y\uparrow}+p^\dagger_{4x\uparrow})+d^\dagger_{x^2-y^2\downarrow}\left[\frac{1}{2}(p^\dagger_{1y\uparrow}+p^\dagger_{2x\uparrow}-p^\dagger_{3y\uparrow}-p^\dagger_{4x\uparrow})-p^\dagger_{5z\uparrow}+p^\dagger_{6z\uparrow}\right]\right)\Biggr\}|0\rangle
 \end{aligned}$%
 }
\label{eq:zrs}
\end{equation}
\\
with an eigenenergy $E_s=\epsilon_p+\epsilon_d-\frac{6t^2(U-J)}{(\epsilon_p-\epsilon_d)(U-J-\epsilon_p+\epsilon_d)}$. We point out the $2 J$ difference in the energy eigenvalue. See the appendix for more details.
\section{Two lattice unit calculation for local electric polarization}
\begin{figure}[ht]
\includegraphics[scale=0.2]{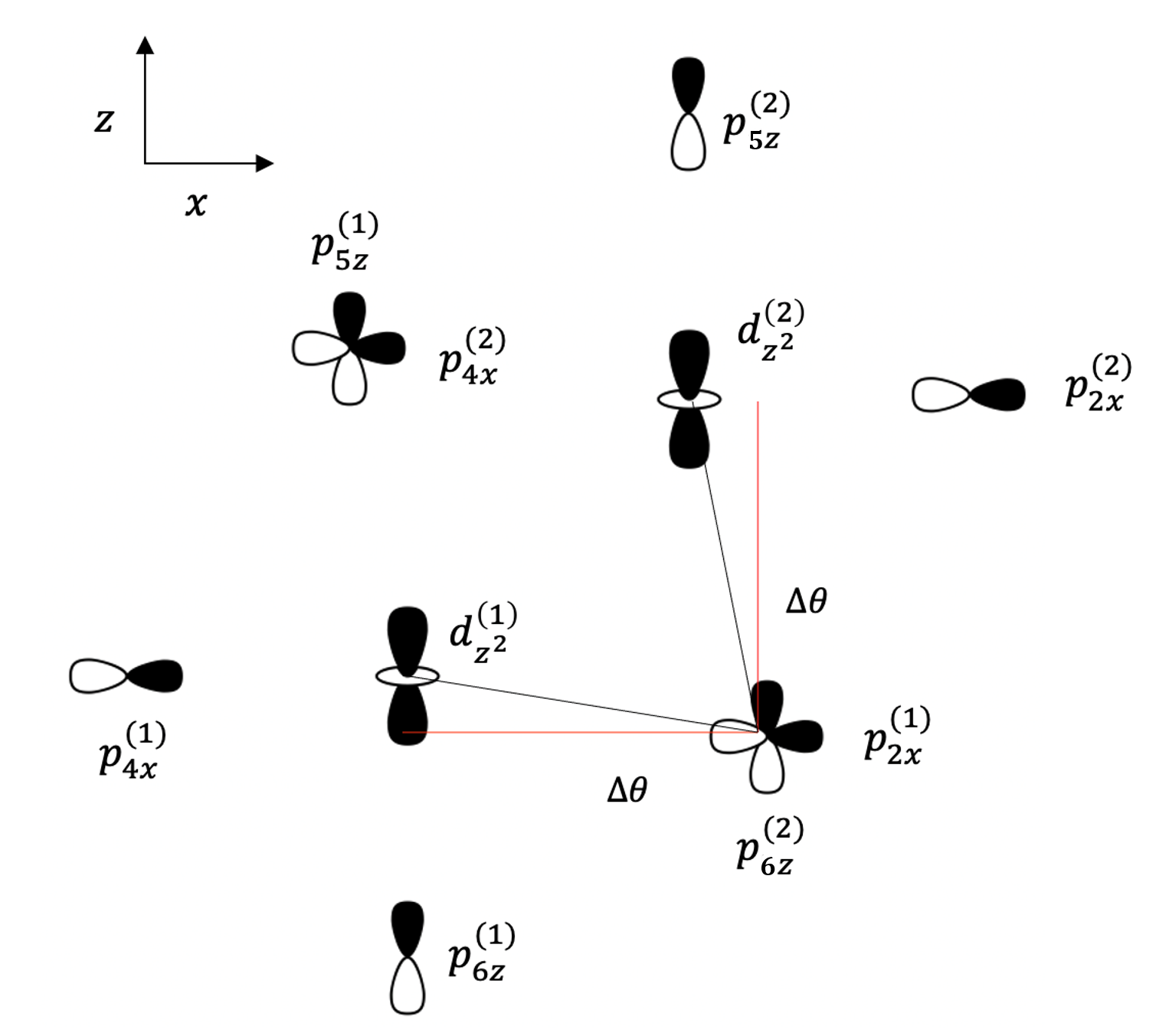}
\centering
\caption{Two NiS\textsubscript{6} lattice units. Two red lines denote $90^{\circ}$ and two black lines show the actual bonding angle $90^{\circ} - 2 \Delta\theta$.}
\label{fig:tl}
\end{figure}

To obtain the local electric polarization, we consider the case of two NiS\textsubscript{6} lattice units as shown in Figure \ref{fig:tl}. It is important to notice that the Ni-P-Ni bond angle is not $90^{\circ}$ but slightly smaller than $90^{\circ}$ \cite{gu2019ni}. See Figure 3. As a result, we obtain a small but finite hopping integral proportional to the deviation angle $\Delta\theta$ from $90^{\circ}$ based on the Slater-Koster table \cite{slater1954simplified} as follows

\begin{equation}
\resizebox{\textwidth}{!}{$%
H_{12}=t\Delta\theta\sum_{\sigma}\left[\frac{\sqrt{3}}{2}d^{(1)\dagger}_{x^2-y^2\sigma}p^{(2)}_{6z\sigma} -\frac{1}{2}d^{(1)\dagger}_{z^2\sigma}p^{(2)}_{6z\sigma}-d^{(2)\dagger}_{z^2\sigma}p^{(1)}_{2x\sigma}-\frac{\sqrt{3}}{2}d^{(2)\dagger}_{x^2-y^2\sigma} p^{(1)}_{5z\sigma}+\frac{1}{2}d^{(2)\dagger}_{z^2\sigma}p^{(1)}_{5z\sigma}+d^{(1)\dagger}_{z^2\sigma}p^{(2)}_{4x\sigma}+h.c.\right] .
\label{eq:hopping12}$%
}
\end{equation}
\\
Here, the superscripts (1) and (2) in the creation and annihilation operators denote the NiS\textsubscript{6} unit number of the two as shown in Figure \ref{fig:tl}.

Now, we consider the perturbation (\ref{eq:hopping12}) with respect to the Zhang-Rice triplet and singlet states of each NiS\textsubscript{6} unit, where the superscripts (1) and (2) are introduced into the creation and annihilation operators (\ref{eq:zrt}).\\
The resulting state is given by
\begin{equation}
|1'\rangle \otimes|2'\rangle = |1\rangle\otimes|2\rangle + \sum_{u} |u\rangle \langle u| \frac{H_{12}}{2(\epsilon_p+\epsilon_d)-E_u} |1\rangle\otimes|2\rangle .
\label{eq:1-2_perturbation}
\end{equation}
Here, $|1\rangle$ ($|2\rangle$) represents the Zhang-Rice triplet and singlet states of the first (second) NiS\textsubscript{6} unit before the perturbation, and $|1'\rangle \otimes |2'\rangle$ denotes the resulting state after the perturbation. $|u \rangle$ are additional excited states besides the Zhang-Rice triplet and singlet states, given by the Hilbert space projected by $P_{0}$ and obtained from the previous Schrodinger equation. Schematically, they are given by $|u\rangle = d^\dagger d^\dagger d^\dagger p^\dagger|0\rangle$, $p^\dagger p^\dagger d^\dagger p^\dagger|0\rangle$, $d^\dagger p^\dagger d^\dagger d^\dagger|0\rangle$, and $p^\dagger p^\dagger d^\dagger p^\dagger|0\rangle$, not shown here. $E_u$ is the energy eigenvalue of the corresponding state $|u \rangle$.
\begin{figure}[ht]
\includegraphics[scale=0.35]{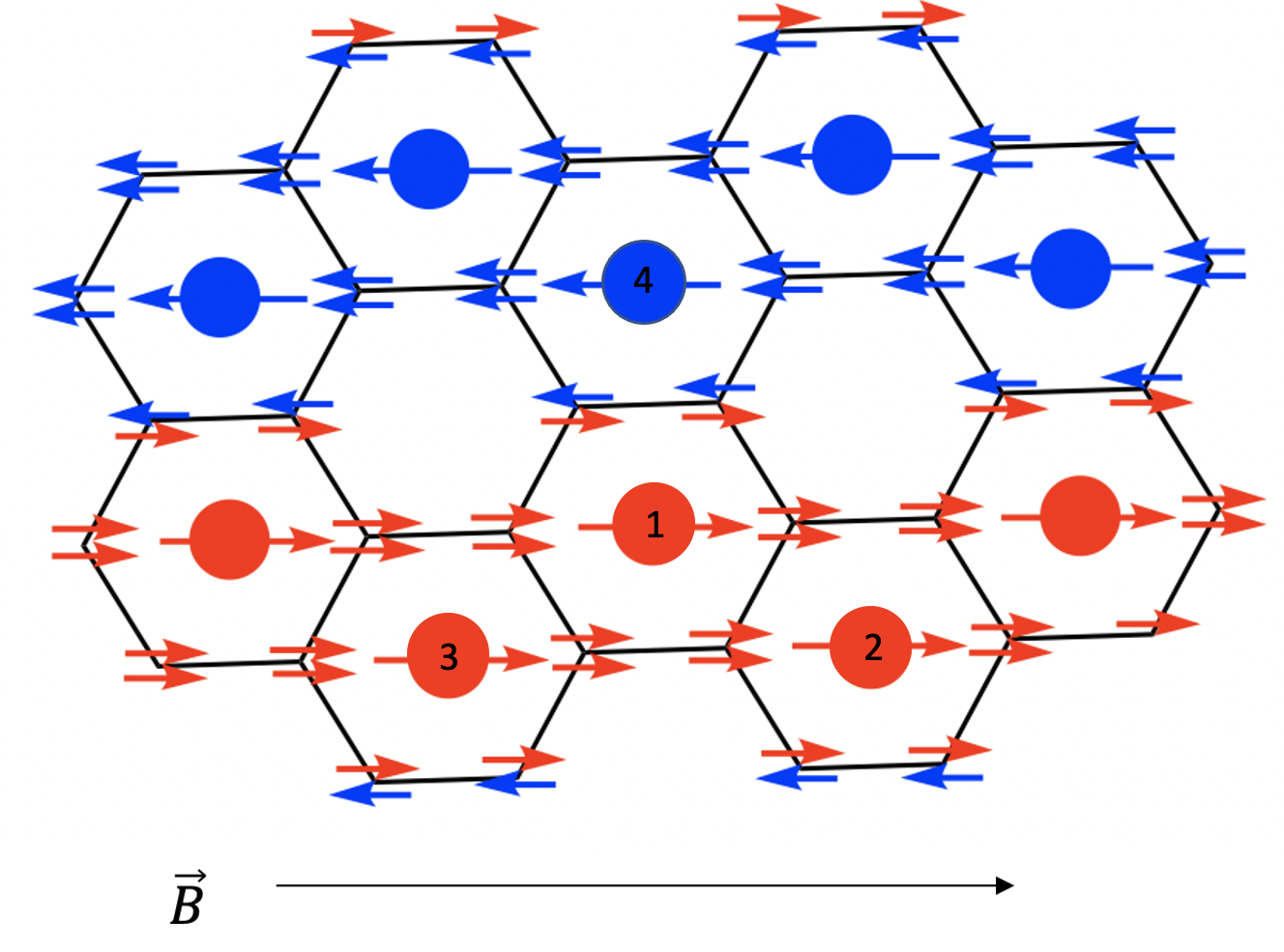}
\centering
\caption{Four lattice units for the calculation of the local electric polarization under external magnetic fields.}
\label{fig:numbering}
\end{figure}

\section{Four lattice unit calculation for local electric polarization in a zigzag antiferromagnetic phase}Based on this two lattice-unit result, we calculate the local electric polarization considering only three nearest neighborhoods. We recall that the ground state of NiPS\textsubscript{3} shows an antiferromagnetic zigzag ordering pattern. In this respect we write down the state of these four sites as $|1_T\rangle_{\uparrow}\otimes|2_T\rangle_{\uparrow}\otimes|3_T\rangle_{\uparrow}\otimes|4_T\rangle_{\downarrow}$, where the position operator of each state is given by $\vec{r}_1=\vec{r}\otimes 1\otimes 1\otimes 1$, $\vec{r}_2=1\otimes \vec{r}\otimes 1\otimes 1$, $\vec{r}_3=1\otimes 1\otimes \vec{r}\otimes 1$, and $\vec{r}_4=1\otimes 1\otimes 1\otimes \vec{r}$, respectively.

Following the same strategy as the previous discussion for $H_{12}$, we construct an effective hopping Hamiltonian between $1$ and $3$ lattice units as

\begin{equation}
\resizebox{\textwidth}{!}{$%
H_{13}=t\Delta\theta\sum_{\sigma}\left[\frac{\sqrt{3}}{2}d^{(1)\dagger}_{x^2-y^2\sigma}p^{(3)}_{5z\sigma} +\frac{1}{2}d^{(1)\dagger}_{z^2\sigma}p^{(3)}_{5z\sigma}-d^{(1)\dagger}_{z^2\sigma}p^{(3)}_{1y\sigma} -\frac{\sqrt{3}}{2}d^{(3)\dagger}_{x^2-y^2\sigma}p^{(1)}_{6z\sigma}-\frac{1}{2}d^{(3)\dagger}_{z^2\sigma}p^{(1)}_{6z\sigma} +d^{(3)\dagger}_{z^2\sigma}p^{(1)}_{3y\sigma}+h.c.\right]$%
}
\label{eq:hopping13}
\end{equation}
and that between $1$ and $4$ lattice units as
\begin{equation}
\resizebox{\textwidth}{!}{$%
\begin{aligned}
H_{14}=t\Delta\theta\sum_{\sigma}\Biggl[\frac{\sqrt{3}}{2}d^{(1)\dagger}_{x^2-y^2\sigma}p^{(4)}_{3y\sigma} -\frac{1}{2}d^{(1)\dagger}_{z^2\sigma}p^{(4)}_{3y\sigma}&-\frac{\sqrt{3}}{2}d^{(4)\dagger}_{x^2-y^2\sigma}p^{(1)}_{1y\sigma} +\frac{1}{2}d^{(4)\dagger}_{z^2\sigma}p^{(1)}_{1y\sigma}\\
&+\frac{\sqrt{3}}{2}d^{(1)\dagger}_{x^2-y^2\sigma}p^{(4)}_{2x\sigma}+\frac{1}{2}d^{(1)\dagger}_{z^2\sigma}p^{(4)}_{2x\sigma} -\frac{\sqrt{3}}{2}d^{(4)\dagger}_{x^2-y^2\sigma}p^{(1)}_{4x\sigma}-\frac{1}{2}d^{(4)\dagger}_{z^2\sigma}p^{(1)}_{4x\sigma}+h.c\Biggr] .
\end{aligned}$%
}
\label{eq:hopping14}
\end{equation}
\\
We also apply magnetic fields to control the electric polarization by the Zeeman effect
\begin{equation}
\resizebox{\textwidth}{!}{$
 H_{ext}^{(i)} = -\frac{\mu_B}{\hbar}\vec{B}\cdot\Biggl[\frac{g\hbar}{2}\sum_{m,\sigma,\sigma'}d^{(i) \dagger}_{m\sigma}\vec{\tau}_{\sigma\sigma'}d_{m\sigma}^{(i)}+\sum_{m,m',\sigma}d^{(i) \dagger}_{m\sigma}\vec{L}_{m m'}d_{m\sigma}^{(i)}+\frac{g\hbar}{2}\sum_{m,\sigma,\sigma'}p^{(i) \dagger}_{m\sigma}\vec{\tau}_{\sigma\sigma'}p_{m\sigma'}^{(i)}+\sum_{m,m',\sigma}p^{(i) \dagger}_{m\sigma}\vec{L}_{m m'}p_{m\sigma}^{(i)}\Biggr] 
\label{eq:zeeman}$
}
\end{equation}
Here, $\mu_B$ is the Bohr magneton and g is the anomalous gyromagnetic ratio of electron ($\approx 2.0023192$). The superscript $(i)$ runs from $1$ to $4$.

To find local electric polarization, we realize that it is essential to consider atomic spin-orbit interaction as
\begin{equation}
H_{LS}=\frac{\lambda_d}{\hbar^2}(\vec{L}_{d}\cdot\vec{S}_{d})+\frac{\lambda_p}{\hbar^2}(\vec{L}_p\cdot\vec{S}_p) .
\label{eq:spin-orbit coupling}
\end{equation}
Here, $\lambda_d$ ($\lambda_p$) represents the spin-orbit coupling constant of the d-orbital (ligand p-orbitals). Accordingly, $\vec{L}_{d(p)}$ and $\vec{S}_{d(p)}$ are angular-momentum and spin operators of each orbital, respectively. This atomic spin-orbit coupling Hamiltonian gives rise to mixing between $t_{2g}$ and $e_g$ orbitals. Considering the energy gap $\Delta = E_{e_g}-E_{t_{2g}}$ between these two orbitals, the perturbation theory results in

\begin{equation}
\begin{split}
&|d_{z^2\uparrow}\rangle\rightarrow |d^{SO}_{z^2\uparrow}\rangle=|d_{z^2\uparrow}\rangle-i\sqrt{\frac{3}{2}}\frac{\lambda_d}{\lambda_d/2+\Delta} \Biggl(\frac{1}{\sqrt{2}}|d_{yz\downarrow}\rangle-\frac{i}{\sqrt{2}}|d_{zx\downarrow}\rangle\Biggr)\\
&|d_{z^2\downarrow}\rangle\rightarrow |d^{SO}_{z^2\downarrow}\rangle=|d_{z^2\downarrow}\rangle+ i\sqrt{\frac{3}{2}}\frac{\lambda_d}{\lambda_d/2+\Delta}\Biggl(-\frac{1}{\sqrt{2}}|d_{yz\uparrow}\rangle -\frac{i}{\sqrt{2}}|d_{zx\uparrow}\rangle\Biggr)\\
&|d_{x^2-y^2\uparrow}\rangle\rightarrow|d^{SO}_{x^2-y^2\uparrow}\rangle=|d_{x^2-y^2\uparrow}\rangle +i\sqrt{\frac{3}{2}}\frac{\lambda_d}{\lambda_d/2+\Delta}\Biggl(-\frac{1}{\sqrt{6}}|d_{yz\downarrow}\rangle -\frac{i}{\sqrt{6}}|d_{zx\downarrow}\rangle+\sqrt{\frac{2}{3}}|d_{xy\uparrow}\rangle\Biggr)\\
&|d_{x^2-y^2\downarrow}\rangle\rightarrow|d^{SO}_{x^2-y^2\downarrow}\rangle=|d_{x^2-y^2\downarrow}\rangle -i\sqrt{\frac{3}{2}}\frac{\lambda_d}{\lambda_d/2+\Delta}\Biggl(\frac{1}{\sqrt{6}}|d_{yz\uparrow}\rangle -\frac{i}{\sqrt{6}}|d_{zx\uparrow}\rangle+\sqrt{\frac{2}{3}}|d_{xy\downarrow}\rangle\Biggr)
\end{split}
\label{eq:soc d}
\end{equation}
\\
for $e_{g}$ orbitals \cite{stamokostas2018mixing}. On the other hand, there is no energy gap in ligand p-orbitals, and the spin-orbit coupling gives rise to mixing between six degenerate states. As a result, we obtain states $p_x$, $p_y$, and $p_z$ as linear combinations of eigenstates of $\frac{\lambda_p}{\hbar^2}(\vec{L}_p\cdot\vec{S}_p)$.

Under the external magnetic field shown in Figure \ref{fig:numbering} and based on the basis (\ref{eq:soc d}), we represent the Zeeman-effect term as follows

\begin{equation}
\begin{split}
H^{(i)}_{ext}&=-\mu_BB
\begin{pNiceMatrix}[margin,last-col]
 1-\frac{3}{2}(\frac{\lambda_d}{\lambda_d/2+\Delta})^2 & 0 & 0 & 0 & d^{SO(i)\dagger }_{z^2\uparrow}\\
 0 & 1+\frac{1}{2}(\frac{\lambda_d}{\lambda_d/2+\Delta})^2 & 0 & 0 & d^{SO(i)\dagger }_{x^2-y^2\uparrow}\\
 0 & 0 & -1+\frac{3}{2}(\frac{\lambda_d}{\lambda_d/2+\Delta})^2 & 0 & d^{SO(i)\dagger }_{z^2\downarrow}\\
 0 & 0 & 0 & -1-\frac{1}{2}(\frac{\lambda_d}{\lambda_d/2+\Delta})^2 & d^{SO(i)\dagger }_{x^2-y^2\downarrow}\\
\end{pNiceMatrix}\\
&-\frac{\mu_BB}{\sqrt{3}}
\begin{pNiceMatrix}[margin,last-col]
    \frac{3}{2}(\frac{\lambda_d}{\lambda_d/2+\Delta})^2 & 0 & \frac{3(1-i)\lambda_d}{\lambda_d/2+\Delta} & \frac{(1+i)\sqrt{3}\lambda_d\Delta}{(\Delta+\lambda_d/2)^2} & d^{SO(i)\dagger}_{z^2\uparrow}\\
    0 & -\frac{1}{2}(\frac{\lambda_d}{\Delta+\lambda_d/2})^2+\frac{4\lambda_d}{\lambda_d/2+\Delta} & \frac{(1+i)\sqrt{3}\lambda_d\Delta}{(\Delta+\lambda_d/2)^2} & \frac{(1-i)\sqrt{3}\lambda_d(\Delta+\frac{3}{2}\lambda_d)}{(\Delta+\lambda_d/2)^2} & d^{SO(i)\dagger}_{x^2-y^2\uparrow}\\
    \frac{3(1+i)\lambda_d}{\lambda_d/2+\Delta} & \frac{(1-i)\sqrt{3}\lambda_d\Delta}{(\Delta+\lambda_d/2)^2} & -\frac{3}{2}(\frac{\lambda_d}{\lambda_d/2+\Delta})^2 & 0 & d^{SO(i)\dagger}_{z^2\downarrow}\\
    \frac{(1-i)\sqrt{3}\lambda_d\Delta}{(\Delta+\lambda_d/2)^2} & \frac{(1+i)\sqrt{3}\lambda_d(\Delta+\frac{3}{2}\lambda_d)}{(\Delta+\lambda_d/2)^2}  & 0 & \frac{1}{2}(\frac{\lambda_d}{\Delta+\lambda_d/2})^2-\frac{4\lambda_d}{\Delta+\lambda_d/2} & d^{SO(i)\dagger}_{x^2-y^2\downarrow}\\ 
    \end{pNiceMatrix} 
\label{eq:zeeman with SO}
\end{split}
\end{equation}
\\
See the appendix for more details.
Now, we repeat the perturbation analysis for $H'=H_{12}+H_{13}+H_{14}+H_{ext}$ as before. As a result, we obtain
\begin{equation}
 |1'\rangle\otimes|2'\rangle\otimes|3'\rangle\otimes|4'\rangle = |1\rangle\otimes|2\rangle\otimes|3\rangle\otimes|4\rangle + \sum_{u} |u \rangle \langle u| \frac{H'}{E_{1234}-E_u} |1\rangle\otimes|2\rangle\otimes|3\rangle\otimes|4\rangle.
\end{equation}
\\
Here, $E_{1234}$ is an energy eigenvalue given by $4(\epsilon_p+\epsilon_d)$ + (diagonal terms of $H'$), not shown explicitly, and $|u\rangle$ are excited states, where the hole of the Ni d-orbital hops into the S p-orbital, and vice versa.

It is straightforward to calculate the electric polarization $\vec{P}_{1}=\langle 1'|\otimes\langle 2'|\otimes\langle 3'|\otimes\langle 4'| \vec{r}_1 |1'\rangle\otimes|2'\rangle\otimes|3'\rangle\otimes|4'\rangle$. Up to the first order in $t\Delta\theta$, we obtain

\begin{equation}
\small
\begin{split}
    \vec{P}_1=&-\frac{t\Delta\theta}{6}\frac{\vec{p}_1}{\epsilon_p-\epsilon_d}-t\Delta\theta\Biggl\{\frac{1}{U-J-\epsilon_p+\epsilon_d}-\frac{1}{2}\frac{1}{U-3J-\epsilon_p+\epsilon_d}\Biggr\}\vec{p}_2+\frac{\sqrt{3}t\Delta\theta B(\lambda_1-\lambda_2)}{2(\epsilon_p-\epsilon_d)^2}\vec{p}_3\\
    &-\frac{t\Delta\theta B(\lambda_1-\lambda_2)}{(U-3J-\epsilon_p+\epsilon_d)^2}\vec{p}_4-\frac{2t\Delta\theta B(\lambda_1-\lambda_2)}{(U-J-\epsilon_p+\epsilon_d)^2}\vec{p}_5 ,
\label{eq:p1}
\end{split}
\end{equation}
where
\begin{equation}
\vec{p}_1=-\langle p^{(2)}_{4x}|x|d^{(1)}_{z^2}\rangle\left(\hat{x}-\hat{y}\right),
\end{equation}
\begin{equation}
\resizebox{\textwidth}{!}{$%
\begin{aligned}
\vec{p}_2=&-2\Biggl(\frac{1}{12}\langle p^{(1)}_{1y}|x|d^{(2)}_{z^2}\rangle-\frac{5}{24}\langle p^{(1)}_{2x}|x|d^{(2)}_{z^2}\rangle+\frac{5}{24}\langle p^{(1)}|x|d^{(2)}_{z^2}\rangle+\frac{1}{4\sqrt{3}}\langle p^{(1)}_{1y}|x|d^{(2)}_{x^2-y^2}\rangle+\frac{1}{8\sqrt{3}}\langle p^{(1)}_{2x}|x|d^{(2)}_{x^2-y^2}\rangle\\
&\ \ \ \ \ \ \ \ \ \ -\frac{1}{8\sqrt{3}}\langle p^{(1)}_{4x}|x|d^{(2)}_{x^2-y^2}-\frac{1}{6}\langle p^{(1)}_{6z}|x|d^{(2)}_{z^2}+\frac{1}{4\sqrt{3}}\langle p^{(1)}_{6z}|x|d^{(2)}_{x^2-y^2}\rangle\Biggr)(\hat{x}-\hat{y}) ,
\end{aligned}$%
}
\end{equation}
\begin{equation}
\begin{split}
\vec{p}_3=&\langle p^{(4)}_{2x}|x|d^{(1)}_{x^2-y^2}\rangle(\hat{x}-\hat{y}) ,
\end{split}
\end{equation}

\begin{equation}
\resizebox{\textwidth}{!}{$%
\begin{aligned}
\vec{p}_4=&\Biggl(\frac{1}{4\sqrt{3}}\langle p^{(1)}_{1y}|x|d^{(2)}_{x^2-y^2}\rangle+\frac{1}{8\sqrt{3}}\langle p^{(1)}_{2x}|x|d^{(2)}_{x^2-y^2}\rangle-\frac{1}{8\sqrt{3}}\langle p^{(1)}_{4x}|x|d^{(2)}_{x^2-y^2}\rangle+\frac{1}{4\sqrt{3}}\langle p^{(1)}_{6z}|x|d^{(2)}_{x^2-y^2}\rangle\Biggr)(\hat{x}-\hat{y}) ,
\end{aligned}$%
}
\end{equation}
and
\begin{equation}
\begin{split}
\vec{p}_5=&\Biggl(\frac{\sqrt{3}}{8}\langle p^{(1)}_{2x}|x|d^{(4)}_{x^2-y^2}\rangle+\frac{\sqrt{3}}{8}\langle p^{(1)}_{3y}|x|d^{(4)}_{x^2-y^2}\rangle-\frac{\sqrt{3}}{8}\langle p^{(1)}_{4x}|x|d^{(4)}_{x^2-y^2}\rangle\Biggr)(\hat{x}-\hat{y}) .
\end{split}
\end{equation}
\\
Here, $\lambda_1=\mu_B(1+\frac{1}{\sqrt{3}}\frac{4\lambda_d}{\lambda_d/2+\Delta})$ and $\lambda_2=\mu_B$ are effective Zeeman coupling constants for d- and p-orbitals, respectively. We emphasize that the central role of the atomic spin-orbit coupling is to give the difference in the effective Zeeman coupling constant. We point out that there exist hopping effects from the atomic spin-orbit coupling in the sub-leading order $t \Delta\theta \lambda_d$. See the appendix for more details.

\begin{figure}[t]
\includegraphics[scale=0.28]{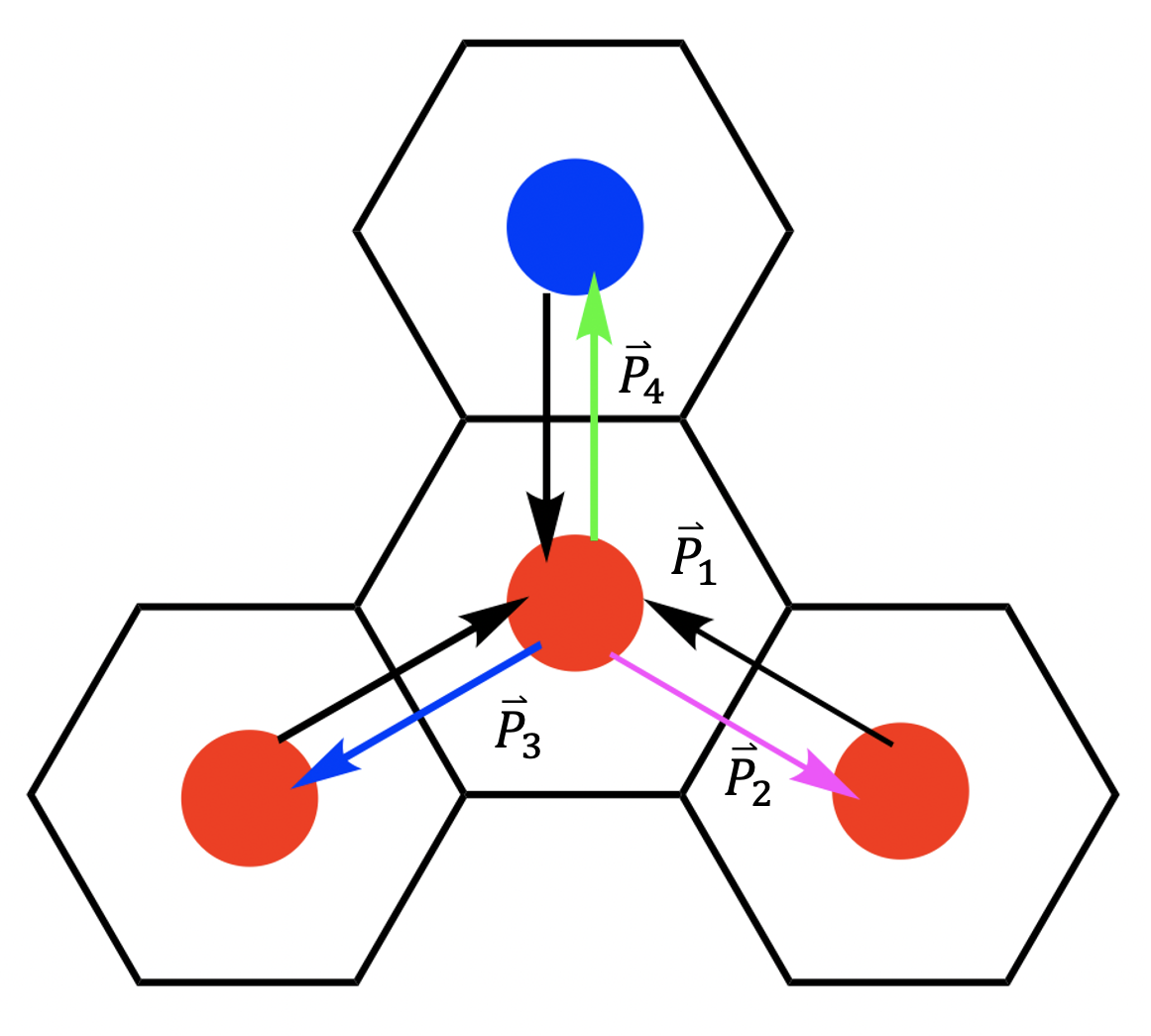}
\centering
\caption{Local electric polarization given by sum of $\vec{P}_1$ (black arrow), $\vec{P}_2$ (magenta arrow), $\vec{P}_3$ (blue arrow), and $\vec{P}_4$ (green arrow).}
\label{fig:plz}
\end{figure}

In the same way, we obtain $\vec{P}_2 = \langle \vec{r}_2 \rangle$, $\vec{P}_3 = \langle \vec{r}_3 \rangle$, and $\vec{P}_4 = \langle \vec{r}_4 \rangle$ as shown in Figure \ref{fig:plz}. The only thing which one should be careful about is to keep the coordinate frame in a consistent way. Electric polarizations represented by the arrows between red circles are all canceled out. On the other hand, electric polarizations given by the arrows between red and blue circles are not canceled. In particular, the Zeeman effect enhances the net polarization. As a result, the net electric polarization of the site $1$ is given by\\
\begin{equation}
    \vec{P}=\frac{\sqrt{3}t\Delta\theta B(\lambda_1-\lambda_2)}{(\epsilon_p-\epsilon_d)^2}\vec{p}_3-\frac{4t\Delta\theta B(\lambda_1-\lambda_2)}{(U-J-\epsilon_p+\epsilon_d)^2}\vec{p}_5.
    \label{eq:lep}
\end{equation}
\\
This local electric polarization shows an antiferromagnetically ordered pattern. 

\begin{figure}[ht]
\includegraphics[scale=0.6]{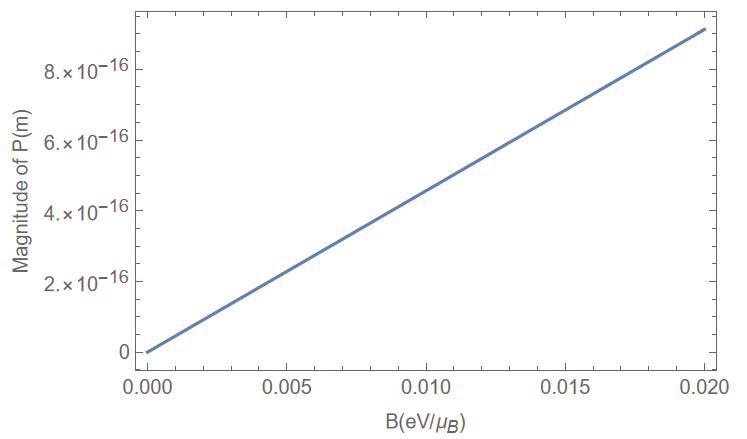}
\centering
\caption{Local electric polarization as a function of applied magnetic fields}
\label{fig:pol}
\end{figure}

For more quantitative analysis, we use the following specific values of parameters: $U = 5eV,\ U_{eff} = U-J = 4eV,\ \epsilon_p-\epsilon_d = 1eV,\ t = -0.74 eV,\ \lambda_d = 0.08eV,\ \Delta = 1.4eV,$ and the distance $a= 2.5\r{A}$ between Ni and S \cite{NiPS3_Nature,gu2019ni,kim2018charge}. Then, we obtain the local electric polarization as
\begin{equation}
\vec{P}=3.23\times 10^{-14}B\mu_B(\hat{x}-\hat{y})[m].
\label{eq:polarization}
\end{equation}
\\
See Figure \ref{fig:pol} and the appendix for more details.
\noindent
More explicitly, we obtain $|e\vec{P}|/V\sim 10^{-1}\mu C/m^{2}$ for the magnetic field $B=10T$. Here, V is a volume of NiS\textsubscript{6}.

When the Zhang-Rice triplet to singlet excitation happens in the central unit, we obtain
\begin{equation}
\vec{P}_1 = -\frac{\sqrt{3}t\Delta\theta B(\lambda_1-\lambda_2)}{2(\epsilon_p-\epsilon_d)^2}\vec{p}_3+\frac{t\Delta\theta B(\lambda_1-\lambda_2)}{(U-3J-\epsilon_p+\epsilon_d)^2}\vec{p}_4 -\frac{2t\Delta\theta B(\lambda_1-\lambda_2)}{(U-J-\epsilon_p+\epsilon_d)^2} \vec{p}_5 .
\label{eq:lep1}
\end{equation}
\\
This local polarization satisfies $\vec{P}_2+\vec{P}_3+\vec{P}_4=-\vec{P}_1$, which means the absence of the local electric polarization in the central unit due to the cancelation. As a result, there appears local net electric polarization around the central unit. 

\section{Effective tight binding lattice model}
To figure out a local Berry curvature picture for the electric polarization, it is necessary to construct an effective tight binding lattice model. Ref. \cite{Spin_Current_Polarization} discussed a real space Berry curvature, being responsible for electric polarization. Here, the authors used specific two degenerate states of t\textsubscript{2g} spin-orbit coupled states, given by
        $$
        |a\rangle = \frac{1}{\sqrt{3}}(|d_{xy,\uparrow}\rangle + |d_{yz,\downarrow}\rangle + i|d_{zx,\downarrow}\rangle),  ~~~~~ \\ 
        |b\rangle = \frac{1}{\sqrt{3}}(|d_{xy,\downarrow}\rangle - |d_{yz,\uparrow}\rangle + i|d_{zx,\uparrow}\rangle).
        $$
       Based on this basis, the authors constructed an effective tight binding model (Fig. 1 of Ref. \cite{Spin_Current_Polarization}), which takes into account an on-site coulomb potential and an effective hopping term between the transition metal d orbital and the ligand p orbital. Within this effective lattice model construction, the authors calculated the resulting ground state and computed the expectation value of the postion vector $\vec{r}$. As a result, they found the local electric polarization in terms of a real space Berry curvature. More specifically, they obtained $|\vec{P}|\sim 10^{4}(\frac{V}{\Delta})^3\mu C/m^{2}$, where V is the hopping parameter and $\Delta$ is an enegy gap between the d and p orbitals.

Benchmarking this strategy, we construct an effective tight binding model (\ref{eq:tbm}) for the site 1 and its three nearest neighbors, wherein we consider the Zhang-Rice triplet states as a building block of our system. 
        Unlike Ref. 3, our model features a many particle wavefunction given by the Zhang-Rice triplet state.

When we regard the ground state of the single NiS\textsubscript{6} cluster (the Zhang-Rice triplet state) as the buliding block, we can construct an effective tight binding model in terms of the Zhang-Rice triplet creation and annihilation operators,
\begin{equation}
\resizebox{\textwidth}{!}{$%
\begin{aligned}
T_{\sigma}^{(i)\dagger}=\frac{1}{\sqrt{6}}&\Biggl\{-\frac{\sqrt{3}}{2}d^{(i)\dagger}_{z^2\sigma} (p^{(i)\dagger}_{1y\sigma}-p^{(i)\dagger}_{2x\sigma}-p^{(i)\dagger}_{3y\sigma}+p^{(i)\dagger}_{4x\sigma})+d^{(i)\dagger}_{x^2-y^2\sigma} \left[\frac{1}{2}(p^{(i)\dagger}_{1y\sigma}+p^{(i)\dagger}_{2x\sigma}-p^{(i)\dagger}_{3y\sigma}-p^{(i)\dagger}_{4x\sigma}) -p^{(i)\dagger}_{5z\sigma}+p^{(i)\dagger}_{6z\sigma}\right]\Biggr\},\\
T_0^{(i)\dagger}=\frac{1}{2\sqrt{3}}& \Biggl\{\left(-\frac{\sqrt{3}}{2}d^{(i)\dagger}_{z^2\uparrow}(p^{(i)\dagger}_{1y\downarrow}-p^{(i)\dagger}_{2x\downarrow} -p^{(i)\dagger}_{3y\downarrow}+p^{(i)\dagger}_{4x\downarrow})+d^{(i)\dagger}_{x^2-y^2\downarrow} \left[\frac{1}{2}(p^{(i)\dagger}_{1y\downarrow}+p^{(i)\dagger}_{2x\downarrow}-p^{(i)\dagger}_{3y\downarrow} -p^{(i)\dagger}_{4x\downarrow})-p^{(i)\dagger}_{5z\downarrow}+p^{(i)\dagger}_{6z\downarrow}\right]\right)\\
&+\left(-\frac{\sqrt{3}}{2}d^{(i)\dagger}_{z^2\downarrow}(p^{(i)\dagger}_{1y\uparrow}-p^{(i)\dagger}_{2x\uparrow}-p^{(i)\dagger}_{3y\uparrow} +p^{(i)\dagger}_{4x\uparrow})+d^{(i)\dagger}_{x^2-y^2\downarrow}\left[\frac{1}{2}(p^{(i)\dagger}_{1y\uparrow} +p^{(i)\dagger}_{2x\uparrow}-p^{(i)\dagger}_{3y\uparrow}-p^{(i)\dagger}_{4x\uparrow}) -p^{(i)\dagger}_{5z\uparrow}+p^{(i)\dagger}_{6z\uparrow}\right]\right)\Biggr\}.
\end{aligned}$%
}
\label{eq:zrto}
\end{equation}
These operators are nothing but those of Eq. (\ref{eq:zrt}) with $\sigma=\uparrow, \downarrow$. In other words, if these operators act on the vacuum state, they result in the Zhang-Rice triplet state. 

It is quite interesting to check out their commutation relations, given by
\begin{equation}
\left[T_{\sigma},T_{\sigma'}\right]=[T_{0},T_{0}]=[T_\uparrow,T^{\dagger}_\downarrow]=0,\ \ \ \ \ \sigma,\sigma'=\uparrow,\downarrow ,
\label{eq:cl1}
\end{equation}

\begin{equation}
\resizebox{\textwidth}{!}{$%
\begin{aligned}
[T_{\sigma},T_{\sigma}^{\dagger}]=1+\frac{1}{6}&\Biggl\{\Biggl[-p^{\dagger}_{1y\sigma}p_{1y\sigma}-p^{\dagger}_{2x\sigma}p_{2x\sigma}-p^{\dagger}_{3y\sigma}p_{3y\sigma}-p^{\dagger}_{4x\sigma}p_{4x\sigma}-p^{\dagger}_{5z\sigma}p_{5z\sigma}-p^{\dagger}_{6z\sigma}p_{6z\sigma}\Biggr]\\
&+\frac{1}{2}\Biggl[p^{\dagger}_{1y\sigma}\Biggl(p_{2x\sigma}-p_{4x\sigma}+p_{5z\sigma}-p_{6z\sigma}\Biggr)+p^{\dagger}_{2x\sigma}\Biggl(p_{1y\sigma}-p_{3x\sigma}+p_{5z\sigma}-p_{6z\sigma}\Biggr)\\
&-p^{\dagger}_{3y\sigma}\Biggl(p_{2x\sigma}-p_{4x\sigma}+p_{5z\sigma}-p_{6z\sigma}\Biggr)-p^{\dagger}_{4x\sigma}\Biggl(p_{1y\sigma}-p_{3x\sigma}+p_{5z\sigma}-p_{6z\sigma}\Biggr)\\
&+p^{\dagger}_{5z\sigma}\Biggl(p_{1y\sigma}+p_{2x\sigma}-p_{3y\sigma}-p_{4x\sigma}\Biggr)-p^{\dagger}_{6z\sigma}\Biggl(p_{1y\sigma}+p_{2x\sigma}-p_{3y\sigma}-p_{4x\sigma}\Biggr)\Biggr]\\
&+p^{\dagger}_{1y\sigma}p_{3y\sigma}+p^{\dagger}_{2x\sigma}p_{4x\sigma}+p^{\dagger}_{3y\sigma}p_{1y\sigma}+p^{\dagger}_{4x\sigma}p_{2x\sigma}+p^{\dagger}_{5z\sigma}p_{6z\sigma}+p^{\dagger}_{6z\sigma}p_{5z\sigma}\\
&-3\Biggl(d^{\dagger}_{z^2\sigma}d_{z^2\sigma}+d^{\dagger}_{x^2-y^2\sigma}d_{x^2-y^2\sigma}\Biggr)+\sqrt{3}\Biggl(d^{\dagger}_{z^2\sigma}d_{x^2-y^2\sigma}+d^{\dagger}_{x^2-y^2\sigma}d_{z^2\sigma}\Biggr)\Biggr\},
\end{aligned}$%
}
\label{eq:cl2}
\end{equation}

\begin{equation}
\resizebox{\textwidth}{!}{$%
\begin{aligned}
[T_{0},T_{0}^{\dagger}]=1+\frac{1}{12}\sum_{\sigma=\uparrow,\downarrow}&\Biggl\{\Biggl[-p^{\dagger}_{1y\sigma}p_{1y\sigma}-p^{\dagger}_{2x\sigma}p_{2x\sigma}-p^{\dagger}_{3y\sigma}p_{3y\sigma}-p^{\dagger}_{4x\sigma}p_{4x\sigma}-p^{\dagger}_{5z\sigma}p_{5z\sigma}-p^{\dagger}_{6z\sigma}p_{6z\sigma}\Biggr]\\
&+\frac{1}{2}\Biggl[p^{\dagger}_{1y\sigma}\Biggl(p_{2x\sigma}-p_{4x\sigma}+p_{5z\sigma}-p_{6z\sigma}\Biggr)+p^{\dagger}_{2x\sigma}\Biggl(p_{1y\sigma}-p_{3x\sigma}+p_{5z\sigma}-p_{6z\sigma}\Biggr)\\
&-p^{\dagger}_{3y\sigma}\Biggl(p_{2x\sigma}-p_{4x\sigma}+p_{5z\sigma}-p_{6z\sigma}\Biggr)-p^{\dagger}_{4x\sigma}\Biggl(p_{1y\sigma}-p_{3x\sigma}+p_{5z\sigma}-p_{6z\sigma}\Biggr)\\
&+p^{\dagger}_{5z\sigma}\Biggl(p_{1y\sigma}+p_{2x\sigma}-p_{3y\sigma}-p_{4x\sigma}\Biggr)-p^{\dagger}_{6z\sigma}\Biggl(p_{1y\sigma}+p_{2x\sigma}-p_{3y\sigma}-p_{4x\sigma}\Biggr)\Biggr]\\
&+p^{\dagger}_{1y\sigma}p_{3y\sigma}+p^{\dagger}_{2x\sigma}p_{4x\sigma}+p^{\dagger}_{3y\sigma}p_{1y\sigma}+p^{\dagger}_{4x\sigma}p_{2x\sigma}+p^{\dagger}_{5z\sigma}p_{6z\sigma}+p^{\dagger}_{6z\sigma}p_{5z\sigma}\\
&-3\Biggl(d^{\dagger}_{z^2\sigma}d_{z^2\sigma}+d^{\dagger}_{x^2-y^2\sigma}d_{x^2-y^2\sigma}\Biggr)+\sqrt{3}\Biggl(d^{\dagger}_{z^2\sigma}d_{x^2-y^2\sigma}+d^{\dagger}_{x^2-y^2\sigma}d_{z^2\sigma}\Biggr)\Biggr\},
\end{aligned}$%
}
\label{eq:cl3}
\end{equation}

\begin{equation}
\resizebox{\textwidth}{!}{$%
\begin{aligned}
[T_{\uparrow},T_{0}^{\dagger}]=\Biggl([T_{\downarrow},T_{0}^{\dagger}]\Biggr)^{\dagger}=\frac{1}{6\sqrt{2}}&\Biggl\{\Biggl[-p^{\dagger}_{1y\downarrow}p_{1y\uparrow}-p^{\dagger}_{2x\downarrow}p_{2x\uparrow}-p^{\dagger}_{3y\downarrow}p_{3y\uparrow}-p^{\dagger}_{4x\downarrow}p_{4x\uparrow}-p^{\dagger}_{5z\downarrow}p_{5z\uparrow}-p^{\dagger}_{6z\downarrow}p_{6z\uparrow}\Biggr]\\
&+\frac{1}{2}\Biggl[p^{\dagger}_{1y\downarrow}\Biggl(p_{2x\uparrow}-p_{4x\uparrow}+p_{5z\uparrow}-p_{6z\uparrow}\Biggr)+p^{\dagger}_{2x\downarrow}\Biggl(p_{1y\uparrow}-p_{3x\uparrow}+p_{5z\uparrow}-p_{6z\uparrow}\Biggr)\\
&-p^{\dagger}_{3y\downarrow}\Biggl(p_{2x\uparrow}-p_{4x\uparrow}+p_{5z\uparrow}-p_{6z\uparrow}\Biggr)-p^{\dagger}_{4x\downarrow}\Biggl(p_{1y\uparrow}-p_{3x\uparrow}+p_{5z\uparrow}-p_{6z\uparrow}\Biggr)\\
&+p^{\dagger}_{5z\downarrow}\Biggl(p_{1y\uparrow}+p_{2x\uparrow}-p_{3y\uparrow}-p_{4x\uparrow}\Biggr)-p^{\dagger}_{6z\downarrow}\Biggl(p_{1y\uparrow}+p_{2x\uparrow}-p_{3y\uparrow}-p_{4x\uparrow}\Biggr)\Biggr]\\
&+p^{\dagger}_{1y\downarrow}p_{3y\uparrow}+p^{\dagger}_{2x\downarrow}p_{4x\uparrow}+p^{\dagger}_{3y\downarrow}p_{1y\uparrow}+p^{\dagger}_{4x\downarrow}p_{2x\uparrow}+p^{\dagger}_{5z\downarrow}p_{6z\uparrow}+p^{\dagger}_{6z\downarrow}p_{5z\uparrow}\\
&-3\Biggl(d^{\dagger}_{z^2\downarrow}d_{z^2\uparrow}+d^{\dagger}_{x^2-y^2\downarrow}d_{x^2-y^2\uparrow}\Biggr)+\sqrt{3}\Biggl(d^{\dagger}_{z^2\downarrow}d_{x^2-y^2\uparrow}+d^{\dagger}_{x^2-y^2\downarrow}d_{z^2\uparrow}\Biggr)\Biggr\}.
\end{aligned}$%
}
\label{eq:cl4}
\end{equation}
These commutation relations verify that the Zhang-Rice triplet creation and annihilation operators are neither fermions nor bosons. 

\begin{figure}[ht]
\includegraphics[scale=0.38]{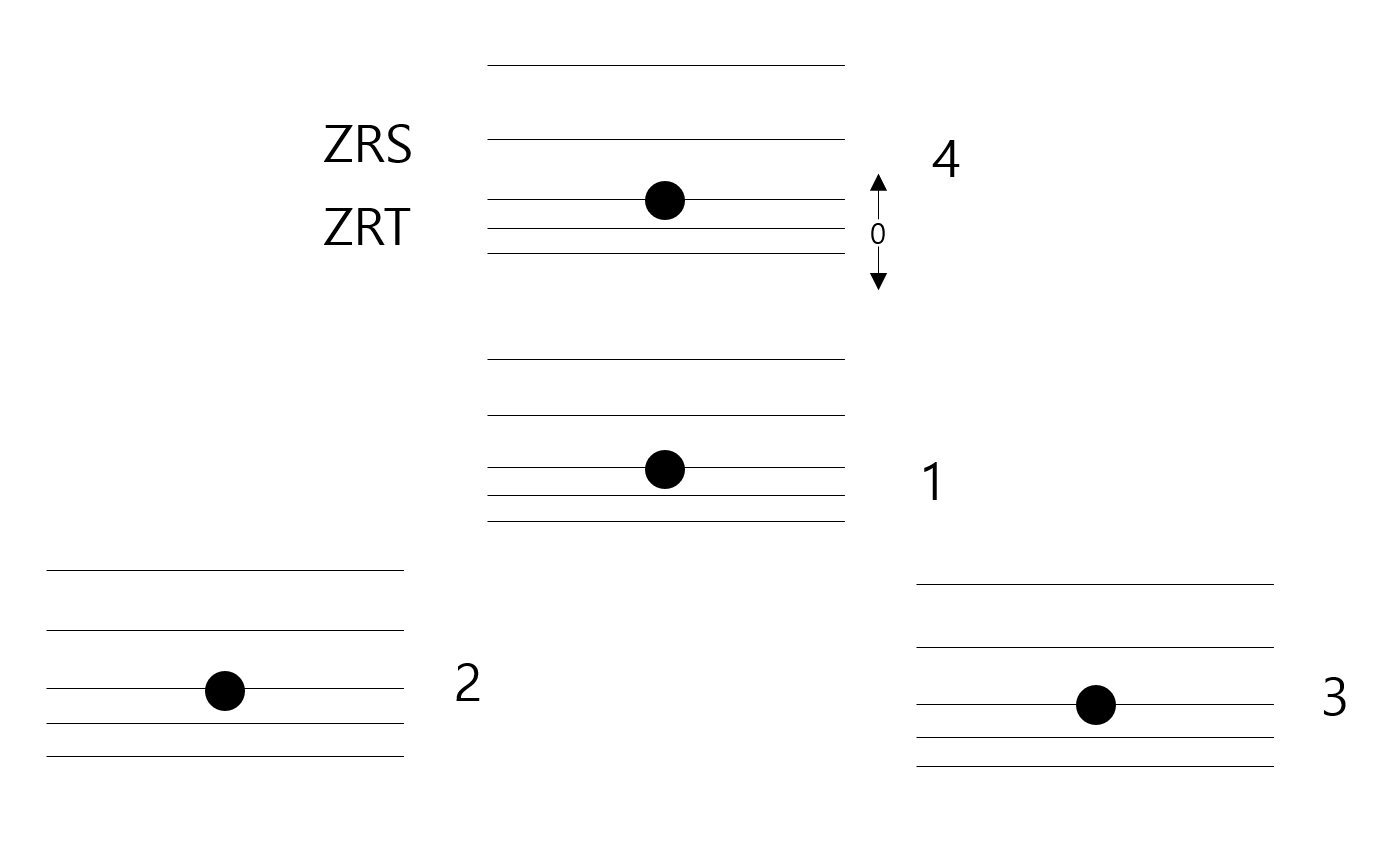}
\centering
\caption{The ground state of the site 1 and its three nearest neighborhoods}
\label{fig:tb1}
\end{figure}

\begin{figure}[ht]
\includegraphics[scale=0.38]{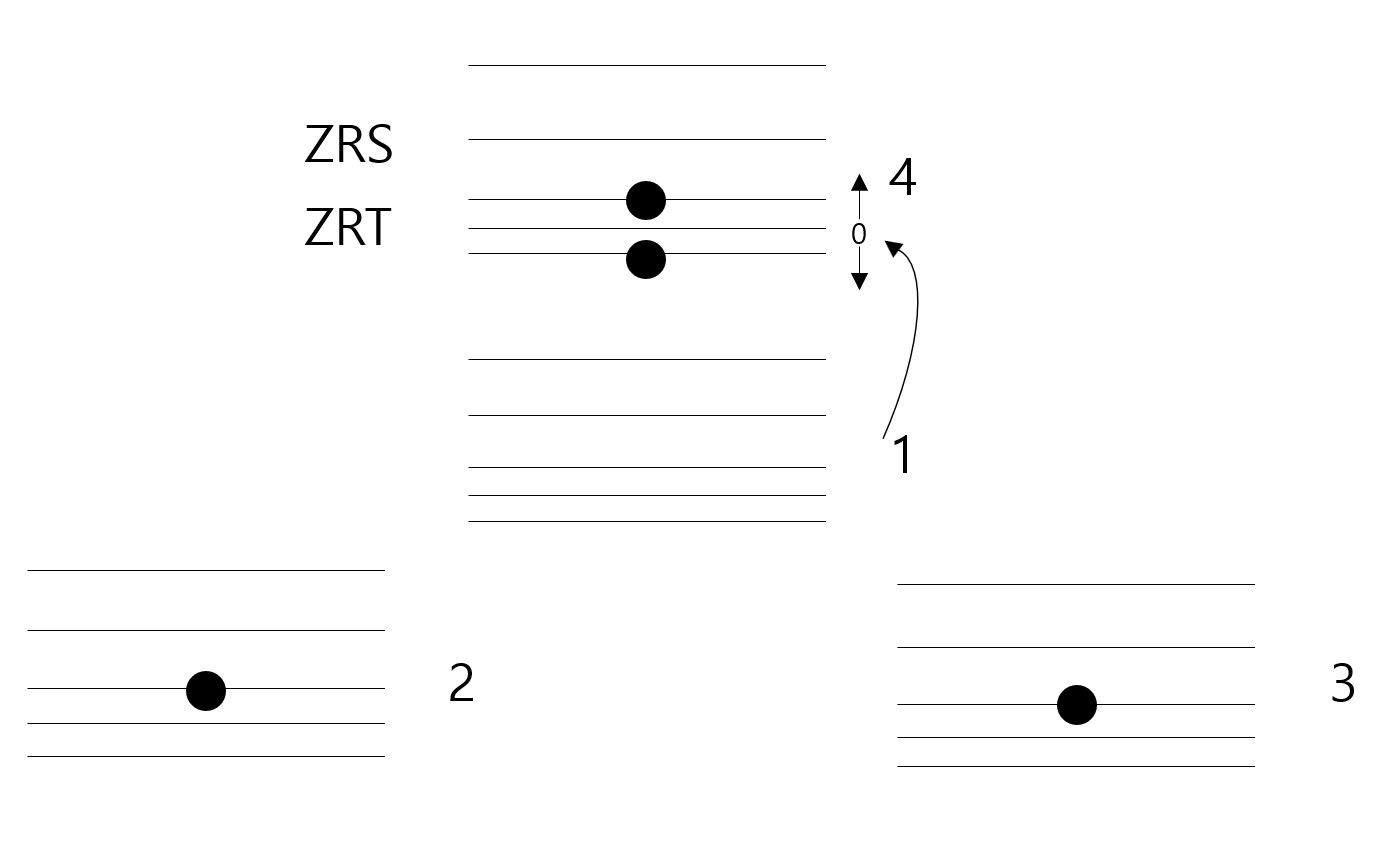}
\centering
\caption{Hopping of the site 1 Zhang-Rice triplet spin-up state to the site 4}
\label{fig:tb2}
\end{figure}

\begin{figure}[ht]
\includegraphics[scale=0.38]{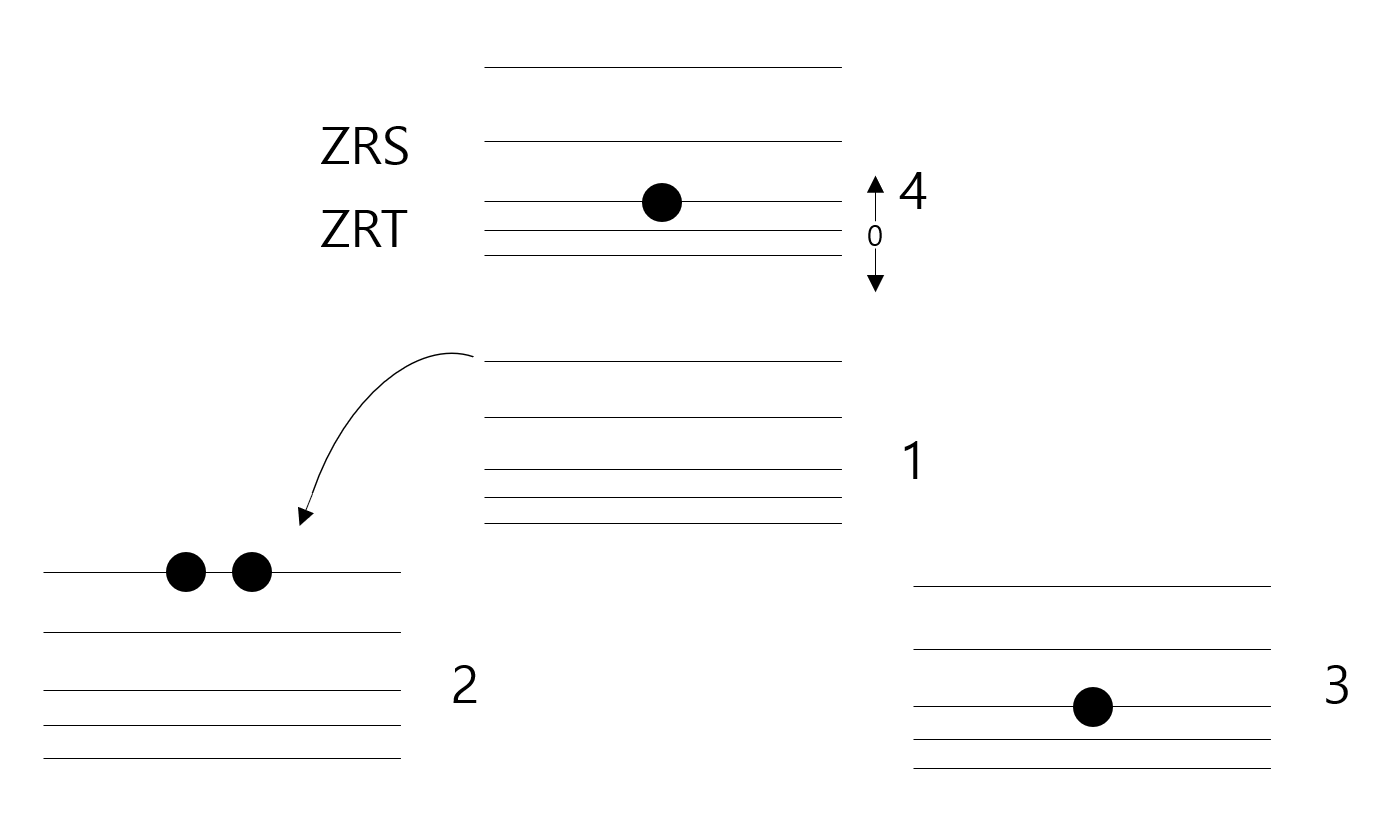}
\centering
\caption{Hopping of the site 1 Zhang-Rice triplet spin-up state to the site 2}
\label{fig:tb3}
\end{figure}

Based on these constructions, we discuss how the microscopic hopping terms of (\ref{eq:hopping12}), (\ref{eq:hopping13}), and (\ref{eq:hopping14}) can be represented by these Zhang-Rice triplet creation and annihilation operators. Applying these microscopic hopping terms twice to the state $|1_T\rangle_{\uparrow}\otimes|2_T\rangle_{\uparrow}\otimes|3_T\rangle_{\uparrow}\otimes|4_T\rangle_{\downarrow}$, we have two possible cases as shown in Figures \ref{fig:tb1}, \ref{fig:tb2}, and \ref{fig:tb3}. Figure \ref{fig:tb2} shows that if there is hopping between conversely spin oriented two Zhang-Rice states, the two Zhang-Rice triplet states can occupy the same site with their opposite spin orientations. On the other hand, Figure \ref{fig:tb3} demonstrates that the Zhang-Rice triplet states with the  same spin orientations cannot be allowed, but one Zhang-Rice state particle goes into another higher energy state because the Zhang-Rice triplet state is neither a boson nor a fermion. 

If we think the ground state of the site 1 and its nearest neighborhood like $|1_T\rangle_{\uparrow}\otimes|2_T\rangle_{\uparrow}\otimes|3_T\rangle_{\uparrow}\otimes|4_T\rangle_{\downarrow}$, the effective tight binding model is given by
\begin{equation}
H_{eff}=E_t\Sigma_{i=1}^{4}T^{(i)\dagger}T^{(i)}+t_{14}\Biggl(T^{(1)\dagger}_\uparrow T^{(4)}_\downarrow+h.c\Biggl).
\label{eq:tbm}
\end{equation}

\noindent
Here, $E_t$ is the energy of the Zhang-Rice triplet state. The hopping integral t\textsubscript{14} is calculated, given by 
\begin{equation}
\resizebox{\textwidth}{!}{$%
\begin{aligned}
t_{14}=&-\frac{(t\Delta\theta)^2}{4}\Biggl[\frac{1}{\epsilon_p-\epsilon_d}\Biggl(1+\frac{1}{2}\frac{B(\lambda_1-\lambda_2)}{\epsilon_p-\epsilon_d}\Biggr)+\frac{1}{U-J-\epsilon_p+\epsilon_d}\Biggl(1+\frac{1}{2}\frac{B(\lambda_1-\lambda_2)}{U-J-\epsilon_p+\epsilon_d}\Biggr)\Biggr]\\
&=\Sigma_{pp}\frac{\Biggl(\langle 4_T|_\uparrow\otimes\langle2_T|_\uparrow\otimes\langle 3_T|_\uparrow\otimes\langle 4_T|_\downarrow H_{14}|pp\rangle\otimes|2_T\rangle_\uparrow\otimes|3_T\rangle_\uparrow\otimes|4_T\rangle_\downarrow\Biggr)\Biggl(\langle pp|\otimes\langle 2_T|_\uparrow\otimes\langle 3_T|_\uparrow\otimes\langle 4_T|_\downarrow H_{14}|1_T\rangle_\uparrow\otimes|2_T\rangle_\uparrow\otimes|3_T\rangle_\uparrow\otimes|4_T\rangle_\downarrow\Biggr)}{E_{pp}-E_t}\\
&+\Sigma_{dd}\frac{\Biggl(\langle 4_T|_\uparrow\otimes\langle2_T|_\uparrow\otimes\langle 3_T|_\uparrow\otimes\langle 4_T|_\downarrow H_{14}|dd\rangle\otimes|2_T\rangle_\uparrow\otimes|3_T\rangle_\uparrow\otimes|4_T\rangle_\downarrow\Biggr)\Biggl(\langle dd|\otimes\langle 2_T|_\uparrow\otimes\langle 3_T|_\uparrow\otimes\langle 4_T|_\downarrow H_{14}|1_T\rangle_\uparrow\otimes|2_T\rangle_\uparrow\otimes|3_T\rangle_\uparrow\otimes|4_T\rangle_\downarrow\Biggr)}{E_{dd}-E_t}.
\end{aligned}$%
}
\end{equation}
\\
Because there are two holes in the NiS\textsubscript{6} single cluster, this hopping parameter is obtained by considering the hopping term between the cluster 1 and 4, i.e., $H_{14}$ twice. The state $|pp\rangle$ ($|dd\rangle$) means that the site 1's d-orbital (p-orbital) hole is hopping to the site 4's p-orbital (d-orbital), where $E_{pp}$ and $E_{dd}$ are their energy eigenvalues.\\

Unfortunately, we could not diagonalize this effective tight binding Hamiltonian and find the corresponding wavefunction due to the many-particle wavefunction nature, where the Zhang-Rice triplet state is neither a boson nor a fermion.
        In spite of this difficulty, we verify that there exists a spin dependent hopping term t\textsubscript{14} that originates from the antiferromagnetic zig-zag pattern of NiPS\textsubscript{3}. In this respect we propose that NiS\textsubscript{6} demonstrates a real space Berry curvature in this effective lattice model, attributable to the presence of a non-zero local electric polarization in our microscopic calcuation. We leave more concrete and detailed calculations for the real space Berry curvature to a future research direction.

\section{Conclusion}
In summary, we proposed one possible microscopic mechanism for appearance of the local electric polarization in NiPS\textsubscript{3}. First, we constructed an effective lattice Hamiltonian for the NiS\textsubscript{6} single unit and showed that local hybridized multi-particle eigenstates appear in the form of Zhang-Rice triplet and singlet states between Ni d- and S p-orbitals. Second, we introduced effective hopping of holes between two NiS\textsubscript{6} units and found an extended state for these two units in the perturbation analysis of the effective hopping Hamiltonian. Finally, we considered four NiS\textsubscript{6} units with an antiferromagnetic zigzag ordering pattern and obtained an extended many-body eigenstate under external magnetic fields, following the strategy of the two-unit case. Spin-exchange interactions turn out to be responsible for charge imbalance in the ligand states in the presence of the atomic spin-orbit interactions of Ni and S. As a result, we found an analytic expression for the local electric polarization, controlled by applied magnetic fields. More quantitatively, we obtain $|e\vec{P}|/V\sim 10^{-1}\mu C/m^{2}$ for the magnetic field $B=10T$, where V is a volume of NiS\textsubscript{6}. 

The analytic expression for the local electric polarization may be naturally reformulated in terms of the emergent real space Berry curvature. In particular, we constructed an effective tight binding lattice Hamiltonian in terms of the Zhang-Rice triplet creation and annihilation operators as emergent effective degrees of freedom. Since these effective many-body degrees of freedom are neither bosons nor fermions, we could not diagonalize the effective lattice Hamiltonian and find the resulting eigenstate. 

We emphasize that spin-orbit interactions are not taken into account in the hopping energy contribution, which turn out to be sub-leading (See the appendix for more details). We recall that the central role of the atomic spin-orbit coupling is to give the difference in the effective Zeeman coupling constant. In this respect the reformulation based on Berry curvature would propose novel mechanism on how the Berry curvature could arise in strongly correlated systems as a result of forming multi-particle entanglement states.

\section*{Acknowlegments}
K.-S. Kim was supported by the Ministry of Education, Science, and Technology (NRF-2021R1A2C1006453 and NRF-2021R1A4A3029839) of the National Research Foundation of Korea (NRF) and by TJ Park Science Fellowship of the POSCO TJ Park Foundation. We appreciate fruitful discussions with B.-H. Kim and Y.-W. Son.

\appendix
\section{Zhang-Rice triplet and singlet states in the NiS\textsubscript{6} single unit}

\setcounter{MaxMatrixCols}{40}
\setcounter{equation}{0}
\setcounter{figure}{0}

As discussed in the main text, the effective lattice Hamiltonian for the NiS\textsubscript{6} single unit is give by

\begin{equation}
\resizebox{\textwidth}{!}{$%
 \begin{aligned}
H_K=U\sum_{m}d^\dagger_{m\uparrow}d_{m\uparrow}d^\dagger_{m\downarrow}d_{m\downarrow}+&(U-2J)\sum_{m\neq m'}d^\dagger_{m\uparrow} d_{m\uparrow} d^\dagger_{m'\downarrow} d_{m'\downarrow}+(U-3J)\sum_{m<m',\sigma}d^\dagger_{m\sigma} d_{m\sigma} d^\dagger_{m'\sigma}d_{m'\sigma}\\
 &-J\sum_{m\neq m'}d^\dagger_{m\uparrow}d_{m\downarrow}d^\dagger_{m'\downarrow}d_{m'\uparrow}+J\sum_{m\neq m'}d^\dagger_{m\uparrow}d^\dagger_{m\downarrow}d_{m'\downarrow}d_{m'\uparrow}+\epsilon_d\sum_{m\sigma}d^{\dagger}_{m\sigma}d_{m\sigma}+\epsilon_p\sum_{m\sigma}p^{\dagger}_{m\sigma}p_{m\sigma}
 \end{aligned}$%
 }
\label{seq:KM}
\end{equation}
 \begin{equation}
 \resizebox{\textwidth}{!}{$%
H_t = t \sum_{\sigma} \Biggl\{\frac{\sqrt{3}}{2}d^\dagger_{x^2-y^2\sigma}(p_{1y\sigma}-p_{2x\sigma}-p_{3y\sigma}+p_{4x\sigma})+d^\dagger_{z^2\sigma} \left[\frac{1}{2}(p_{1y\sigma}+p_{2x\sigma}-p_{3y\sigma}-p_{4x\sigma})-p_{5z\sigma}+p_{6z\sigma}\right]\Biggr\}+h.c.$%
}
\label{seq:hopping}
\end{equation}
First, let's consider the case when spins of both d- and p-orbital are parallel. Then, we represent this effective Hamiltonian $H=H_K+H_t$ as follows
\begin{equation}
\resizebox{1\hsize}{!}{$
H\doteq
\begin{pNiceMatrix}[margin,last-col]
 \epsilon_p+\epsilon_d & 0 & 0 & 0 & 0 & 0 & 0 & 0 & 0 & 0&| & \frac{\sqrt{3}}{2}t & -\frac{1}{2}t & \frac{1}{2}t & \frac{1}{2}t & t & -t & 0 & 0 & 0 & 0 & 0 & 0 & 0 & 0 & 0& d_{z^2\sigma} p_{1y\sigma}\\
 0 & \epsilon_p+\epsilon_d & 0 & 0 & 0 & 0 & 0 & 0 & 0 & 0&| & -\frac{\sqrt{3}}{2}t & \frac{1}{2}t & 0 & 0 & 0 & 0 & \frac{1}{2}t & \frac{1}{2}t & t & -t & 0 & 0 & 0 & 0 & 0& d_{z^2\sigma} p_{2x\sigma}\\
 0 & 0 & \epsilon_p+\epsilon_d & 0 & 0 & 0 & 0 & 0 & 0 & 0&| & -\frac{\sqrt{3}}{2}t & 0 & \frac{1}{2}t & 0 & 0 & 0 & \frac{1}{2}t & 0 & 0 & 0 & \frac{1}{2}t & t & -t & 0 & 0& d_{z^2\sigma} p_{3y\sigma}\\
 0 & 0 & 0 & \epsilon_p+\epsilon_d & 0 & 0 & 0 & 0 & 0 & 0&| & \frac{\sqrt{3}}{2}t & 0 & 0 & \frac{1}{2}t & 0 & 0 & 0 & \frac{1}{2}t & 0 & 0 & -\frac{1}{2}t & 0 & 0 & t & -t& d_{z^2\sigma} p_{4x\sigma}\\
 0 & 0 & 0 & 0 & \epsilon_p+\epsilon_d & 0 & 0 & 0 & 0 & 0&| & -\frac{1}{2}t & \frac{\sqrt{3}}{2}t & \frac{\sqrt{3}}{2}t & -\frac{\sqrt{3}}{2}t & 0 & 0 & 0 & 0 & 0 & 0 & 0 & 0 & 0 & 0 & 0& d_{x^2-y^2\sigma}p_{1y\sigma}\\
 0 & 0 & 0 & 0 & 0 & \epsilon_p+\epsilon_d & 0 & 0 & 0 & 0&| & -\frac{1}{2}t & \frac{\sqrt{3}}{2}t & 0 & 0 & 0 & 0 & \frac{\sqrt{3}}{2}t & -\frac{\sqrt{3}}{2}t & 0 & 0 & 0 & 0 & 0 & 0 & 0& d_{x^2-y^2\sigma}p_{2x\sigma}\\
 0 & 0 & 0 & 0 & 0 & 0 & \epsilon_p+\epsilon_d & 0 & 0 & 0&| & \frac{1}{2}t & 0 & \frac{\sqrt{3}}{2}t & 0 & 0 & 0 & -\frac{\sqrt{3}}{2}t & 0 & 0 & 0 & -\frac{\sqrt{3}}{2}t & 0 & 0 & 0 & 0& d_{x^2-y^2\sigma}p_{3y\sigma}\\
 0 & 0 & 0 & 0 & 0 & 0 & 0 & \epsilon_p+\epsilon_d & 0 & 0&| & \frac{1}{2}t & 0 & 0 & \frac{\sqrt{3}}{2}t & 0 & 0 & 0 & -\frac{\sqrt{3}}{2}t & 0 & 0 & -\frac{\sqrt{3}}{2}t & 0 & 0 & 0 & 0& d_{x^2-y^2\sigma}p_{4x\sigma}\\
 0 & 0 & 0 & 0 & 0 & 0 & 0 & 0 & \epsilon_p+\epsilon_d & 0&| & t & 0 & 0 & 0 & \frac{\sqrt{3}}{2}t & 0 & 0 & 0 & -\frac{\sqrt{3}}{2}t & 0 & 0 & -\frac{\sqrt{3}}{2}t & 0 & \frac{\sqrt{3}}{2}t & 0& d_{x^2-y^2\sigma}p_{5z\sigma}\\
 0 & 0 & 0 & 0 & 0 & 0 & 0 & 0 & 0 & \epsilon_p+\epsilon_d&| & -t & 0 & 0 & 0 & 0 & \frac{\sqrt{3}}{2}t & 0 & 0 & 0 & -\frac{\sqrt{3}}{2}t & 0 & 0 & -\frac{\sqrt{3}}{2}t & 0 & \frac{\sqrt{3}}{2}t& d_{x^2-y^2\sigma}p_{6z\sigma}\\
\hline
 \frac{\sqrt{3}}{2}t & -\frac{\sqrt{3}}{2}t & -\frac{\sqrt{3}}{2}t & \frac{\sqrt{3}}{2}t & -\frac{1}{2}t & -\frac{1}{2}t & \frac{1}{2}t & \frac{1}{2}t & t & -t&| &  U-3J+2\epsilon_d &0 & 0 & 0 & 0 & 0 & 0 & 0 & 0 & 0 & 0 & 0 & 0 & 0 & 0& d_{z^2\sigma}d_{x^2-y^2\sigma}\\
 -\frac{1}{2}t & \frac{1}{2}t & 0 & 0 & \frac{\sqrt{3}}{2}t & \frac{\sqrt{3}}{2}t & 0 & 0 & 0 & 0&| & 0 &  2\epsilon_p &  0 & 0 & 0 & 0 & 0 & 0 & 0 & 0 & 0 & 0 & 0 & 0 & 0& p_{1y\sigma}p_{2x\sigma}\\
 \frac{1}{2}t & 0 & \frac{1}{2}t & 0 & \frac{\sqrt{3}}{2}t & 0 & \frac{\sqrt{3}}{2}t & 0 & 0 & 0&| & 0 & 0 &  2\epsilon_p &  0 & 0 & 0 & 0 & 0 & 0 & 0 & 0 & 0 & 0 & 0 & 0& p_{1y\sigma} p_{3y\sigma}\\
 \frac{1}{2}t & 0 & 0 & \frac{1}{2}t & -\frac{\sqrt{3}}{2}t & 0 & 0 & \frac{\sqrt{3}}{2}t & 0 & 0&| & 0 & 0 & 0 &  2\epsilon_p &  0 & 0 & 0 & 0 & 0 & 0 & 0 & 0 & 0 & 0 & 0& p_{1y\sigma} p_{4x\sigma}\\
 t & 0 & 0 & 0 & 0 & 0 & 0 & 0 &  \frac{\sqrt{3}}{2}t & 0&| & 0 & 0 & 0 & 0 &  2\epsilon_p &  0 & 0 & 0 & 0 & 0 & 0 & 0 & 0 & 0 & 0& p_{1y\sigma} p_{5z\sigma}\\
 -t & 0 & 0 & 0 & 0 & 0 & 0 & 0 & 0 &  \frac{\sqrt{3}}{2}t&| & 0 & 0 & 0 & 0 & 0 &  2\epsilon_p &  0 & 0 & 0 & 0 & 0 & 0 & 0 & 0 & 0& p_{1y\sigma} p_{6z\sigma}\\
 0 & \frac{1}{2}t & \frac{1}{2}t & 0 & 0 &  \frac{\sqrt{3}}{2}t & - \frac{\sqrt{3}}{2}t & 0 & 0 & 0&| & 0 & 0 & 0 & 0 & 0 & 0 &  2\epsilon_p &  0 & 0 & 0 & 0 & 0 & 0 & 0 & 0& p_{2x\sigma} p_{3y\sigma}\\
 0 & \frac{1}{2}t & 0 & \frac{1}{2}t & 0 & - \frac{\sqrt{3}}{2}t & 0 & - \frac{\sqrt{3}}{2}t & 0 & 0&| & 0 & 0 & 0 & 0 & 0 & 0 & 0 &  2\epsilon_p &  0 & 0 & 0 & 0 & 0 & 0 & 0&  p_{2x\sigma} p_{4x\sigma}\\
 0 & t & 0 & 0 & 0 & 0 & 0 & 0 & - \frac{\sqrt{3}}{2}t & 0&| & 0 & 0 & 0 & 0 & 0 & 0 & 0 & 0 &  2\epsilon_p &  0 & 0 & 0 & 0 & 0 & 0&  p_{2x\sigma} p_{5z\sigma}\\
 0 & -t & 0 & 0 & 0 & 0 & 0 & 0 & 0 & - \frac{\sqrt{3}}{2}t&| & 0 & 0 & 0 & 0 & 0 & 0 & 0 & 0 & 0 &  2\epsilon_p &  0 & 0 & 0 & 0 & 0& p_{2x\sigma} p_{6z\sigma}\\
 0 & 0 & \frac{1}{2}t & -\frac{1}{2}t & 0 & 0 & - \frac{\sqrt{3}}{2}t & - \frac{\sqrt{3}}{2}t & 0 & 0&| & 0 & 0 & 0 & 0 & 0 & 0 & 0 & 0 & 0 & 0 &  2\epsilon_p &  0 & 0 & 0 & 0&  p_{3y\sigma} p_{4x\sigma}\\
 0 & 0 & t & 0 & 0 & 0 & 0 & 0 & - \frac{\sqrt{3}}{2}t & 0&| & 0 & 0 & 0 & 0 & 0 & 0 & 0 & 0 & 0 & 0 &  0 & 2\epsilon_p &  0 & 0 & 0& p_{3y\sigma} p_{5z\sigma}\\
 0 & 0 & -t & 0 & 0 & 0 & 0 & 0 & 0 & - \frac{\sqrt{3}}{2}t&| & 0 & 0 & 0 & 0 & 0 & 0 & 0 & 0 & 0 & 0 & 0 & 0 &  2\epsilon_p &  0 & 0& p_{3y\sigma} p_{6z\sigma}\\
 0 & 0 & 0 & t & 0 & 0 & 0 & 0 &  \frac{\sqrt{3}}{2}t & 0&| & 0 & 0 & 0 & 0 & 0 & 0 & 0 & 0 & 0 & 0 & 0 & 0 & 0 &  2\epsilon_p &  0&  p_{4x\sigma} p_{5z\sigma}\\
 0 & 0 & 0 & -t & 0 & 0 & 0 & 0 & 0 &  \frac{\sqrt{3}}{2}t&| & 0 & 0 & 0 & 0 & 0 & 0 & 0 & 0 & 0 & 0 & 0 & 0 & 0 & 0 &  2\epsilon_p & p_{4x\sigma} p_{6z\sigma}\\
\end{pNiceMatrix} .
\label{seq:pl}
$}
\end{equation}

Let $P_0$ be a projection operator to $d^\dagger p^\dagger$, i.e., $P_0=|dp\rangle\langle dp|$, and $P_1$ be a projection operator to $d^\dagger d^\dagger$ and $p^\dagger p^\dagger$, i.e., $P_1=|pp\rangle\langle pp| + |dd\rangle\langle dd|$. Then, $P_0+P_1$ makes a complete Hilbert space for the effective hamiltonian $H$. Using these projection operators, we represent the effective Hamiltonian in the following way
\begin{equation}
H\doteq
\begin{pNiceMatrix}[last-col,vlines]
\\
 P_0HP_0  &  P_0HP_1  & d^\dagger p^\dagger \\
 \\
 \hline
 \\
 P_1HP_0  &  P_1HP_1  & d^\dagger d^\dagger, p^\dagger p^\dagger .\\
 \\
 \end{pNiceMatrix}
 \label{seq:mt}
\end{equation}
\\
Consider the Schrodinger equation with an eigenstate $\Psi$ and an eigenvalue $E$ as
$$
    H\Psi=E\Psi = H(P_0+P_1)\Psi=E(P_0+P_1)\Psi ,
$$
where $P_0+P_1 = I$ was introduced. Applying the projection operator $P_1$ to both sides, we obtain
$$
(P_1HP_0+P_1HP_1)\Psi=EP_1\Psi \longrightarrow P_1\Psi=\frac{1}{E-P_1HP_1}P_1HP_0\Psi .
$$
\noindent
Applying the projection operator $P_0$ to both sides of the first equation and replacing $P_1\Psi$ with the above, we obtain

$$
\left(P_0HP_0+P_0HP_1\frac{1}{E-P_1HP_1}P_1HP_0\right)P_0\Psi=EP_0\Psi .
$$

Introducing (\ref{seq:pl}) into the above, we obtain eigenvalues as\\
\resizebox{\textwidth}{!}{$
E=(\epsilon_p+\epsilon_d)+\Biggl\{ 0, 0, -\frac{6t^2(U-3J)}{(\epsilon_p-\epsilon_d)(U-3J+\epsilon_d-\epsilon_p)}, -\frac{3t^2}{\epsilon_p-\epsilon_d}, -\frac{3t^2}{\epsilon_p-\epsilon_d}, -\frac{3t^2}{\epsilon_p-\epsilon_d}, -\frac{3t^2}{\epsilon_p-\epsilon_d}, -\frac{3t^2}{\epsilon_p-\epsilon_d}, -\frac{3t^2}{\epsilon_p-\epsilon_d}, -\frac{2t^2}{\epsilon_p-\epsilon_d}\Biggr\} .
$}
The ground-state energy is given by $(\epsilon_p+\epsilon_d)-\frac{6t^2(U-3J)}{(\epsilon_p-\epsilon_d)(U-3J+\epsilon_d-\epsilon_p)}$ with its eigenstate
\begin{equation}
\resizebox{\textwidth}{!}{$%
|ZRT, \sigma \rangle = \frac{1}{\sqrt{6}}\Biggl\{-\frac{\sqrt{3}}{2}d^\dagger_{z^2\sigma} (p^\dagger_{1y\sigma}-p^\dagger_{2x\sigma}-p^\dagger_{3y\sigma}+p^\dagger_{4x\sigma})+d^\dagger_{x^2-y^2\sigma} \left[\frac{1}{2}(p^\dagger_{1y\sigma}+p^\dagger_{2x\sigma}-p^\dagger_{3y\sigma}-p^\dagger_{4x\sigma})-p^\dagger_{5z\sigma}+p^\dagger_{6z\sigma}\right] \Biggr\}|0\rangle .$
}
\label{seq:zrt1}
\end{equation}
\noindent
Second, let's consider the case when spins of both d- and p-orbital are anti-parallel. Then, the effective Hamiltonian $H$ can be represented as (\ref{seq:apl1}) and (\ref{seq:apl2}) with $P_0HP_0=(\epsilon_p+\epsilon_d)\ 1_{20\times20}$. Repeating the same procedure as the above, we obtain the energy eigenvalue $E=(\epsilon_p+\epsilon_d)-\frac{6t^2(U-3J)}{(\epsilon_p-\epsilon_d)(U-3J+\epsilon_d-\epsilon_p)}$ with the corresponding eigenstate

\begin{equation}
\resizebox{\textwidth}{!}{$%
\begin{aligned}
|ZRT, 0 \rangle = \frac{1}{2\sqrt{3}}& \Biggl\{\left(-\frac{\sqrt{3}}{2}d^\dagger_{z^2\uparrow}(p^\dagger_{1y\downarrow}-p^\dagger_{2x\downarrow} -p^\dagger_{3y\downarrow}+p^\dagger_{4x\downarrow})+d^\dagger_{x^2-y^2\downarrow} \left[\frac{1}{2}(p^\dagger_{1y\downarrow}+p^\dagger_{2x\downarrow}-p^\dagger_{3y\downarrow}-p^\dagger_{4x\downarrow}) -p^\dagger_{5z\downarrow}+p^\dagger_{6z\downarrow}\right]\right)\\
&+\left(-\frac{\sqrt{3}}{2}d^\dagger_{z^2\downarrow} (p^\dagger_{1y\uparrow}-p^\dagger_{2x\uparrow}-p^\dagger_{3y\uparrow}+p^\dagger_{4x\uparrow}) +d^\dagger_{x^2-y^2\downarrow} \left[\frac{1}{2}(p^\dagger_{1y\uparrow}+p^\dagger_{2x\uparrow}-p^\dagger_{3y\uparrow} -p^\dagger_{4x\uparrow})-p^\dagger_{5z\uparrow}+p^\dagger_{6z\uparrow}\right]\right)\Biggr\}|0\rangle .
\end{aligned}$%
}
\label{seq:zrt2}
\end{equation}

The Zhang-Rice singlet state with an energy eigenvalue $E=(\epsilon_p+\epsilon_d)-\frac{6t^2(U-J)}{(\epsilon_p-\epsilon_d)(U-J+\epsilon_d-\epsilon_p)}$ is given by
\begin{equation}
\resizebox{\textwidth}{!}{$%
\begin{aligned}
|ZRS\rangle=\frac{1}{2\sqrt{3}}&\Biggl\{\left(\frac{\sqrt{3}}{2}d^\dagger_{z^2\uparrow}(p^\dagger_{1y\downarrow}-p^\dagger_{2x\downarrow} -p^\dagger_{3y\downarrow}+p^\dagger_{4x\downarrow})+d^\dagger_{x^2-y^2\downarrow}\left[\frac{1}{2}(p^\dagger_{1y\downarrow} +p^\dagger_{2x\downarrow}-p^\dagger_{3y\downarrow}-p^\dagger_{4x\downarrow})-p^\dagger_{5z\downarrow}+p^\dagger_{6z\downarrow}\right]\right)\\
&-\left(\frac{\sqrt{3}}{2}d^\dagger_{z^2\downarrow}(p^\dagger_{1y\uparrow}-p^\dagger_{2x\uparrow}-p^\dagger_{3y\uparrow} +p^\dagger_{4x\uparrow})+d^\dagger_{x^2-y^2\downarrow}\left[\frac{1}{2}(p^\dagger_{1y\uparrow}+p^\dagger_{2x\uparrow} -p^\dagger_{3y\uparrow}-p^\dagger_{4x\uparrow})-p^\dagger_{5z\uparrow}+p^\dagger_{6z\uparrow}\right]\right)\Biggr\}|0\rangle .
\end{aligned}$%
}
\label{seq:zrs}
\end{equation}

\newpage
~~~~~~~~~~~~~~~~~~~~~~~~~~~~~~~~~~~~~~~~~~~~~~~~
\\
~
\\
\newline
~
\newline
\begin{turn}{90}
\begin{minipage}{\linewidth}

\begin{equation}\label{seq:apl1}
\hspace*{-4cm}
\resizebox{1.35\hsize}{!}{$%
P_0HP_1=
\begin{pNiceMatrix}[margin,last-col,first-row]
d_{z^2\uparrow}d_{z^2\downarrow} & d_{x^2-y^2\uparrow}d_{x^2-y^2\downarrow} & p_{1y\uparrow}p_{1y\downarrow} & p_{2x\uparrow}p_{2x\downarrow} & p_{3y\uparrow}p_{3y\downarrow} & p_{4x\uparrow}p_{4x\downarrow} & d_{x^2-y^2\downarrow}d_{z^2\uparrow} & p_{1y\downarrow}p_{2x\uparrow} & p_{1y\downarrow}p_{3y\uparrow} & p_{1y\downarrow}p_{4x\uparrow} & p_{1y\downarrow}p_{5z\uparrow} & p_{1y\downarrow}p_{6z\uparrow} & p_{2x\downarrow}p_{3y\uparrow} & p_{2x\downarrow}p_{4x\uparrow} & p_{2x\downarrow}p_{5z\uparrow} & p_{2x\downarrow}p_{6z\uparrow} & p_{3y\downarrow}p_{4x\uparrow} & p_{3y\downarrow}p_{5z\uparrow} & p_{3y\downarrow}p_{6z\uparrow} & p_{4x\downarrow}p_{5z\uparrow} & p_{4x\downarrow}p_{6z\uparrow} & d_{x^2-y^2\uparrow}d_{z^2\downarrow} & p_{1y \uparrow}p_{2x \downarrow} & p_{1y\uparrow}p_{3y \downarrow} & p_{1y \uparrow}p_{4x  \downarrow} & p_{1y \uparrow}p_{5z  \downarrow} & p_{1y \uparrow}p_{6z  \downarrow} & p_{2x \uparrow}p_{3y  \downarrow} & p_{2x \uparrow}p_{4x  \downarrow} & p_{2x \uparrow}p_{5z  \downarrow} & p_{2x \uparrow}p_{6z  \downarrow} & p_{3y \uparrow}p_{4x  \downarrow} & p_{3y\uparrow}p_{5z \downarrow} & p_{3y \uparrow}p_{6z  \downarrow} & p_{4x \uparrow}p_{5z  \downarrow} & p_{4x \uparrow}p_{6z  \downarrow}\\
\frac{1}{2}t & 0 & \frac{1}{2}t & 0 & 0 & 0 & -\frac{\sqrt{3}}{2}t & -\frac{1}{2}t & \frac{1}{2}t & \frac{1}{2}t & t & -t & 0 & 0 & 0 & 0 & 0 & 0 & 0 & 0 & 0 & 0 & 0 & 0 & 0 & 0 & 0 & 0 & 0 & 0 & 0 & 0 & 0 & 0 & 0 & 0 & d_{z^2\uparrow}p_{1y\downarrow}\\
-\frac{1}{2}t & 0 & -\frac{1}{2}t & 0 & 0 & 0 & 0 & 0 & 0 & 0 & 0 & 0 & 0 & 0 & 0 & 0 & 0 & 0 & 0 & 0 & 0 & -\frac{\sqrt{3}}{2}t & -\frac{1}{2}t & \frac{1}{2}t & \frac{1}{2}t & t & -t & 0 & 0 & 0 & 0 & 0 & 0 & 0 & 0 & 0 &d_{z^2 \downarrow}p_{1y \uparrow}\\
\frac{1}{2}t & 0 & 0 & \frac{1}{2}t & 0 & 0 & \frac{\sqrt{3}}{2}t & 0 & 0 & 0 & 0 & 0 & \frac{1}{2}t & \frac{1}{2}t & t & -t & 0 & 0 & 0 & 0 & 0 & 0 & 0 & 0 & 0 & 0 & 0 & 0 & 0 & 0 & 0 & 0 & 0 & 0 & 0 & 0 & d_{z^2\uparrow}p_{2x\downarrow}\\
-\frac{1}{2}t & 0 & 0 & -\frac{1}{2}t & 0 & 0 & 0 & \frac{1}{2}t & 0 & 0 & 0 & 0 & 0 & 0 & 0 & 0 & 0 & 0 & 0 & 0 & 0 & \frac{\sqrt{3}}{2}t & 0 & 0 & 0 & 0 & 0 & \frac{1}{2}t & \frac{1}{2}t & t & -t & 0 & 0 & 0 & 0 & 0 & d_{z^2 \downarrow}p_{2x \uparrow}\\
-\frac{1}{2}t & 0 & 0 & 0 & -\frac{1}{2}t & 0 & \frac{\sqrt{3}}{2}t & 0 & 0 & 0 & 0 & 0 & 0 & 0 & 0 & 0 & \frac{1}{2}t & t & -t & 0 & 0 & 0 & 0 &  \frac{1}{2}t & 0 & 0 & 0 &  \frac{1}{2}t & 0 & 0 & 0 & 0 & 0 & 0 & 0 & 0 & d_{z^2\uparrow}p_{3y\downarrow}\\
\frac{1}{2}t & 0 & 0 & 0 & \frac{1}{2}t & 0 & 0 & 0 &  \frac{1}{2}t & 0 & 0 & 0 & \frac{1}{2}t & 0 & 0 & 0 & 0 & 0 & 0 & 0 & 0 & \frac{\sqrt{3}}{2}t & 0 & 0 & 0 & 0 & 0 & 0 & 0 & 0 & 0 & \frac{1}{2}t & t & -t & 0 & 0 & d_{z^2 \downarrow}p_{3y \uparrow}\\
-\frac{1}{2}t & 0 & 0 & 0 & 0 & -\frac{1}{2}t & -\frac{\sqrt{3}}{2}t & 0 & 0 & 0 & 0 & 0 & 0 & 0 & 0 & 0 & 0 & 0 & 0 & t & -t & 0 & 0 & 0 & \frac{1}{2}t & 0 & 0 & 0 & \frac{1}{2}t & 0 & 0 & -\frac{1}{2}t & 0 & 0 & 0 & 0 &d_{z^2\uparrow}p_{4x\downarrow}\\
\frac{1}{2}t & 0 & 0 & 0 & 0 & \frac{1}{2}t & 0 & 0 & 0 & \frac{1}{2}t & 0 & 0 & 0 & \frac{1}{2}t & 0 & 0 & -\frac{1}{2}t & 0 & 0 & 0 & 0 & - \frac{\sqrt{3}}{2}t & 0 & 0 & 0 & 0 & 0 & 0 & 0 & 0 & 0 & 0 & 0 & 0 & t & -t & d_{z^2 \downarrow}p_{4x \uparrow}\\
0 & \frac{\sqrt{3}}{2}t & \frac{\sqrt{3}}{2}t & 0 & 0 & 0 & 0 & \frac{\sqrt{3}}{2}t & \frac{\sqrt{3}}{2}t & -\frac{\sqrt{3}}{2}t & 0 & 0 & 0 & 0 & 0 & 0 & 0 & 0 & 0 & 0 & 0 & \frac{1}{2}t & 0 & 0 & 0 & 0 & 0 & 0 & 0 & 0 & 0 & 0 & 0 & 0 & 0 & 0 & d_{x^2-y^2\uparrow}p_{1y\downarrow}\\
0 & -\frac{\sqrt{3}}{2}t & -\frac{\sqrt{3}}{2}t & 0 & 0 & 0 & \frac{1}{2}t & 0 & 0 & 0 & 0 & 0 & 0 & 0 & 0 & 0 & 0 & 0 & 0 & 0 & 0 & 0 & \frac{\sqrt{3}}{2}t & \frac{\sqrt{3}}{2}t & -\frac{\sqrt{3}}{2}t & 0 & 0 & 0 & 0 & 0 & 0 & 0 & 0 & 0 & 0 & 0 & d_{x^2-y^2 \downarrow}p_{1y \uparrow}\\
0 & -\frac{\sqrt{3}}{2}t & 0 & -\frac{\sqrt{3}}{2}t & 0 & 0 & 0 & 0 & 0 & 0 & 0 & 0 & \frac{\sqrt{3}}{2}t & -\frac{\sqrt{3}}{2}t  & 0 & 0 & 0 & 0 & 0 & 0 & 0 & \frac{1}{2}t & \frac{\sqrt{3}}{2}t  & 0 & 0 & 0 & 0 & 0 & 0 & 0 & 0 & 0 & 0 & 0 & 0 & 0 & d_{x^2-y^2\uparrow}p_{2x\downarrow}\\
0 & \frac{\sqrt{3}}{2}t & 0 & \frac{\sqrt{3}}{2}t & 0 & 0 & \frac{1}{2}t & \frac{\sqrt{3}}{2}t & 0 & 0 & 0 & 0 & 0 & 0 & 0 & 0 & 0 & 0 & 0 & 0 & 0 & 0 & 0 & 0 & 0 & 0 & 0 & \frac{\sqrt{3}}{2}t & -\frac{\sqrt{3}}{2}t & 0 & 0 & 0 & 0 & 0 & 0 & 0 & d_{x^2-y^2 \downarrow}p_{2x \uparrow}\\
0 & -\frac{\sqrt{3}}{2}t & 0 & 0 & -\frac{\sqrt{3}}{2}t & 0 & 0 & 0 & 0 & 0 & 0 & 0 & 0 & 0 & 0 & 0 & -\frac{\sqrt{3}}{2}t & 0 & 0 & 0 & 0 & -\frac{1}{2}t & 0 & \frac{\sqrt{3}}{2}t & 0 & 0 & 0 & -\frac{\sqrt{3}}{2}t & 0 & 0 & 0 & 0 & 0 & 0 & 0 & 0 & d_{x^2-y^2\uparrow}p_{3y\downarrow}\\
0 & \frac{\sqrt{3}}{2}t & 0 & 0 & \frac{\sqrt{3}}{2}t & 0 & -\frac{1}{2}t & 0 & \frac{\sqrt{3}}{2}t & 0 & 0 & 0 & -\frac{\sqrt{3}}{2}t & 0 & 0 & 0 & 0 & 0 & 0 & 0 & 0 & 0 & 0 & 0 & 0 & 0 & 0 & 0 & 0 & 0 & 0 & -\frac{\sqrt{3}}{2}t & 0 & 0 & 0 & 0 & d_{x^2-y^2 \downarrow}p_{3y \uparrow}\\
0 & \frac{\sqrt{3}}{2}t & 0 & 0 & 0 & \frac{\sqrt{3}}{2}t & 0 & 0 & 0 & 0 & 0 & 0 & 0 & 0 & 0 & 0 & 0 & 0 & 0 & 0 & 0 & -\frac{1}{2}t & 0 & 0 & \frac{\sqrt{3}}{2}t & 0 & 0 & 0 & -\frac{\sqrt{3}}{2}t & 0 & 0 & -\frac{\sqrt{3}}{2}t & 0 & 0 & 0 & 0 & d_{x^2-y^2\uparrow}p_{4x\downarrow}\\
0 & -\frac{\sqrt{3}}{2}t & 0 & 0 & 0 & -\frac{\sqrt{3}}{2}t & -\frac{1}{2}t & 0 & 0 & \frac{\sqrt{3}}{2}t & 0 & 0 & 0 & -\frac{\sqrt{3}}{2}t & 0 & 0 & -\frac{\sqrt{3}}{2}t & 0 & 0 & 0 & 0 & 0 & 0 & 0 & 0 & 0 & 0 & 0 & 0 & 0 & 0 & 0 & 0 & 0 & 0 & 0 & d_{x^2-y^2 \downarrow}p_{4x \uparrow}\\
0 & 0 & 0 & 0 & 0 & 0 & 0 & 0 & 0 & 0 & 0 & 0 & 0 & 0 & 0 & 0 & 0 & 0 & 0 & 0 & 0 & -t & 0 & 0 & 0 & \frac{\sqrt{3}}{2}t & 0 & 0 & 0 & -\frac{\sqrt{3}}{2}t & 0 & 0 & -\frac{\sqrt{3}}{2}t & 0 & \frac{\sqrt{3}}{2}t & 0 & d_{x^2-y^2\uparrow}p_{5z\downarrow}\\
0 & 0 & 0 & 0 & 0 & 0 & -t & 0 & 0 & 0 & \frac{\sqrt{3}}{2}t & 0 & 0 & 0 & -\frac{\sqrt{3}}{2}t & 0 & 0 & -\frac{\sqrt{3}}{2}t & 0 & \frac{\sqrt{3}}{2}t & 0 & 0 & 0 & 0 & 0 & 0 & 0 & 0 & 0 & 0 & 0 & 0 & 0 & 0 & 0 & 0 &d_{x^2-y^2 \downarrow}p_{5z\uparrow}\\
0 & 0 & 0 & 0 & 0 & 0 & 0 & 0 & 0 & 0 & 0 & 0 & 0 & 0 & 0 & 0 & 0 & 0 & 0 & 0 & 0 & t & 0 & 0 & 0 & 0 & \frac{\sqrt{3}}{2}t & 0 & 0 & 0 & -\frac{\sqrt{3}}{2}t & 0 & 0 & -\frac{\sqrt{3}}{2}t & 0 & \frac{\sqrt{3}}{2}t & d_{x^2-y^2\uparrow}p_{6z\downarrow}\\
0 & 0 & 0 & 0 & 0 & 0 & t & 0 & 0 & 0 & 0 & \frac{\sqrt{3}}{2}t & 0 & 0 & 0 & -\frac{\sqrt{3}}{2}t & 0 & 0 & -\frac{\sqrt{3}}{2}t & 0 & \frac{\sqrt{3}}{2}t & 0 & 0 & 0 & 0 & 0 & 0 & 0 & 0 & 0 & 0 & 0 & 0 & 0 & 0 & 0 & d_{x^2-y^2 \downarrow}p_{6z\uparrow}\\
\end{pNiceMatrix}~~~
$%
}%
\end{equation}

\begin{equation}\label{seq:apl2}
\hspace*{-4cm}
\resizebox{1.3\hsize}{!}{$%
P_1HP_1=
\begin{pNiceMatrix}[margin,last-col]
U+2\epsilon_d & J & 0 & 0 & 0 & 0 & 0 & 0 & 0 & 0 & 0 & 0 & 0 & 0 & 0 & 0 & 0 & 0 & 0 & 0 & 0 & 0 & 0 & 0 & 0 & 0& 0 & 0 & 0 & 0 & 0 & 0 & 0 & 0 & 0 & 0 &d_{z^2\uparrow}d_{z^2\downarrow}\\
J & U+2\epsilon_d & 0 & 0 & 0 & 0 & 0 & 0 & 0 & 0 & 0 & 0 & 0 & 0 & 0 & 0 & 0 & 0 & 0 & 0 & 0 & 0 & 0 & 0 & 0 & 0& 0 & 0 & 0 & 0 & 0 & 0 & 0 & 0 & 0 & 0 &d_{x^2-y^2\uparrow}d_{x^2-y^2\downarrow}\\
0 & 0 & 2\epsilon_p & 0 & 0 & 0 & 0 & 0 & 0 & 0 & 0 & 0 & 0 & 0 & 0 & 0 & 0 & 0 & 0 & 0 & 0 & 0 & 0 & 0 & 0 & 0& 0 & 0 & 0 & 0 & 0 & 0 & 0 & 0 & 0 & 0 &p_{1x\uparrow}p_{1x\downarrow}\\
0 & 0 & 0 & 2\epsilon_p & 0 & 0 & 0 & 0 & 0 & 0 & 0 & 0 & 0 & 0 & 0 & 0 & 0 & 0 & 0 & 0 & 0 & 0 & 0 & 0 & 0 & 0& 0 & 0 & 0 & 0 & 0 & 0 & 0 & 0 & 0 & 0 &p_{2x\uparrow}p_{2x\downarrow}\\
0 & 0 & 0 & 0 & 2\epsilon_p & 0 & 0 & 0 & 0 & 0 & 0 & 0 & 0 & 0 & 0 & 0 & 0 & 0 & 0 & 0 & 0 & 0 & 0 & 0 & 0 & 0& 0 & 0 & 0 & 0 & 0 & 0 & 0 & 0 & 0 & 0 & p_{3y\uparrow}p_{3y\downarrow}\\
0 & 0 & 0 & 0 & 0 & 2\epsilon_p & 0 & 0 & 0 & 0 & 0 & 0 & 0 & 0 & 0 & 0 & 0 & 0 & 0 & 0 & 0 & 0 & 0 & 0 & 0 & 0& 0 & 0 & 0 & 0 & 0 & 0 & 0 & 0 & 0 & 0 &p_{4x\uparrow}p_{4x\downarrow}\\
0 & 0 & 0 & 0 & 0 & 0 & U-2J+2\epsilon_d & 0 & 0 & 0 & 0 & 0 & 0 & 0 & 0 & 0 & 0 & 0 & 0 & 0 & 0 & -J & 0 & 0 & 0 & 0& 0 & 0 & 0 & 0 & 0 & 0 & 0 & 0 & 0 & 0 &d_{x^2-y^2\downarrow}d_{z^2\uparrow} \\
0 & 0 & 0 & 0 & 0 & 0 & 0 & 2\epsilon_p & 0 & 0 & 0 & 0 & 0 & 0 & 0 & 0 & 0 & 0 & 0 & 0 & 0 & 0 & 0 & 0 & 0 & 0& 0 & 0 & 0 & 0 & 0 & 0 & 0 & 0 & 0 & 0 & p_{1y\downarrow}p_{2x\uparrow} \\
0 & 0 & 0 & 0 & 0 & 0 & 0 & 0 & 2\epsilon_p & 0 & 0 & 0 & 0 & 0 & 0 & 0 & 0 & 0 & 0 & 0 & 0 & 0 & 0 & 0 & 0 & 0& 0 & 0 & 0 & 0 & 0 & 0 & 0 & 0 & 0 & 0 & p_{1y\downarrow}p_{3y\uparrow}\\
0 & 0 & 0 & 0 & 0 & 0 & 0 & 0 & 0 & 2\epsilon_p & 0 & 0 & 0 & 0 & 0 & 0 & 0 & 0 & 0 & 0 & 0 & 0 & 0 & 0 & 0 & 0& 0 & 0 & 0 & 0 & 0 & 0 & 0 & 0 & 0 & 0 & p_{1y\downarrow}p_{4x\uparrow}\\
0 & 0 & 0 & 0 & 0 & 0 & 0 & 0 & 0 & 0 & 2\epsilon_p & 0 & 0 & 0 & 0 & 0 & 0 & 0 & 0 & 0 & 0 & 0 & 0 & 0 & 0 & 0& 0 & 0 & 0 & 0 & 0 & 0 & 0 & 0 & 0 & 0 & p_{1y\downarrow}p_{5z\uparrow}\\
0 & 0 & 0 & 0 & 0 & 0 & 0 & 0 & 0 & 0 & 0 & 2\epsilon_p & 0 & 0 & 0 & 0 & 0 & 0 & 0 & 0 & 0 & 0 & 0 & 0 & 0 & 0& 0 & 0 & 0 & 0 & 0 & 0 & 0 & 0 & 0 & 0 & p_{1y\downarrow}p_{6z\uparrow}\\
0 & 0 & 0 & 0 & 0 & 0 & 0 & 0 & 0 & 0 & 0 & 0 & 2\epsilon_p & 0 & 0 & 0 & 0 & 0 & 0 & 0 & 0 & 0 & 0 & 0 & 0 & 0& 0 & 0 & 0 & 0 & 0 & 0 & 0 & 0 & 0 & 0 &  p_{2x\downarrow}p_{3y\uparrow}\\
0 & 0 & 0 & 0 & 0 & 0 & 0 & 0 & 0 & 0 & 0 & 0 & 0 & 2\epsilon_p & 0 & 0 & 0 & 0 & 0 & 0 & 0 & 0 & 0 & 0 & 0 & 0& 0 & 0 & 0 & 0 & 0 & 0 & 0 & 0 & 0 & 0 & p_{2x\downarrow}p_{4x\uparrow}\\
0 & 0 & 0 & 0 & 0 & 0 & 0 & 0 & 0 & 0 & 0 & 0 & 0 & 0 & 2\epsilon_p & 0 & 0 & 0 & 0 & 0 & 0 & 0 & 0 & 0 & 0 & 0& 0 & 0 & 0 & 0 & 0 & 0 & 0 & 0 & 0 & 0 & p_{2x\downarrow}p_{5z\uparrow} \\
0 & 0 & 0 & 0 & 0 & 0 & 0 & 0 & 0 & 0 & 0 & 0 & 0 & 0 & 0 & 2\epsilon_p & 0 & 0 & 0 & 0 & 0 & 0 & 0 & 0 & 0 & 0& 0 & 0 & 0 & 0 & 0 & 0 & 0 & 0 & 0 & 0 & p_{2x\downarrow}p_{6z\uparrow}\\
0 & 0 & 0 & 0 & 0 & 0 & 0 & 0 & 0 & 0 & 0 & 0 & 0 & 0 & 0 & 0 & 2\epsilon_p & 0 & 0 & 0 & 0 & 0 & 0 & 0 & 0 & 0& 0 & 0 & 0 & 0 & 0 & 0 & 0 & 0 & 0 & 0 & p_{3y\downarrow}p_{4x\uparrow}\\
0 & 0 & 0 & 0 & 0 & 0 & 0 & 0 & 0 & 0 & 0 & 0 & 0 & 0 & 0 & 0 & 0 & 2\epsilon_p & 0 & 0 & 0 & 0 & 0 & 0 & 0 & 0& 0 & 0 & 0 & 0 & 0 & 0 & 0 & 0 & 0 & 0 & p_{3y\downarrow}p_{5z\uparrow}\\
0 & 0 & 0 & 0 & 0 & 0 & 0 & 0 & 0 & 0 & 0 & 0 & 0 & 0 & 0 & 0 & 0 & 0 & 2\epsilon_p & 0 & 0 & 0 & 0 & 0 & 0 & 0& 0 & 0 & 0 & 0 & 0 & 0 & 0 & 0 & 0 & 0 &  p_{3y\downarrow}p_{6z\uparrow}\\
0 & 0 & 0 & 0 & 0 & 0 & 0 & 0 & 0 & 0 & 0 & 0 & 0 & 0 & 0 & 0 & 0 & 0 & 0 & 2\epsilon_p & 0 & 0 & 0 & 0 & 0 & 0& 0 & 0 & 0 & 0 & 0 & 0 & 0 & 0 & 0 & 0 & p_{4x\downarrow}p_{5z\uparrow}\\
0 & 0 & 0 & 0 & 0 & 0 & 0 & 0 & 0 & 0 & 0 & 0 & 0 & 0 & 0 & 0 & 0 & 0 & 0 & 0 & 2\epsilon_p & 0 & 0 & 0 & 0 & 0& 0 & 0 & 0 & 0 & 0 & 0 & 0 & 0 & 0 & 0 &  p_{4x\downarrow}p_{6z\uparrow}\\
0 & 0 & 0 & 0 & 0 & 0 & -J & 0 & 0 & 0 & 0 & 0 & 0 & 0 & 0 & 0 & 0 & 0 & 0 & 0 & 0 & U-2J+2\epsilon_d & 0 & 0 & 0 & 0& 0 & 0 & 0 & 0 & 0 & 0 & 0 & 0 & 0 & 0 & d_{x^2-y^2\uparrow}d_{z^2\downarrow}\\
0 & 0 & 0 & 0 & 0 & 0 & 0 & 0 & 0 & 0 & 0 & 0 & 0 & 0 & 0 & 0 & 0 & 0 & 0 & 0 & 0 & 0 & 2\epsilon_p & 0 & 0 & 0& 0 & 0 & 0 & 0 & 0 & 0 & 0 & 0 & 0 & 0 & p_{1y \uparrow}p_{2x\downarrow}\\
0 & 0 & 0 & 0 & 0 & 0 & 0 & 0 & 0 & 0 & 0 & 0 & 0 & 0 & 0 & 0 & 0 & 0 & 0 & 0 & 0 & 0 & 0 & 2\epsilon_p & 0 & 0& 0 & 0 & 0 & 0 & 0 & 0 & 0 & 0 & 0 & 0 & p_{1y\uparrow}p_{3y\downarrow}\\
0 & 0 & 0 & 0 & 0 & 0 & 0 & 0 & 0 & 0 & 0 & 0 & 0 & 0 & 0 & 0 & 0 & 0 & 0 & 0 & 0 & 0 & 0 & 0 & 2\epsilon_p & 0 & 0 & 0 & 0 & 0 & 0 & 0 & 0 & 0 & 0 & 0 &  p_{1y \uparrow}p_{4x\downarrow}\\
0 & 0 & 0 & 0 & 0 & 0 & 0 & 0 & 0 & 0 & 0 & 0 & 0 & 0 & 0 & 0 & 0 & 0 & 0 & 0 & 0 & 0 & 0 & 0 & 0 & 2\epsilon_p & 0 & 0 & 0 & 0 & 0 & 0 & 0 & 0 & 0 & 0 &  p_{1y\uparrow}p_{5z\downarrow}\\
0 & 0 & 0 & 0 & 0 & 0 & 0 & 0 & 0 & 0 & 0 & 0 & 0 & 0 & 0 & 0 & 0 & 0 & 0 & 0 & 0 & 0 & 0 & 0 & 0 & 0 & 2\epsilon_p & 0 & 0 & 0 & 0 & 0 & 0 & 0 & 0 & 0 & p_{1y\uparrow}p_{6z\downarrow}\\
0 & 0 & 0 & 0 & 0 & 0 & 0 & 0 & 0 & 0 & 0 & 0 & 0 & 0 & 0 & 0 & 0 & 0 & 0 & 0 & 0 & 0 & 0 & 0 & 0 & 0 & 0 & 2\epsilon_p & 0 & 0 & 0 & 0 & 0 & 0 & 0 & 0 & p_{2x \uparrow}p_{3y\downarrow}\\
0 & 0 & 0 & 0 & 0 & 0 & 0 & 0 & 0 & 0 & 0 & 0 & 0 & 0 & 0 & 0 & 0 & 0 & 0 & 0 & 0 & 0 & 0 & 0 & 0 & 0 & 0 & 0 & 2\epsilon_p & 0 & 0 & 0 & 0 & 0 & 0 & 0 &p_{2x \uparrow}p_{4x\downarrow}\\
0 & 0 & 0 & 0 & 0 & 0 & 0 & 0 & 0 & 0 & 0 & 0 & 0 & 0 & 0 & 0 & 0 & 0 & 0 & 0 & 0 & 0 & 0 & 0 & 0 & 0 & 0 & 0 & 0 & 2\epsilon_p & 0 & 0 & 0 & 0 & 0 & 0 &  p_{2x \uparrow}p_{5z  \downarrow}\\
0 & 0 & 0 & 0 & 0 & 0 & 0 & 0 & 0 & 0 & 0 & 0 & 0 & 0 & 0 & 0 & 0 & 0 & 0 & 0 & 0 & 0 & 0 & 0 & 0 & 0 & 0 & 0 & 0 & 0 & 2\epsilon_p & 0 & 0 & 0 & 0 & 0 & p_{2x \uparrow}p_{6z\downarrow}\\
0 & 0 & 0 & 0 & 0 & 0 & 0 & 0 & 0 & 0 & 0 & 0 & 0 & 0 & 0 & 0 & 0 & 0 & 0 & 0 & 0 & 0 & 0 & 0 & 0 & 0 & 0 & 0 & 0 & 0 & 0 & 2\epsilon_p & 0 & 0 & 0 & 0 & p_{3y \uparrow}p_{4x\downarrow}\\
0 & 0 & 0 & 0 & 0 & 0 & 0 & 0 & 0 & 0 & 0 & 0 & 0 & 0 & 0 & 0 & 0 & 0 & 0 & 0 & 0 & 0 & 0 & 0 & 0 & 0 & 0 & 0 & 0 & 0 & 0 & 0 & 2\epsilon_p & 0 & 0 & 0 & p_{3y \uparrow}p_{5z\downarrow}\\
0 & 0 & 0 & 0 & 0 & 0 & 0 & 0 & 0 & 0 & 0 & 0 & 0 & 0 & 0 & 0 & 0 & 0 & 0 & 0 & 0 & 0 & 0 & 0 & 0 & 0 & 0 & 0 & 0 & 0 & 0 & 0 & 0 & 2\epsilon_p & 0 & 0 & p_{3y \uparrow}p_{6z\downarrow}\\
0 & 0 & 0 & 0 & 0 & 0 & 0 & 0 & 0 & 0 & 0 & 0 & 0 & 0 & 0 & 0 & 0 & 0 & 0 & 0 & 0 & 0 & 0 & 0 & 0 & 0 & 0 & 0 & 0 & 0 & 0 & 0 & 0 & 0 & 2\epsilon_p & 0 &  p_{4x\uparrow}p_{5z\downarrow}\\
0 & 0 & 0 & 0 & 0 & 0 & 0 & 0 & 0 & 0 & 0 & 0 & 0 & 0 & 0 & 0 & 0 & 0 & 0 & 0 & 0 & 0 & 0 & 0 & 0 & 0 & 0 & 0 & 0 & 0 & 0 & 0 & 0 & 0 & 0 & 2\epsilon_p & p_{4x\uparrow}p_{6z\downarrow}\\
\end{pNiceMatrix}~~~
$%
}%
\end{equation}
\end{minipage}
\end{turn}
\\
~~
\\
~~
\\
~~
\\
~~
\\
~~
\\
~~
\\
~~
\\
~~
\\
~~
\\
~~
\\
\section{Electric Polarization}
\makeatletter

\renewcommand \thefigure{B\@arabic\c@figure}
\renewcommand \theequation{B\@arabic\c@equation}
\makeatother
We recall Zhang-Rice triplet and singlet states as follows

\begin{equation}
\resizebox{\textwidth}{!}{$%
\begin{aligned}
|i_T\rangle_\sigma=\frac{1}{\sqrt{6}}&\Biggl\{-\frac{\sqrt{3}}{2}d^{(i)\dagger}_{z^2\sigma} (p^{(i)\dagger}_{1y\sigma}-p^{(i)\dagger}_{2x\sigma}-p^{(i)\dagger}_{3y\sigma}+p^{(i)\dagger}_{4x\sigma})+d^{(i)\dagger}_{x^2-y^2\sigma} \left[\frac{1}{2}(p^{(i)\dagger}_{1y\sigma}+p^{(i)\dagger}_{2x\sigma}-p^{(i)\dagger}_{3y\sigma}-p^{(i)\dagger}_{4x\sigma}) -p^{(i)\dagger}_{5z\sigma}+p^{(i)\dagger}_{6z\sigma}\right]\Biggr\}|0\rangle,
\\|i_T\rangle_0=\frac{1}{2\sqrt{3}}& \Biggl\{\left(-\frac{\sqrt{3}}{2}d^{(i)\dagger}_{z^2\uparrow}(p^{(i)\dagger}_{1y\downarrow}-p^{(i)\dagger}_{2x\downarrow} -p^{(i)\dagger}_{3y\downarrow}+p^{(i)\dagger}_{4x\downarrow})+d^{(i)\dagger}_{x^2-y^2\downarrow} \left[\frac{1}{2}(p^{(i)\dagger}_{1y\downarrow}+p^{(i)\dagger}_{2x\downarrow}-p^{(i)\dagger}_{3y\downarrow} -p^{(i)\dagger}_{4x\downarrow})-p^{(i)\dagger}_{5z\downarrow}+p^{(i)\dagger}_{6z\downarrow}\right]\right)\\
&+\left(-\frac{\sqrt{3}}{2}d^{(i)\dagger}_{z^2\downarrow}(p^{(i)\dagger}_{1y\uparrow}-p^{(i)\dagger}_{2x\uparrow}-p^{(i)\dagger}_{3y\uparrow} +p^{(i)\dagger}_{4x\uparrow})+d^{(i)\dagger}_{x^2-y^2\downarrow}\left[\frac{1}{2}(p^{(i)\dagger}_{1y\uparrow} +p^{(i)\dagger}_{2x\uparrow}-p^{(i)\dagger}_{3y\uparrow}-p^{(i)\dagger}_{4x\uparrow}) -p^{(i)\dagger}_{5z\uparrow}+p^{(i)\dagger}_{6z\uparrow}\right]\right)\Biggr\}|0\rangle,\\
|i_S\rangle=\frac{1}{2\sqrt{3}}&\Biggl\{\left(\frac{\sqrt{3}}{2}d^{(i)\dagger}_{z^2\uparrow} (p^{(i)\dagger}_{1y\downarrow}-p^{(i)\dagger}_{2x\downarrow}-p^{(i)\dagger}_{3y\downarrow} +p^{(i)\dagger}_{4x\downarrow})+d^{(i)\dagger}_{x^2-y^2\downarrow}[\frac{1}{2}(p^{(i)\dagger}_{1y\downarrow} +p^{(i)\dagger}_{2x\downarrow}-p^{(i)\dagger}_{3y\downarrow}-p^{(i)\dagger}_{4x\downarrow}) -p^{(i)\dagger}_{5z\downarrow}+p^{(i)\dagger}_{6z\downarrow}]\right)\\
&-\left(\frac{\sqrt{3}}{2}d^{(i)\dagger}_{z^2\downarrow}(p^{(i)\dagger}_{1y\uparrow} -p^{(i)\dagger}_{2x\uparrow}-p^{(i)\dagger}_{3y\uparrow}+p^{(i)\dagger}_{4x\uparrow}) +d^{(i)\dagger}_{x^2-y^2\downarrow}\left[\frac{1}{2}(p^{(i)\dagger}_{1y\uparrow} +p^{(i)\dagger}_{2x\uparrow}-p^{(i)\dagger}_{3y\uparrow}-p^{(i)\dagger}_{4x\uparrow}) -p^{(i)\dagger}_{5z\uparrow}+p^{(i)\dagger}_{6z\uparrow}\right]\right)\Biggr\}|0\rangle .
\end{aligned}$%
}
\label{seq:isite}
\end{equation}

Here, $i$ stands for the unit index from $1$ to $4$ in the four NiS\textsubscript{6} units.

\begin{figure}[ht]
\includegraphics[scale=0.28]{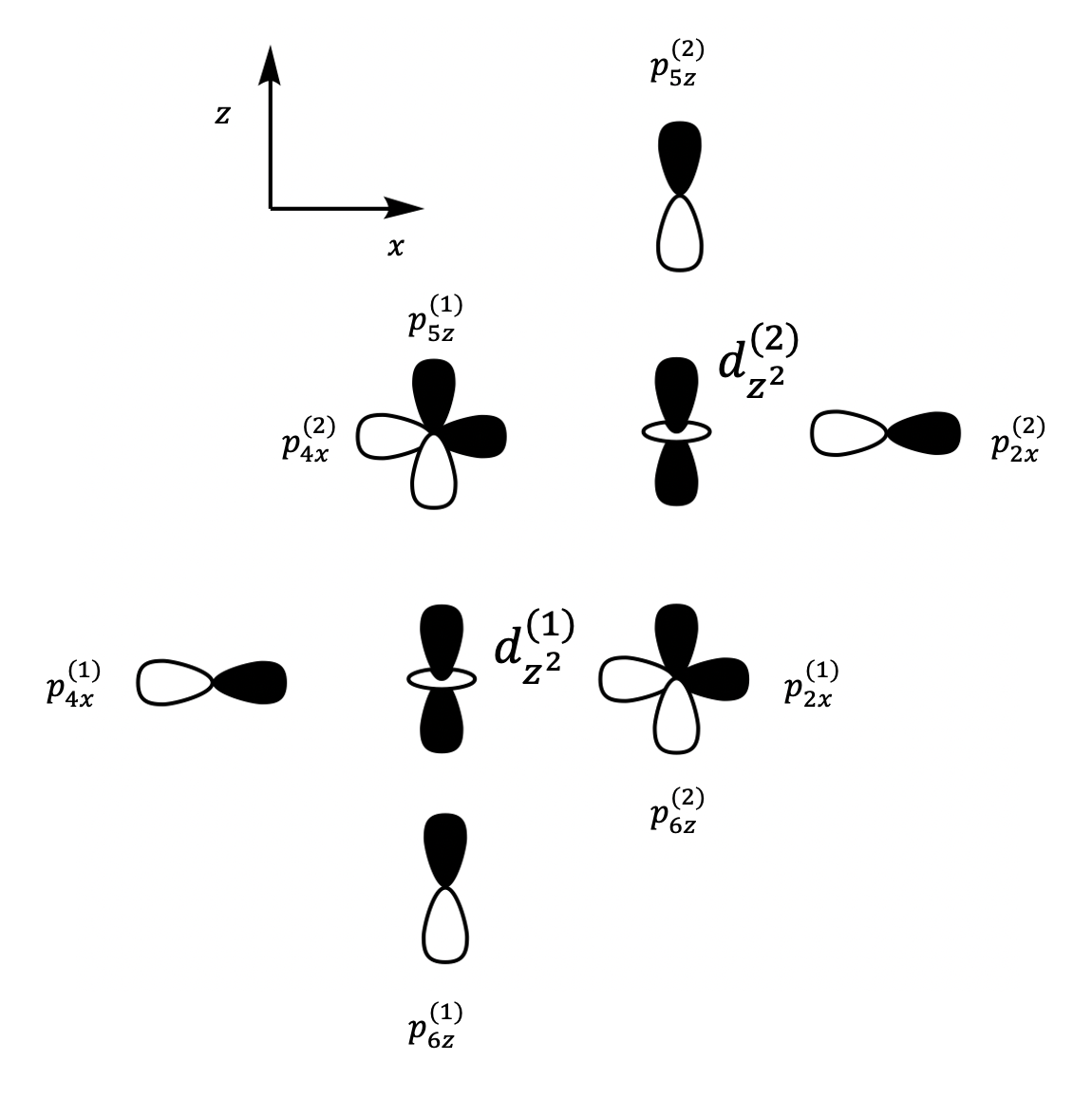}
\centering
\includegraphics[scale=0.28]{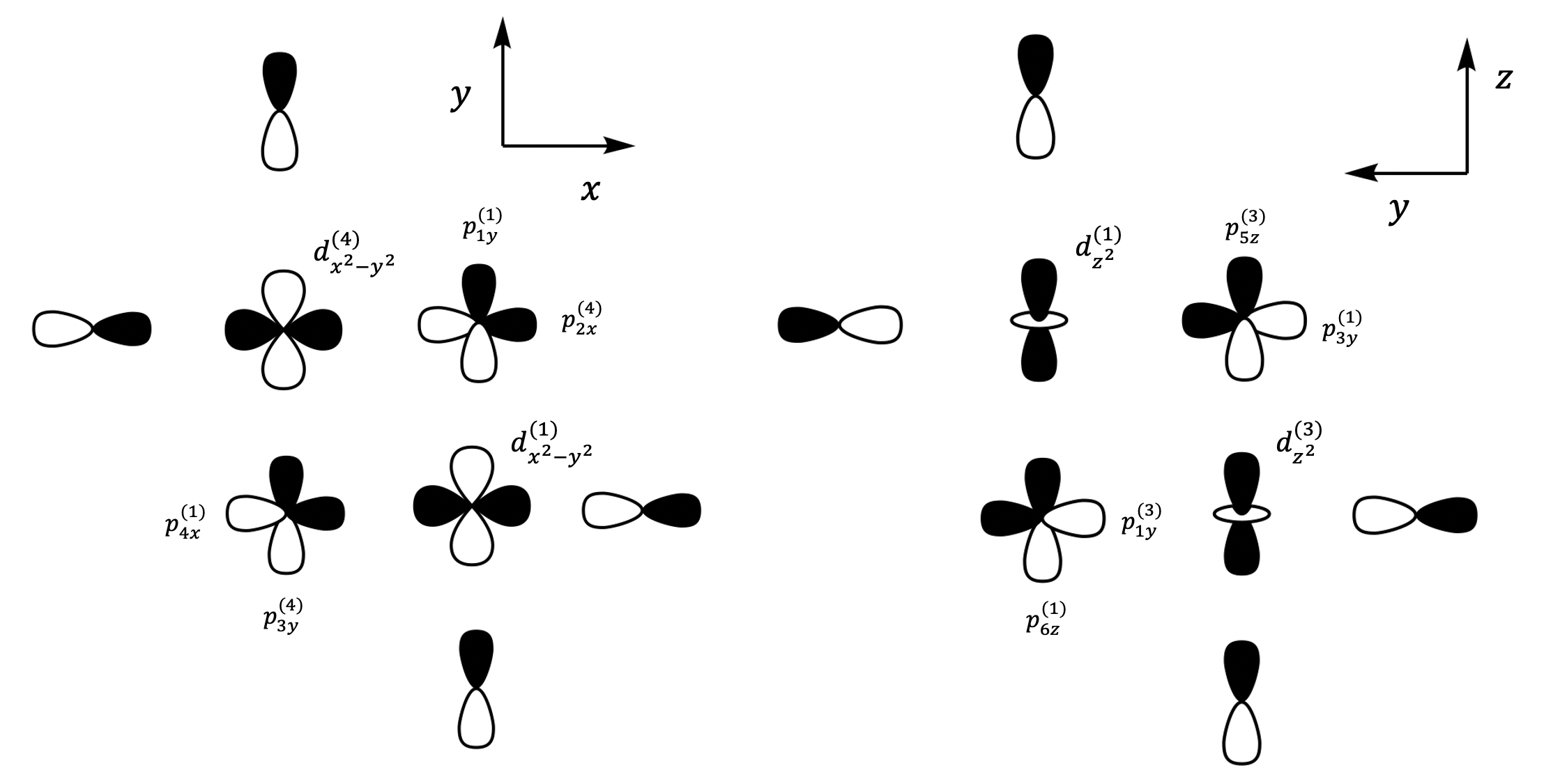}
\centering
\caption{Proper coordinate choice near the center unit of NiS\textsubscript{6}.}
\label{sfig:combination}
\end{figure}

We introduce our proper coordinate choice near the center unit of NiS\textsubscript{6} as shown in Figure {\ref{sfig:combination}}. It is important to notice that the Ni-P-Ni bond angle is not $90^{\circ}$ but slightly smaller than $90^{\circ}$, as pointed out in the main text. As a result, we construct an effective hopping Hamiltonian between the center NiS\textsubscript{6} unit and the nearest-neighbor NiS\textsubscript{6} unit as follows

\begin{equation}
\resizebox{\textwidth}{!}{$%
H_{12}=t\Delta\theta\sum_{\sigma}\left[\frac{\sqrt{3}}{2}d^{(1)\dagger}_{x^2-y^2\sigma}p^{(2)}_{6z\sigma} -\frac{1}{2}d^{(1)\dagger}_{z^2\sigma}p^{(2)}_{6z\sigma}-d^{(2)\dagger}_{z^2\sigma}p^{(1)}_{2x\sigma} -\frac{\sqrt{3}}{2}d^{(2)\dagger}_{x^2-y^2\sigma}p^{(1)}_{5z\sigma}+\frac{1}{2}d^{(2)\dagger}_{z^2\sigma}p^{(1)}_{5z\sigma} +d^{(1)\dagger}_{z^2\sigma}p^{(2)}_{4x\sigma}+h.c.\right] ,$%
}
\label{seq:hopping12}
\end{equation}
\begin{equation}
\resizebox{\textwidth}{!}{$%
H_{13}=t\Delta\theta\sum_{\sigma}\left[\frac{\sqrt{3}}{2}d^{(1)\dagger}_{x^2-y^2\sigma}p^{(3)}_{5z\sigma} +\frac{1}{2}d^{(1)\dagger}_{z^2\sigma}p^{(3)}_{5z\sigma}-d^{(1)\dagger}_{z^2\sigma}p^{(3)}_{1y\sigma} -\frac{\sqrt{3}}{2}d^{(3)\dagger}_{x^2-y^2\sigma}p^{(1)}_{6z\sigma}-\frac{1}{2}d^{(3)\dagger}_{z^2\sigma}p^{(1)}_{6z\sigma} +d^{(3)\dagger}_{z^2\sigma}p^{(1)}_{3y\sigma}+h.c.\right] ,$%
}
\label{seq:hopping13}
\end{equation}
\begin{equation}
\resizebox{\textwidth}{!}{$%
\begin{aligned}
H_{14}=t\Delta\theta\sum_{\sigma}\Biggl[\frac{\sqrt{3}}{2}d^{(1)\dagger}_{x^2-y^2\sigma}p^{(4)}_{3y\sigma}-\frac{1}{2}d^{(1)\dagger}_{z^2\sigma}p^{(4)}_{3y\sigma}&-\frac{\sqrt{3}}{2}d^{(4)\dagger}_{x^2-y^2\sigma}p^{(1)}_{1y\sigma}+\frac{1}{2}d^{(4)\dagger}_{z^2\sigma}p^{(1)}_{1y\sigma}\\
&+\frac{\sqrt{3}}{2}d^{(1)\dagger}_{x^2-y^2\sigma}p^{(4)}_{2x\sigma}+\frac{1}{2}d^{(1)\dagger}_{z^2\sigma}p^{(4)}_{2x\sigma} -\frac{\sqrt{3}}{2}d^{(4)\dagger}_{x^2-y^2\sigma}p^{(1)}_{4x\sigma}-\frac{1}{2}d^{(4)\dagger}_{z^2\sigma}p^{(1)}_{4x\sigma}+h.c\Biggr] .
\end{aligned}$%
}
\label{seq:hopping14}
\end{equation}

To control the local electric polarization, we apply an external magnetic field, described by the Zeeman-effect term

\begin{equation}
\resizebox{\textwidth}{!}{$%
    H_{ext}^{(i)} = -\frac{\mu_B}{\hbar}\vec{B}\cdot\Biggl[\frac{g\hbar}{2}\sum_{m,\sigma,\sigma'}d^{(i) \dagger}_{m\sigma}\vec{\tau}_{\sigma\sigma'}d_{m\sigma}^{(i)}+\sum_{m,m',\sigma}d^{(i) \dagger}_{m\sigma}\vec{L}_{m m'}d_{m\sigma}^{(i)}+\frac{g\hbar}{2}\sum_{m,\sigma,\sigma'}p^{(i) \dagger}_{m\sigma}\vec{\tau}_{\sigma\sigma'}p_{m\sigma'}^{(i)}+\sum_{m,m',\sigma}p^{(i) \dagger}_{m\sigma}\vec{L}_{m m'}p_{m\sigma}^{(i)}\Biggr] .$%
    }
\label{seq:zeeman}
\end{equation}
\noindent
Here, $\mu_B$ is the Bohr magneton and g is the anomalous gyromagnetic ratio of electron ($\approx 2.0023192$). The superscript $(i)$ runs from $1$ to $4$.

To find local electric polarizations, we realize that it is essential to consider atomic spin-orbit interaction as
\begin{equation}
H_{LS}=\frac{\lambda_d}{\hbar^2}(\vec{L}_{d}\cdot\vec{S}_{d})+\frac{\lambda_p}{\hbar^2}(\vec{L}_p\cdot\vec{S}_p) .
\label{seq:spin-orbit coupling}
\end{equation}
Here, $\lambda_d$ ($\lambda_p$) represents the spin-orbit coupling constant of the d-orbital (ligand p-orbitals). Accordingly, $\vec{L}_{d(p)}$ and $\vec{S}_{d(p)}$ are angular-momentum and spin operators of each orbital, respectively. Then, we obtain \\

\resizebox{\textwidth}{!}{$%
\begin{aligned}
 H_K+H_{LS}=&
 \begin{pNiceMatrix}[margin,last-col]
 E_{t_{2g}} & -\frac{i}{2}\lambda_d & \frac{i}{2}\lambda_d & \frac{\sqrt{3}}{2}\lambda_d & -\frac{1}{2}\lambda_d & 0 & 0 & 0 & 0 & 0 & d^{\dagger}_{zx\uparrow}\\
 \frac{i}{2}\lambda_d & E_{t_{2g}} & -\frac{1}{2}\lambda_d & -i\frac{\sqrt{3}}{2}\lambda_d & -\frac{i}{2}\lambda_d & 0 & 0 & 0 & 0 & 0 & d^{\dagger}_{yz\uparrow}\\
 -\frac{i}{2}\lambda_d & -\frac{1}{2}\lambda_d & E_{t_{2g}} & 0 & -i\lambda_d & 0 & 0 & 0 & 0 & 0 & d^{\dagger}_{xy\uparrow}\\
 \frac{\sqrt{3}}{2}\lambda_d & i\frac{\sqrt{3}}{2}\lambda_d & 0 & \epsilon_d & 0 & 0 & 0 & 0 & 0 & 0 & d^{\dagger}_{z^2\downarrow}\\
 -\frac{1}{2}\lambda_d & \frac{i}{2}\lambda_d & i\lambda_d & 0 & \epsilon_d & 0 & 0 & 0 & 0 & 0 & d^{\dagger}_{x^2-y^2\downarrow}\\
 0 & 0 & 0 & 0 & 0 & E_{t_{2g}} & \frac{i}{2}\lambda_d & \frac{i}{2}\lambda_d & -\frac{\sqrt{3}}{2}\lambda_d & \frac{1}{2}\lambda_d & d^{\dagger}_{zx\uparrow}\\
 0 & 0 & 0 & 0 & 0 & -\frac{i}{2}\lambda_d & E_{t_{2g}} & \frac{1}{2}\lambda_d & -i\frac{\sqrt{3}}{2}\lambda_d & -\frac{i}{2}\lambda_d & d^{\dagger}_{yz\downarrow}\\
 0 & 0 & 0 & 0 & 0 & -\frac{i}{2}\lambda_d & \frac{1}{2}\lambda_d & E_{t_{2g}} & 0 & i\lambda_d & d^{\dagger}_{xy\downarrow}\\
 0 & 0 & 0 & 0 & 0 & -\frac{\sqrt{3}}{2}\lambda_d & i\frac{\sqrt{3}}{2}\lambda_d & 0 & \epsilon_d & 0 & d^{\dagger}_{z^2\uparrow}\\
 0 & 0 & 0 & 0 & 0 & \frac{1}{2}\lambda_d & \frac{i}{2}\lambda_d & -i\lambda_d & 0 & \epsilon_d & d^{\dagger}_{x^2-y^2\uparrow}.\\
 \end{pNiceMatrix}\\
 &+\begin{pNiceMatrix}[margin,last-col]
 \epsilon_p & -i\frac{\lambda_p}{2} & 0 & 0 & 0 & \frac{\lambda_p}{2} & p^{\dagger}_{x\uparrow}\\
 i\frac{\lambda_p}{2} & \epsilon_p & 0 & 0 & 0 & \frac{\lambda_p}{2} & p^{\dagger}_{x\downarrow}\\
 0 & 0 & \epsilon_p & -\frac{\lambda_p}{2} & i\frac{\lambda_p}{2} & 0 & p^{\dagger}_{y\uparrow}\\
 0 & 0 & -\frac{\lambda_p}{2} & \epsilon_p & i\frac{\lambda_p}{2} & 0 & p^{\dagger}_{y\downarrow}\\
 0 & 0 & -i\frac{\lambda_p}{2} & -i\frac{\lambda_p}{2} & \epsilon_p & 0 & p^{\dagger}_{z\uparrow}\\
 \frac{\lambda_p}{2} & i\frac{\lambda_p}{2} & 0 & 0 & 0 & \epsilon_p & p^{\dagger}_{z\downarrow},\\
 \end{pNiceMatrix}
\end{aligned}$%
 }
\noindent
where $E_{t_{2g}}$ is the energy of the $t_{2g}$ orbital. This atomic spin-orbit coupling Hamiltonian gives rise to mixing between $t_{2g}$ and $e_g$ orbitals.

Considering the energy gap between these two orbitals, the perturbation theory results in

\begin{equation}
\resizebox{0.9\hsize}{!}{$
\begin{pNiceMatrix}[margin,last-col]
 E_{t_{2g}}+\lambda_d & 0 & 0 & 0 & 0 & 0 & 0 & 0 & 0 & 0 & \frac{1}{\sqrt{3}}(d^{\dagger}_{yz\uparrow}-id^{\dagger}_{zx\uparrow}-d^{\dagger}_{xy\downarrow})\\
 0 & E_{t_{2g}}-\frac{1}{2}\lambda_d & i\frac{\sqrt{6}}{2}\lambda_d & 0 & 0 & 0 & 0 & 0 & 0 & 0 & -\frac{1}{\sqrt{2}}(d^{\dagger}_{yz\uparrow}+id^{\dagger}_{zx\uparrow})\\
 0 & -i\frac{\sqrt{6}}{2}\lambda_d & \epsilon_d & 0 & 0 & 0 & 0 & 0 & 0 & 0 & d^{\dagger}_{z^2\downarrow}\\
 0 & 0 & 0 & E_{t_{2g}}-\frac{1}{2}\lambda_d & -i\frac{\sqrt{6}}{2}\lambda_d & 0 & 0 & 0 & 0 & 0 & \frac{1}{\sqrt{6}}(d^{\dagger}_{yz\uparrow}-id^{\dagger}_{zx\uparrow}+\sqrt{\frac{2}{3}}d^{\dagger}_{xy\downarrow})\\
 0 & 0 & 0 & -i\frac{\sqrt{6}}{2}\lambda_d & \epsilon_d & 0 & 0 & 0 & 0 & 0 & d^{\dagger}_{x^2-y^2\downarrow}\\
 0 & 0 & 0 & 0 & 0 & E_{t_{2g}}+\lambda_d & 0 & 0 & 0 & 0 & \frac{1}{\sqrt{3}}(d^{\dagger}_{yz\downarrow}+id^{\dagger}_{zx\downarrow}+d^{\dagger}_{xy\uparrow})\\
 0 & 0 & 0 & 0 & 0 & 0 & E_{t_2g}-\frac{1}{2}\lambda_d & -i\frac{\sqrt{6}}{2}\lambda_d & 0 & 0 & -\frac{1}{\sqrt{2}}(d^{\dagger}_{yz\downarrow}-id^{\dagger}_{zx\downarrow})\\
 0 & 0 & 0 & 0 & 0 & 0 & i\frac{\sqrt{6}}{2}\lambda_d & \epsilon_d & 0 & 0 & d^{\dagger}_{z^2\uparrow}\\
 0 & 0 & 0 & 0 & 0 & 0 & 0 & 0 & E_{t_2g}-\frac{1}{2}\lambda_d & i\frac{\sqrt{6}}{2}\lambda_d & -\frac{1}{\sqrt{6}}(d^{\dagger}_{yz\downarrow}-id^{\dagger}_{zx\downarrow}-\sqrt{\frac{2}{3}}d^{\dagger}_{xy\uparrow})\\
 0 & 0 & 0 & 0 & 0 & 0 & 0 & 0 & -i\frac{\sqrt{6}}{2}\lambda_d & \epsilon_d & d^{\dagger}_{x^2-y^2\uparrow}.\\
 \end{pNiceMatrix}
 $}
 \end{equation}

Accordingly, the resulting eigenstates are given by
\begin{equation}
\begin{split}
&|d_{z^2\uparrow}\rangle\rightarrow |d^{SO}_{z^2\uparrow}\rangle=|d_{z^2\uparrow}\rangle-i\sqrt{\frac{3}{2}}\frac{\lambda_d}{\lambda_d/2+\Delta} \Biggl(\frac{1}{\sqrt{2}}|d_{yz\downarrow}\rangle-\frac{i}{\sqrt{2}}|d_{zx\downarrow}\rangle\Biggr)\\
&|d_{z^2\downarrow}\rangle\rightarrow |d^{SO}_{z^2\downarrow}\rangle=|d_{z^2\downarrow}\rangle+ i\sqrt{\frac{3}{2}}\frac{\lambda_d}{\lambda_d/2+\Delta}\Biggl(-\frac{1}{\sqrt{2}}|d_{yz\uparrow}\rangle -\frac{i}{\sqrt{2}}|d_{zx\uparrow}\rangle\Biggr)\\
&|d_{x^2-y^2\uparrow}\rangle\rightarrow|d^{SO}_{x^2-y^2\uparrow}\rangle=|d_{x^2-y^2\uparrow}\rangle +i\sqrt{\frac{3}{2}}\frac{\lambda_d}{\lambda_d/2+\Delta}\Biggl(-\frac{1}{\sqrt{6}}|d_{yz\downarrow}\rangle -\frac{i}{\sqrt{6}}|d_{zx\downarrow}\rangle+\sqrt{\frac{2}{3}}|d_{xy\uparrow}\rangle\Biggr)\\
&|d_{x^2-y^2\downarrow}\rangle\rightarrow|d^{SO}_{x^2-y^2\downarrow}\rangle=|d_{x^2-y^2\downarrow}\rangle -i\sqrt{\frac{3}{2}}\frac{\lambda_d}{\lambda_d/2+\Delta}\Biggl(\frac{1}{\sqrt{6}}|d_{yz\uparrow}\rangle -\frac{i}{\sqrt{6}}|d_{zx\uparrow}\rangle+\sqrt{\frac{2}{3}}|d_{xy\downarrow}\rangle\Biggr)
\end{split}
\label{seq:soc d}
\end{equation}
where $\Delta=\epsilon_d-E_{t_{2g}}$.

Introducing an external magnetic field $\vec{B}=\frac{B}{\sqrt{3}}(\hat{x}+\hat{y}+\hat{z})$, we obtain an effective Zeeman coupling term as follows \\ 

\begin{equation}
\resizebox{\textwidth}{!}{$
-\frac{\mu_Bg}{\hbar}\vec{S}\cdot\vec{B}=-\mu_B\frac{g}{2}\vec{B}\cdot\vec{\tau}\doteq-\mu_BB
\begin{pNiceMatrix}[margin,last-col]
   1-\frac{3}{2}(\frac{\lambda_d}{\lambda_d/2+\Delta})^2 & 0 & 0 & 0 & d^{\dagger SO}_{z^2\uparrow}\\
   0 & 1+\frac{1}{2}(\frac{\lambda_d}{\lambda_d/2+\Delta})^2 & 0 & 0 & d^{\dagger SO}_{x^2-y^2\uparrow}\\
   0 & 0 & -1+\frac{3}{2}(\frac{\lambda_d}{\lambda_d/2+\Delta})^2 & 0 & d^{\dagger SO}_{z^2\downarrow}\\
   0 & 0 & 0 & -1-\frac{1}{2}(\frac{\lambda_d}{\lambda_d/2+\Delta})^2 & d^{\dagger SO}_{x^2-y^2\downarrow}\\
   \end{pNiceMatrix}-\mu_BB\begin{pNiceMatrix}[margin,last-col]
   1 & 0 & 0 & 0 & 0 & 0 & p^{\dagger}_{x\uparrow} \\
   0 & 1 & 0 & 0 & 0 & 0 & p^{\dagger}_{y\uparrow}\\
   0 & 0 & 1 & 0 & 0 & 0 & p^{\dagger}_{z\uparrow}\\
   0 & 0 & 0 & -1 & 0 & 0 & p^{\dagger}_{x\downarrow}\\
   0 & 0 & 0 & 0 & -1 & 0 & p^{\dagger}_{y\downarrow}\\
   0 & 0 & 0 & 0 & 0 & -1 & p^{\dagger}_{z\downarrow}\\
   \end{pNiceMatrix}
    $} ,
    \label{seq:Zeeman with SO1}
    \end{equation}
\begin{equation}
\resizebox{\textwidth}{!}{$
-\frac{\mu_B}{\hbar}\vec{L}\cdot\vec{B}=-\frac{\mu_BB}{\sqrt{3}}
    \begin{pNiceMatrix}[margin,last-col]
    \frac{3}{2}(\frac{\lambda_d}{\lambda_d/2+\Delta})^2 & 0 & \frac{3(1-i)\lambda_d}{\lambda_d/2+\Delta} & \frac{(1+i)\sqrt{3}\lambda_d\Delta}{(\Delta+\lambda_d/2)^2} & d^{\dagger SO}_{z^2\uparrow}\\
    0 & -\frac{1}{2}(\frac{\lambda_d}{\Delta+\lambda_d/2})^2+\frac{4\lambda_d}{\lambda_d/2+\Delta} & \frac{(1+i)\sqrt{3}\lambda_d\Delta}{(\Delta+\lambda_d/2)^2} & \frac{(1-i)\sqrt{3}\lambda_d(\Delta+\frac{3}{2}\lambda_d)}{(\Delta+\lambda_d/2)^2} & d^{\dagger SO}_{x^2-y^2\uparrow}\\
    \frac{3(1+i)\lambda_d}{\lambda_d/2+\Delta} & \frac{(1-i)\sqrt{3}\lambda_d\Delta}{(\Delta+\lambda_d/2)^2} & -\frac{3}{2}(\frac{\lambda_d}{\lambda_d/2+\Delta})^2 & 0 & d^{\dagger SO}_{z^2\downarrow}\\
    \frac{(1-i)\sqrt{3}\lambda_d\Delta}{(\Delta+\lambda_d/2)^2} & \frac{(1+i)\sqrt{3}\lambda_d(\Delta+\frac{3}{2}\lambda_d)}{(\Delta+\lambda_d/2)^2}  & 0 & \frac{1}{2}(\frac{\lambda_d}{\Delta+\lambda_d/2})^2-\frac{4\lambda_d}{\lambda_d/2+\Delta} & d^{\dagger SO}_{x^2-y^2\downarrow}\\
    \end{pNiceMatrix}-\frac{\mu_BB}{\sqrt{3}}\begin{pNiceMatrix}[margin,last-col]
     0 & -i & i & 0 & 0 & 0 & p^{\dagger}_{x\uparrow} \\
    i & 0 & -i & 0 & 0 & 0 & p^{\dagger}_{y\uparrow}\\
    -i & i & 0 & 0 & 0 & 0 & p^{\dagger}_{z\uparrow}\\
    0 & 0 & 0 & 0 & -i & i & p^{\dagger}_{x\downarrow}\\
    0 & 0 & 0 & i & 0 & -i & p^{\dagger}_{y\downarrow}\\
    0 & 0 & 0 & -i & i & 0 & p^{\dagger}_{z\downarrow}\\
    \end{pNiceMatrix}
    $} .
    \label{seq:Zeeman with SO2}
    \end{equation}

The resulting state in the first order with respect to $H'=H_{12}+H_{13}+H_{14}+H_{ext}$ is given by\\
~

\resizebox{\textwidth}{!}{$%
\begin{aligned}
&|1_T\rangle_{\uparrow}\otimes|2_T\rangle_{\uparrow}\otimes|3_T\rangle_{\uparrow}\otimes|4_T\rangle_{\downarrow}\longrightarrow \\
&(|1_T\rangle_{\uparrow}\otimes|2_T\rangle_{\uparrow}\otimes|3_T\rangle_{\uparrow}\otimes|4_T\rangle_{\downarrow})'\approx |1_T\rangle_{\uparrow}\otimes|2_T\rangle_{\uparrow}\otimes|3_T\rangle_{\uparrow}\otimes|4_T\rangle_{\downarrow} + \sum_{u} |u\rangle \langle u| \frac{H'}{E_{1234}-E_u} |1_T\rangle_{\uparrow}\otimes|2_T\rangle_{\uparrow}\otimes|3_T\rangle_{\uparrow}\otimes|4_T\rangle_{\downarrow} .
\end{aligned}
$%
}
\noindent
Here, $E_{1234}$ is an energy eigenvalue given by $4(\epsilon_p+\epsilon_d)$ + (diagonal terms of $H'$).

Considering that hopping terms act on two nearest-neighbor sites only,
%
%
we obtain

$$
\resizebox{\textwidth}{!}{$%
\begin{aligned}
&\sum_{u} |u\rangle \langle u|\frac{H_{12}}{E_{1234}-E_u} |1_T\rangle_{\uparrow}\otimes|2_T\rangle_{\uparrow}\otimes|3_T\rangle_{\uparrow}\otimes|4_T\rangle_{\downarrow} \approx \\
&-\frac{t\Delta\theta}{2\sqrt{2}}\Biggl\{\frac{[p^{(2)\dagger}_{6z\uparrow}(p^{(1)\dagger}_{1y\uparrow}-p^{(1)\dagger}_{3y\uparrow}-p^{(1)\dagger}_{5z\uparrow}+p^{(1)\dagger}_{6z\uparrow})-p^{(2)\dagger}_{4x\uparrow}(p^{(1)\dagger}_{1y\uparrow}-p^{(1)\dagger}_{2x\uparrow}-p^{(1)\dagger}_{3y\uparrow}+p^{(1)\dagger}_{4x\uparrow})]}{\epsilon_p-\epsilon_d}-\frac{B(\lambda_1-\lambda_2)}{(\epsilon_p-\epsilon_d)^2}p^{(2)\dagger}_{6z\uparrow}\Biggl[\frac{1}{2}(p^{(1)\dagger}_{1y\uparrow}+p^{(1)\dagger}_{2x\uparrow}-p^{(1)\dagger}_{3y\uparrow}-p^{(1)\dagger}_{4x\uparrow})-p^{(1)\dagger}_{5z\uparrow}+p^{(1)\dagger}_{6z\uparrow}\Biggr]\\
&\ \ \ \ \ \ \ \ \ \ \ \ \ \ \ \ \ \ \ \ \ \ \ \ \ \ \ \ \ \ \ \ \ \ \ \ \ \ \ \  +\frac{[d^{(2)\dagger}_{z^2\uparrow}d^{(1)\dagger}_{z^2\uparrow}+\frac{2}{\sqrt{3}}d^{(2)\dagger}_{z^2\uparrow}d^{(1)\dagger}_{x^2-y^2\uparrow}-d^{(2)\dagger}_{x^2-y^2\uparrow}d^{(1)\dagger}_{x^2-y^2\uparrow}]}{U-3J-\epsilon_p+\epsilon_d}-\frac{B(\lambda_1-\lambda_2)}{[U-3J-\epsilon_p+\epsilon_d]^2}d^{(2)\dagger}_{x^2-y^2\uparrow}d^{(1)\dagger}_{x^2-y^2\uparrow}\Biggr\}\otimes|2_T\rangle_{\uparrow}\otimes|3_T\rangle_{\uparrow}\otimes|4_T\rangle_{\downarrow}\\
&+|1_T\rangle_{\uparrow}\otimes\frac{t\Delta\theta}{2\sqrt{2}}\Biggl\{\frac{[p^{(1)\dagger}_{5z\uparrow}(p^{(2)\dagger}_{1y\uparrow}-p^{(2)\dagger}_{3y\uparrow}-p^{(2)\dagger}_{5z\uparrow}+p^{(2)\dagger}_{6z\uparrow})-p^{(1)\dagger}_{2x\uparrow}(p^{(2)\dagger}_{1y\uparrow}-p^{(2)\dagger}_{2x\uparrow}-p^{(2)\dagger}_{3y\uparrow}+p^{(2)\dagger}_{4x\uparrow})]}{\epsilon_p-\epsilon_d}-\frac{B(\lambda_1-\lambda_2)}{(\epsilon_p-\epsilon_d)^2}p^{(1)\dagger}_{5z\uparrow}\Biggl[\frac{1}{2}(p^{(2)\dagger}_{1y\uparrow}+p^{(2)\dagger}_{2x\uparrow}-p^{(2)\dagger}_{3y\uparrow}-p^{(2)\dagger}_{4x\uparrow})-p^{(2)\dagger}_{5z\uparrow}+p^{(2)\dagger}_{6z\uparrow}\Biggr]\\
&\ \ \ \ \ \ \ \ \ \ \ \ \ \ \ \ \ \ \ \ \ \ \ \ \ \ \ \ \ \ \ \ \ \ \ \ \ \ \ \ -\frac{[d^{(1)\dagger}_{z^2\uparrow}d^{(2)\dagger}_{z^2\uparrow}+\frac{2}{\sqrt{3}}d^{(1)\dagger}_{z^2\uparrow}d^{(2)\dagger}_{x^2-y^2\uparrow}-d^{(1)\dagger}_{x^2-y^2\uparrow}d^{(2)\dagger}_{x^2-y^2\uparrow}]}{U-3J-\epsilon_p+\epsilon_d}+\frac{B(\lambda_1-\lambda_2)}{[U-3J-\epsilon_p+\epsilon_d]^2}d^{(1)\dagger}_{x^2-y^2\uparrow}d^{(2)\dagger}_{x^2-y^2\uparrow}\Biggr\}\otimes |3_T\rangle_{\uparrow}\otimes|4_T\rangle_{\downarrow}+O\Biggl(\frac{t\Delta\theta}{E_{1234}-E_u}\frac{\lambda_d}{\lambda_d/2+\Delta}\Biggl) .
\end{aligned}~~~
$%
}%
$$
\noindent
Here, $\lambda_1=\mu_B(1+\frac{1}{\sqrt{3}}\frac{4\lambda_d}{\Delta+\lambda_d/2})$ and $\lambda_2=\mu_B$ are effective Zeeman coupling constants for d- and p-orbitals, respectively. We emphasize that the central role of the atomic spin-orbit coupling is to give the difference in the effective Zeeman coupling constant. We point out that there exist hopping effects from the atomic spin-orbit coupling in the sub-leading order $t \Delta\theta \lambda_d$. In the same way, we have

$$
\resizebox{\textwidth}{!}{$%
\begin{aligned}
&\sum_{u} |u\rangle \langle u| \frac{H_{13}}{E_{1234}-E_u} |1_T\rangle_{\uparrow}\otimes|2_T\rangle_{\uparrow}\otimes|3_T\rangle_{\uparrow}\otimes|4_T\rangle_{\downarrow} \approx \\
&-\frac{t\Delta\theta}{2\sqrt{2}}\Biggl\{\frac{[p^{(3)\dagger}_{5z\uparrow}(p^{(1)\dagger}_{2x\uparrow}-p^{(1)\dagger}_{4x\uparrow}-p^{(1)\dagger}_{5z\uparrow}+p^{(1)\dagger}_{6z\uparrow})+p^{(3)\dagger}_{1y\uparrow}(p^{(1)\dagger}_{1y\uparrow}-p^{(1)\dagger}_{2x\uparrow}-p^{(1)\dagger}_{3y\uparrow}+p^{(1)\dagger}_{4x\uparrow})]}{\epsilon_p-\epsilon_d}-\frac{B(\lambda_1-\lambda_2)}{(\epsilon_p-\epsilon_d)^2}p^{(3)\dagger}_{5z\uparrow}\Biggl[\frac{1}{2}(p^{(1)\dagger}_{1y\uparrow}+p^{(1)\dagger}_{2x\uparrow}-p^{(1)\dagger}_{3y\uparrow}-p^{(1)\dagger}_{4x\uparrow})-p^{(1)\dagger}_{5z\uparrow}+p^{(1)\dagger}_{6z\uparrow}\Biggr]\\
&\ \ \ \ \ \ \ \ \ \ \ \ \ \ \ \ \ \ \ \ \ \ \ \ \ \ \ \ \ \ \ \ \ \ \ \ \ \ \ \ +\frac{[-d^{(3)\dagger}_{z^2\uparrow}d^{(1)\dagger}_{z^2\uparrow}+\frac{2}{\sqrt{3}}d^{(3)\dagger}_{z^2\uparrow}d^{(1)\dagger}_{x^2-y^2\uparrow}+d^{(3)\dagger}_{x^2-y^2\uparrow}d^{(1)\dagger}_{x^2-y^2\uparrow}]}{U-3J-\epsilon_p+\epsilon_d}+\frac{B(\lambda_1-\lambda_2)}{[U-3J-\epsilon_p+\epsilon_d]^2}d^{(3)\dagger}_{x^2-y^2\uparrow}d^{(1)\dagger}_{x^2-y^2\uparrow}\Biggr\}\otimes|2_T\rangle_{\uparrow}\otimes|3_T\rangle_{\uparrow}\otimes|4_T\rangle_{\downarrow}\\
&+|1_T\rangle_{\uparrow}\otimes |2_T\rangle_{\uparrow}\otimes\frac{t\Delta\theta}{2\sqrt{2}}\Biggl\{\frac{[p^{(1)\dagger}_{6z\uparrow}(p^{(3)\dagger}_{2x\uparrow}-p^{(3)\dagger}_{4x\uparrow}-p^{(3)\dagger}_{5z\uparrow}+p^{(3)\dagger}_{6z\uparrow})+p^{(1)\dagger}_{3y\uparrow}(p^{(3)\dagger}_{1y\uparrow}-p^{(3)\dagger}_{2x\uparrow}-p^{(3)\dagger}_{3y\uparrow}+p^{(3)\dagger}_{4x\uparrow})]}{\epsilon_p-\epsilon_d}-\frac{B(\lambda_1-\lambda_2)}{(\epsilon_p-\epsilon_d)^2}p^{(1)\dagger}_{6z\uparrow}\Biggl[\frac{1}{2}(p^{(3)\dagger}_{1y\uparrow}+p^{(3)\dagger}_{2x\uparrow}-p^{(3)\dagger}_{3y\uparrow}-p^{(3)\dagger}_{4x\uparrow})-p^{(3)\dagger}_{5z\uparrow}+p^{(3)\dagger}_{6z\uparrow}\Biggr]\\
&\ \ \ \ \ \ \ \ \ \ \ \ \ \ \ \ \ \ \ \ \ \ \ \ \ \ \ \ \ \ \ \ \ \ \ \ \ \ \ \ -\frac{[-d^{(1)\dagger}_{z^2\uparrow}d^{(3)\dagger}_{z^2\uparrow}+\frac{2}{\sqrt{3}}d^{(1)\dagger}_{z^2\uparrow}d^{(3)\dagger}_{x^2-y^2\uparrow}+d^{(1)\dagger}_{x^2-y^2\uparrow}d^{(3)\dagger}_{x^2-y^2\uparrow}]}{U-3J-\epsilon_p+\epsilon_d}-\frac{B(\lambda_1-\lambda_2)}{[U-3J-\epsilon_p+\epsilon_d]^2}d^{(1)\dagger}_{x^2-y^2\uparrow}d^{(3)\dagger}_{x^2-y^2\uparrow}\Biggr\}\otimes|4_T\rangle_{\downarrow}
\end{aligned}~~~
$%
}%
$$
and
$$
\resizebox{\textwidth}{!}{$%
\begin{aligned}
&\sum_{u} |u\rangle \langle u|\frac{H_{14}}{E_{1234}-E_u} |1_T\rangle_{\uparrow}\otimes|2_T\rangle_{\uparrow}\otimes|3_T\rangle_{\uparrow}\otimes|4_T\rangle_{\downarrow} \approx \\
&-\frac{t\Delta\theta}{2\sqrt{2}}\Biggl\{\frac{p^{(4)\dagger}_{3y\uparrow}(p^{(1)\dagger}_{1y\uparrow}-p^{(1)\dagger}_{3y\uparrow}-p^{(1)\dagger}_{5z\uparrow}+p^{(1)\dagger}_{6z\uparrow})+p^{(4)\dagger}_{2x\uparrow}(p^{(1)\dagger}_{2x\uparrow}-p^{(1)\dagger}_{4x\uparrow}-p^{(1)\dagger}_{5z\uparrow}+p^{(1)\dagger}_{6z\uparrow})}{\epsilon_p-\epsilon_d}-\frac{B(\lambda_1-\lambda_2)}{(\epsilon_p-\epsilon_d)^2}(p^{(4)\dagger}_{2x\uparrow}+p^{(4)\dagger}_{3y\uparrow})\Biggl[\frac{1}{2}(p^{(1)\dagger}_{1y\uparrow}+p^{(1)\dagger}_{2x\uparrow}-p^{(1)\dagger}_{3y\uparrow}-p^{(1)\dagger}_{4x\uparrow})-p^{(1)\dagger}_{5z\uparrow}+p^{(1)\dagger}_{6z\uparrow}\Biggr]\\
&\ \ \ \ \ \ \ \ \ \ \ \ \ \ \ \ \ \ \ \ \ \ \ \ \ \ \ \ \ \ \ \ \ \ \ \ \ \ \ \ -\frac{[\frac{1}{\sqrt{3}}d^{(4)\dagger}_{z^2\uparrow}d^{(1)\dagger}_{x^2-y^2\uparrow}+\sqrt{3}d^{(4)\dagger}_{x^2-y^2\uparrow}d^{(1)\dagger}_{z^2\uparrow}]}{U-J-\epsilon_p+\epsilon_d}-\frac{\sqrt{3}B(\lambda_1-\lambda_2)}{[U-J-\epsilon_p+\epsilon_d]^2}d^{(4)\dagger}_{x^2-y^2\uparrow}d^{(1)\dagger}_{z^2\uparrow}\Biggr\}\otimes|2_T\rangle_{\uparrow}\otimes|3_T\rangle_{\uparrow}\otimes|4_T\rangle_{\downarrow}\\
&+|1_T\rangle_{\uparrow}\otimes |2_T\rangle_{\uparrow}\otimes |3_T\rangle_{\uparrow}\otimes\frac{t\Delta\theta}{2\sqrt{2}}\Biggl\{\frac{[p^{(1)\dagger}_{1y\downarrow}(p^{(4)\dagger}_{1y\downarrow}-p^{(4)\dagger}_{3y\downarrow}-p^{(4)\dagger}_{5z\downarrow}+p^{(4)\dagger}_{6z\downarrow})+p^{(1)\dagger}_{4x\downarrow}(p^{(4)\dagger}_{2x\downarrow}-p^{(4)\dagger}_{4x\downarrow}-p^{(4)\dagger}_{5z\downarrow}+p^{(4)\dagger}_{6z\downarrow})]}{\epsilon_p-\epsilon_d}+\frac{B(\lambda_1-\lambda_2)}{(\epsilon_p-\epsilon_d)^2}(p^{(1)\dagger}_{1y\downarrow}+p^{(1)\dagger}_{4x\downarrow})\Biggl[\frac{1}{2}(p^{(4)\dagger}_{1y\downarrow}+p^{(4)\dagger}_{2x\downarrow}-p^{(4)\dagger}_{3y\downarrow}-p^{(4)\dagger}_{4x\downarrow})-p^{(4)\dagger}_{5z\downarrow}+p^{(4)\dagger}_{6z\downarrow}\Biggr]\\
&\ \ \ \ \ \ \ \ \ \ \ \ \ \ \ \ \ \ \ \ \ \ \ \ \ \ \ \ \ \ \ \ \ \ \ \ \ \ \ \ +\frac{[\frac{1}{\sqrt{3}}d^{(1)\dagger}_{z^2\downarrow}d^{(4)\dagger}_{x^2-y^2\downarrow}+\sqrt{3}d^{(1)\dagger}_{x^2-y^2\downarrow}d^{(4)\dagger}_{z^2\downarrow}]}{U-J-\epsilon_p+\epsilon_d}-\frac{\sqrt{3}B(\lambda_1-\lambda_2)}{[U-J-\epsilon_p+\epsilon_d]^2}d^{(1)\dagger}_{x^2-y^2\downarrow}d^{(4)\dagger}_{z^2\downarrow}\Biggr\}
\end{aligned}~~~
$%
}%
$$

Taking $\vec{r_1} = \vec{r}\otimes\ 1_{2\times 2}\otimes 1_{2\times 2}\otimes 1_{2\times 2}$ with $\vec{r}=\vec{r}\otimes 1_{1\times 1} + 1_{1\times 1}\otimes\vec{r}$, we find\\
~
\\
\resizebox{\textwidth}{!}{$%
\begin{aligned}
\vec{P}_1&=(\langle 1_T|_{\uparrow}\otimes \langle 2_T|_{\uparrow}\otimes\langle 3_T|_{\uparrow}\otimes\langle 4_T|_{\downarrow})'\vec{r}_1(|1_T\rangle_{\uparrow}\otimes|2_T\rangle_{\uparrow}\otimes|3_T\rangle_{\uparrow}\otimes|4_T\rangle_{\downarrow})'\\
=-\hat{x}&\Biggl\{\frac{2t\Delta\theta}{\epsilon_p-\epsilon_d}\Biggl[\frac{5}{12}\langle p^{(2)}_{4x}|x|d^{(1)}_{z^2}\rangle+\frac{1}{4}\langle p^{(4)}_{2x}|x|d^{(1)}_{z^2}\rangle+\frac{\sqrt{3}}{4}\Biggl(1-\frac{B(\lambda_1-\lambda_2)}{(\epsilon_p-\epsilon_d)}\Biggr)\langle p^{(4)}_{2x}|x|d^{(1)}_{x^2-y^2}\rangle\Biggr]\\
&+\frac{1}{2}\frac{2t\Delta\theta}{U-3J-\epsilon_p+\epsilon_d}\Biggl[\frac{1}{12}\langle p^{(1)}_{1y}|x|d^{(2)}_{z^2}\rangle-\frac{5}{24}\langle p^{(1)}_{2x}|x|d^{(2)}_{z^2}\rangle+\frac{5}{24}\langle p^{(1)}_{4x}|x|d^{(2)}_{z^2}\rangle-\frac{1}{6}\langle p^{(1)}_{6z}|x|d^{(2)}_{z^2}\rangle\\
&\ \ \ \ +\Biggl(1+\frac{B(\lambda_1-\lambda_2)}{U-3J-\epsilon_p+\epsilon_d}\Biggr)\Biggl(\frac{1}{4\sqrt{3}}\langle p^{(1)}_{1y}|x|d^{(2)}_{x^2-y^2}\rangle+\frac{1}{8\sqrt{3}}\langle p^{(1)}_{2x}|x|d^{(2)}_{x^2-y^2}\rangle-\frac{1}{8\sqrt{3}}\langle p^{(1)}_{4x}|x|d^{(2)}_{x^2-y^2}\rangle+\frac{1}{4\sqrt{3}}\langle p^{(1)}_{6z}|x|d^{(2)}_{x^2-y^2}\rangle\Biggr) \Biggr]\\
&+\frac{2t\Delta\theta}{U-J-\epsilon_p+\epsilon_d}\Biggl[\Biggl(1+\frac{B(\lambda_1-\lambda_2)}{U-J-\epsilon_p+\epsilon_d}\Biggr)\Biggl(\frac{\sqrt{3}}{8}\langle p^{(1)}_{2x}|x|d^{(4)}_{x^2-y^2}\rangle+\frac{\sqrt{3}}{8}\langle p^{(1)}_{3y}|x|d^{(4)}_{x^2-y^2}\rangle-\frac{\sqrt{3}}{8}\langle p^{(1)}_{4x}|x|d^{(4)}_{x^2-y^2}\rangle\Biggr)\\
&\ \ \ \ +\frac{1}{24}\langle p^{(1)}_{2x}|x|d^{(4)}_{z^2}\rangle-\frac{1}{24}\langle p^{(1)}_{3y}|x|d^{(4)}_{z^2}\rangle-\frac{1}{24}\langle p^{(1)}_{4x}|x|d^{(4)}_{z^2}\rangle-\frac{1}{6}\langle p^{(1)}_{5z}|x|d^{(4)}_{z^2}\rangle\Biggr]
\Biggr\}\\
-\hat{y}&\Biggl\{\frac{2t\Delta\theta}{\epsilon_p-\epsilon_d}\Biggl[-\frac{5}{12}\langle p^{(3)}_{1y}|y|d^{(1)}_{z^2}\rangle-\frac{1}{4}\langle p^{(4)}_{3y}|y|d^{(1)}_{z^2}\rangle+\frac{\sqrt{3}}{4}\Biggl(1-\frac{B(\lambda_1-\lambda_2)}{(\epsilon_p-\epsilon_d)}\Biggr)\langle p^{(4)}_{3y}|y|d^{(1)}_{x^2-y^2}\rangle\Biggr]\\
&+\frac{1}{2}\frac{2t\Delta\theta}{U-3J-\epsilon_p+\epsilon_d}\Biggl[-\frac{5}{24}\langle p^{(1)}_{1y}|y|d^{(3)}_{z^2}\rangle +\frac{5}{24}\langle p^{(1)}_{3y}|y|d^{(3)}_{z^2}\rangle+\frac{1}{12}\langle p^{(1)}_{2x}|y|d^{(3)}_{z^2}\rangle+\frac{1}{6}\langle p^{(1)}_{5z}|y|d^{(3)}_{z^2}\rangle\\
&\ \ \ +\Biggl(1+\frac{B(\lambda_1-\lambda_2)}{U-3J-\epsilon_p+\epsilon_d}\Biggr)\Biggl(-\frac{1}{8\sqrt{3}}\langle p^{(1)}_{1y}|y|d^{(3)}_{x^2-y^2}\rangle+\frac{1}{8\sqrt{3}} \langle p^{(1)}_{3y}|y|d^{(3)}_{x^2-y^2}\rangle-\frac{1}{4\sqrt{3}} \langle p^{(1)}_{2x}|y|d^{(3)}_{x^2-y^2}\rangle+\frac{1}{4\sqrt{3}}\langle p^{(1)}_{5z}|y|d^{(3)}_{x^2-y^2}\rangle\Biggr)\Biggr]\\
&+\frac{2t\Delta\theta}{U-J-\epsilon_p+\epsilon_d}\Biggl[\Biggl(1+\frac{B(\lambda_1-\lambda_2)}{U-J-\epsilon_p+\epsilon_d}\Biggr)\Biggl(-\frac{\sqrt{3}}{8}\langle p^{(1)}_{1y}|y|d^{(4)}_{x^2-y^2}\rangle +\frac{\sqrt{3}}{8}\langle p^{(1)}_{2x}|y|d^{(4)}_{x^2-y^2}\rangle +\frac{\sqrt{3}}{8}\langle p^{(1)}_{3y}|y|d^{(4)}_{x^2-y^2}\rangle\Biggr)\\
&\ \ \ +\frac{1}{24}\langle p^{(1)}_{1y}|y|d^{(4)}_{z^2}\rangle+\frac{1}{24}\langle p^{(1)}_{2x}|y|d^{(4)}_{z^2}\rangle-\frac{1}{24}\langle p^{(1)}_{3y}|y|d^{(4)}_{z^2}\rangle-\frac{1}{6}\langle p^{(1)}_{5z}|y|d^{(4)}_{z^2}\rangle\Biggr]
\Biggr\}\\
-\hat{z}&\Biggl\{\frac{2t\Delta\theta}{\epsilon_p-\epsilon_d}\frac{5}{6}\Biggl[-\frac{1}{4}\Biggl(\langle p^{(2)}_{6z}|z|d^{(1)}_{z^2}\rangle-\langle p^{(3)}_{5z}|z|d^{(1)}_{z^2}\rangle\Biggr)+\frac{\sqrt{3}}{4}\Biggl(1-\frac{B(\lambda_1-\lambda_2)}{(\epsilon_p-\epsilon_d)}\Biggr)\Biggl(\langle p^{(2)}_{6z}|z|d^{(1)}_{x^2-y^2}\rangle+\langle p^{(3)}_{5z}|z|d^{(1)}_{x^2-y^2}\rangle\Biggr)\Biggr]\\
&+\frac{1}{2}\frac{2t\Delta\theta}{U-3J-\epsilon_p+\epsilon_d}\Biggl[\frac{1}{12}\langle p^{(1)}_{1y}|z|d^{(2)}_{z^2}\rangle+\frac{5}{24}\langle p^{(1)}_{4x}|z|d^{(2)}_{z^2}\rangle+\frac{1}{6}\langle p^{(1)}_{5z}|z|d^{(2)}_{z^2}\rangle-\frac{1}{6}\langle p^{(1)}_{6z}|z|d^{(2)}_{z^2}\rangle\\
&\ \ \ \ \ \ \ \ \ \ \ \ \ -\frac{5}{24}\langle p^{(1)}_{1y}|z|d^{(3)}_{z^2}\rangle+\frac{1}{12}\langle p^{(1)}_{2x}|z|d^{(3)}_{z^2}\rangle+\frac{1}{6}\langle p^{(1)}_{5z}|z|d^{(3)}_{z^2}\rangle-\frac{1}{6}\langle p^{(1)}_{6z}|z|d^{(3)}_{z^2}\rangle\\
&\ \ \ \ +\Biggl(1+\frac{B(\lambda_1-\lambda_2)}{U-3J-\epsilon_p+\epsilon_d}\Biggr)\Biggl(\frac{1}{4\sqrt{3}}\langle p^{(1)}_{1y}|z|d^{(2)}_{x^2-y^2}\rangle-\frac{1}{8\sqrt{3}} \langle p^{(1)}_{4x}|z|d^{(2)}_{x^2-y^2}\rangle-\frac{1}{4\sqrt{3}} \langle p^{(1)}_{5z}|z|d^{(2)}_{x^2-y^2}\rangle+\frac{1}{4\sqrt{3}}\langle p^{(1)}_{6z}|x|d^{(2)}_{x^2-y^2}\rangle\\
&\ \ \ \ \ \ \ \ \ \ \ \ \ -\frac{1}{8\sqrt{3}}\langle p^{(1)}_{1y}|z|d^{(3)}_{x^2-y^2}\rangle-\frac{1}{4\sqrt{3}} \langle p^{(1)}_{2x}|z|d^{(3)}_{x^2-y^2}\rangle+\frac{1}{4\sqrt{3}} \langle p^{(1)}_{5z}|z|d^{(3)}_{x^2-y^2}\rangle-\frac{1}{4\sqrt{3}}\langle p^{(1)}_{6z}|z|d^{(3)}_{x^2-y^2}\rangle\Biggr)\Biggr]\Biggr\} .
\end{aligned}$%
}

Introducing a proper change for the coordinate x, y, and z (e.g. $x\rightarrow y$, $y\rightarrow -x$, and etc.), we obtain

\begin{equation}
\small
\begin{split}
    \vec{P}_1=&-\frac{t\Delta\theta}{6}\frac{\vec{p}_1}{\epsilon_p-\epsilon_d}-t\Delta\theta\Biggl\{\frac{1}{U-J-\epsilon_p+\epsilon_d}-\frac{1}{2}\frac{1}{U-3J-\epsilon_p+\epsilon_d}\Biggr\}\vec{p}_2+\frac{\sqrt{3}t\Delta\theta B(\lambda_1-\lambda_2)}{2(\epsilon_p-\epsilon_d)^2}\vec{p}_3\\
    &-\frac{t\Delta\theta B(\lambda_1-\lambda_2)}{(U-3J-\epsilon_p+\epsilon_d)^2}\vec{p}_4-\frac{2t\Delta\theta B(\lambda_1-\lambda_2)}{(U-J-\epsilon_p+\epsilon_d)^2}\vec{p}_5 ,
\label{seq:p1}
\end{split}
\end{equation}
where
\begin{equation}
\vec{p}_1=-\langle p^{(2)}_{4x}|x|d^{(1)}_{z^2}\rangle\left(\hat{x}-\hat{y}\right),
\end{equation}
\begin{equation}
\resizebox{\textwidth}{!}{$%
\begin{aligned}
\vec{p}_2=&-2\Biggl(\frac{1}{12}\langle p^{(1)}_{1y}|x|d^{(2)}_{z^2}\rangle-\frac{5}{24}\langle p^{(1)}_{2x}|x|d^{(2)}_{z^2}\rangle+\frac{5}{24}\langle p^{(1)}|x|d^{(2)}_{z^2}\rangle+\frac{1}{4\sqrt{3}}\langle p^{(1)}_{1y}|x|d^{(2)}_{x^2-y^2}\rangle+\frac{1}{8\sqrt{3}}\langle p^{(1)}_{2x}|x|d^{(2)}_{x^2-y^2}\rangle\\
&\ \ \ \ \ \ \ \ \ \ -\frac{1}{8\sqrt{3}}\langle p^{(1)}_{4x}|x|d^{(2)}_{x^2-y^2}\rangle-\frac{1}{6}\langle p^{(1)}_{6z}|x|d^{(2)}_{z^2}\rangle+\frac{1}{4\sqrt{3}}\langle p^{(1)}_{6z}|x|d^{(2)}_{x^2-y^2}\rangle\Biggr)(\hat{x}-\hat{y}) ,
\end{aligned}$%
}
\end{equation}
\begin{equation}
\vec{p}_3=\langle p^{(4)}_{2x}|x|d^{(1)}_{x^2-y^2}\rangle(\hat{x}-\hat{y}) ,
\end{equation}

\begin{equation}
\resizebox{\textwidth}{!}{$%
\begin{aligned}
\vec{p}_4=&\Biggl(\frac{1}{4\sqrt{3}}\langle p^{(1)}_{1y}|x|d^{(2)}_{x^2-y^2}\rangle+\frac{1}{8\sqrt{3}}\langle p^{(1)}_{2x}|x|d^{(2)}_{x^2-y^2}\rangle-\frac{1}{8\sqrt{3}}\langle p^{(1)}_{4x}|x|d^{(2)}_{x^2-y^2}\rangle+\frac{1}{4\sqrt{3}}\langle p^{(1)}_{6z}|x|d^{(2)}_{x^2-y^2}\rangle\Biggr)(\hat{x}-\hat{y}) ,
\end{aligned}$%
}
\end{equation}
and
\begin{equation}
\begin{split}
\vec{p}_5=&\Biggl(\frac{\sqrt{3}}{8}\langle p^{(1)}_{2x}|x|d^{(4)}_{x^2-y^2}\rangle+\frac{\sqrt{3}}{8}\langle p^{(1)}_{3y}|x|d^{(4)}_{x^2-y^2}\rangle-\frac{\sqrt{3}}{8}\langle p^{(1)}_{4x}|x|d^{(4)}_{x^2-y^2}\rangle\Biggr)(\hat{x}-\hat{y}) .
\end{split}
\end{equation}
\noindent
Accordingly, we consider expectation values of $\vec{r}_2=1_{2\times 2}\otimes\vec{r}\otimes\ 1_{2\times 2}\otimes 1_{2\times 2}$, $\vec{r}_3=1_{2\times 2}\otimes 1_{2\times 2}\otimes\vec{r}\otimes 1_{2\times 2}$, and $\vec{r}_4=1_{2\times 2}\otimes 1_{2\times 2}\otimes 1_{2\times 2}\otimes\vec{r}$, which result in

\begin{equation}
\resizebox{\textwidth}{!}{$%
\begin{aligned}
    \vec{P}_2+\vec{P}_3+\vec{P}_4=&\frac{t\Delta\theta}{6}\frac{\vec{p}_1}{\epsilon_p-\epsilon_d}+t\Delta\theta\Biggl\{\frac{1}{U-J-\epsilon_p+\epsilon_d}-\frac{1}{2}\frac{1}{U-3J-\epsilon_p+\epsilon_d}\Biggr\}\vec{p}_2+\frac{\sqrt{3}t\Delta\theta B(\lambda_1-\lambda_2)}{2(\epsilon_p-\epsilon_d)^2}\vec{p}_3\\
    &+\frac{t\Delta\theta B(\lambda_1-\lambda_2)}{(U-3J-\epsilon_p+\epsilon_d)^2}\vec{p}_4-\frac{2t\Delta\theta B(\lambda_1-\lambda_2)}{(U-J-\epsilon_p+\epsilon_d)^2}\vec{p}_5 .
\label{seq:p234}
\end{aligned}$%
}
\end{equation}
\noindent
Finally, we obtain the electric polarization of the center NiS\textsubscript{6} unit as
\begin{equation}
    \vec{P}=\frac{\sqrt{3}t\Delta\theta B(\lambda_1-\lambda_2)}{(\epsilon_p-\epsilon_d)^2}\vec{p}_3-\frac{4t\Delta\theta B(\lambda_1-\lambda_2)}{(U-J-\epsilon_p+\epsilon_d)^2}\vec{p}_5 .
\label{seq:lep}
\end{equation}

To get a specific value of the polarization given by (\ref{seq:lep}), we calculate $\vec{p}_3$ and $\vec{p}_5$. First, we introduce the d- and p-orbital wavefunctions to compute $\vec{p}_3$,
\begin{equation}
\psi_{3d^{(1)}_{x^2-y^2}}=\frac{1}{81\sqrt{2\pi}}\Biggl(\frac{Z_{Ni}}{a_0}\Biggr)^{\frac{7}{2}}r^2 e^{-\frac{Z_{Ni}r}{3a_0}}\frac{x^2-y^2}{r^2},
\label{seq:3d}
\end{equation}

\begin{equation}
\resizebox{\textwidth}{!}{$%
\begin{aligned}
\psi_{3p^{(4)}_{2x}}=&\frac{\sqrt{2}}{81\sqrt{\pi}}\Biggl(\frac{Z_{S}}{a_0}\Biggr)^{\frac{3}{2}}e^{-\frac{Z_{S}\sqrt{x^2+(y-a)^2+z^2}}{3a_0}}\Biggl[6\Biggl(\frac{Z_{S}\sqrt{x^2+(y-a)^2+z^2}}{a_0}\Biggr)-\Biggl(\frac{Z_{S}\sqrt{x^2+(y-a)^2+z^2}}{a_0}\Biggr)^2\Biggr]\\
&\times\frac{x}{\sqrt{x^2+(y-a)^2+z^2}}.
\end{aligned}$%
}
\label{seq:3p}
\end{equation}
Here, $Z_S$ and $Z_{Ni}$ are the atomic number of Sulfur and Nickel, respectively. $a_0$ is the Bohr radius. $a=2.5\times 10^{-10}\ m$ is the distance between Ni and S \cite{gu2019ni}. Then, we obtain
\begin{equation}
\vec{p}_3=4.13\times 10^{-11}(-\hat{x}+\hat{y})[m].
\label{seq:cp3}
\end{equation}
With the same procedure, we obtain
\begin{equation}
\vec{p}_5=0.
\label{seq:cp5}
\end{equation}


\section*{References}

\begin{thebibliography}{9}
\bibitem{Zhang_Rice_Singlet_CuO} F. C. Zhang and T. M. Rice, Phys. Rev. B {\bf 37}, 3759(R) (1988).
\bibitem{Kitaev-type_Models_IrO} G. Jackeli and G. Khaliullin, Phys. Rev. Lett. {\bf 102}, 017205 (2009).
\bibitem{Spin_Current_Polarization} Hosho Katsura, Naoto Nagaosa, and Alexander V. Balatsky, Phys. Rev. Lett. {\bf 95}, 057205 (2005).
\bibitem{Zhang_Rice_Singlet_CuO_Observation} L. H. Tjeng, B. Sinkovic, N. B. Brookes, J. B. Goedkoop, R. Hesper, E. Pellegrin, F. M. F. de Groot, S. Altieri, S. L. Hulbert, E. Shekel, and G. A. Sawatzky, Phys. Rev. Lett. {\bf 78}, 1126 (1997).

\bibitem{brec1986review} Raymond Brec, \textit{Intercalation in Layered Materials}, Springer, C\'edex, 1986.
\bibitem{grasso1988electronic} V. Grasso, F. Neri, and S. Santangelo, and L. Silipigni, Phys. Rev. B {\bf 37}, 4419 (1988).
\bibitem{foot1980optical} P. J. S. Foot, J. Suradi, and P. A. Lee., Material Research Bulletin {\bf 15}, 189 (1980).
\bibitem{kurita1989band1} Noriyuki Kurita and Kenji Nakao, Journal of the Physical Society of Japan {\bf 58}, 610 (1989).
\bibitem{kurita1989band2} Noriyuki Kurita and Kenji Nakao, Journal of the Physical Society of Japan {\bf 58}, 232 (1989).

\bibitem{NiPS3_Nature} Soonmin Kang, Kangwon Kim, Beom Hyun Kim, Jonghyeon Kim, Kyung Ik Sim, Jae-Ung Lee, Sungmin Lee, Kisoo Park, Seokhwan Yun, Taehun Kim, Abhishek Nag, Andrew Walters, Mirian Garcia-Fernandez, Jiemin Li, Laurent Chapon, Ke-Jin Zhou, Young-Woo Son, Jae Hoon Kim, Hyeonsik Cheong, and Je-Geun Park, Nature {\bf 583}, 785 (2020).

\bibitem{gu2019ni} Yuhao Gu, Qiang Zhang, Congcong Le, Yinxiang Li, Tao Xiang, and Jiangping Hu, Phys. Rev. B {\bf 100}, 165405 (2019).

\bibitem{CoPs3} T. Matsuoka, R. Rao, M. A. Susner, B. S. Conner, D. Zhang, and D. Mandrus, Phys. Rev. B {\bf 107}, 165125 (2023).

\bibitem{coleman2011two} Jonathan N. Coleman, Mustafa Lotya, Arlene O Neill, Shane D. Bergin, Paul J. King, Umar Khan, Karen Young, Alexandre Gaucher, Sukanta De, Ronan J. Smith, and Igor V. Shvets, Science {\bf 331}, 568 (2011).
\bibitem{kuo2016exfoliation} Cheng-Tai Kuo, Michael Neumann, Karuppannan Balamurugan, Hyun Ju Park, Soonmin Kang, Hung Wei Shiu, Jin-Hyoun Kang, Byung Hee Hong, Moonsup Han, Tae Won Noh, and Je-Geun Park, Scientific reports {\bf 6}, 1 (2016).

\bibitem{sivadas2015magnetic} Nikhil Sivadas, Matthew W. Daniels, Robert H. Swendsen, Satoshi Okamoto, and Di Xiao, Phys. Rev. B {\bf 91}, 235425 (2015).
\bibitem{wildes2015magnetic} A. R. Wildes, V. Simonet, E. Ressouche, G. J. Mcintyre, M. Avdeev, E. Suard, S. A. J. Kimber, D. Lan{\c{c}}on, G. Pepe, B. Moubaraki, and T. J. Hicks, Phys. Rev. B {\bf 92}, 224408 (2015).

\bibitem{chittari2016electronic} Bheema Lingam Chittari, Youngju Park, Dongkyu Lee, Moonsup Han, Allan H. MacDonald, Euyheon Hwang, and Jeil Jung, Phys. Rev. B {\bf 94}, 184428 (2016).

\bibitem{kim2018charge} So Yeun Kim, Tae Yun Kim, Luke J. Sandilands, Soobin Sinn, Min-Cheol Lee, Jaeseok Son, Sungmin Lee, Ki-Young Choi, Wondong Kim, Byeong-Gyu Park, C. Jeon, Hyeong-Do Kim, Cheol-Hwan Park, Je-Geun Park, S. J. Moon, and T. W. Noh, Phys. Rev. Lett. {\bf 120}, 136402 (2018).

\bibitem{georges2013strong} Antoine Georges, Luca de' Medici, and Jernej Mravlje, Annu. Rev. Condens. Matter Phys. {\bf 4}, 137 (2013).

\bibitem{slater1954simplified} John C. Slater and George F. Koster, Phys. Rev. {\bf 94}, 1498 (1954).

\bibitem{stamokostas2018mixing} Georgois L. Stamokostas and Greory A. Fiete, Phys. Rev. B {\bf 97}, 085150 (2018).


\end{thebibliography}
\end{document}